\documentclass{emulateapj}
\slugcomment{{\sc Accepted to ApJ} }

\usepackage{longtable}
\usepackage[percent]{overpic}
\usepackage{color}

\newcommand{\Msun}[0]{\mbox{ M}_\odot}

\newcommand{\etal}[0]{et al.}

\newcommand{\AMRNC}[0]{{\sf AMR-NC}}
\newcommand{\AMRFID}[0]{{\sf AMR-FID}}
\newcommand{\AMRQSO}[0]{{\sf AMR-QSO}}
\newcommand{\SPHNC}[0]{{\sf SPH-NC}}
\newcommand{\SPHFID}[0]{{\sf SPH-FID}}
\newcommand{\SPHQSO}[0]{{\sf SPH-QSO}}
\newcommand{\NC}[0]{{\sf NC}}
\newcommand{\FID}[0]{{\sf FID}}
\newcommand{\QSO}[0]{{\sf QSO}}

\newcommand{\cc}[0]{\mbox{ cm}^{-3}}
\newcommand{\kevcs}[0]{keV cm$^2$}
\newcommand{\gcc}[0]{\mbox{ g}\mbox{ cm}^{-3}}
\newcommand{\eqn}[1]{Equation~(\ref{#1})}	
\newcommand{\fig}[1]{Figure~\ref{#1}}		
\newcommand{\sect}[1]{\S\ref{#1}}			
\newcommand{\tabl}[1]{Table~\ref{#1}}		

\newcommand{\hydra}[0]{{\tt HYDRA}}
\newcommand{\ramses}[0]{{\tt RAMSES}}

\definecolor{grey}{rgb}{0.75,0.75,0.75}
\definecolor{Orange}{rgb}{1.0,0.5,0.15}
\definecolor{brown}{rgb}{0.7,0.25,0.0}
\definecolor{pink}{rgb}{1.0,0.5,0.5}
\definecolor{darkerred}{rgb}{0.8,0,0}
\definecolor{darkerblue}{rgb}{0,0,0.8}
\definecolor{Blue}{rgb}{0,0.08,0.65}
\definecolor{Red}{rgb}{0.65,0.08,0.05}
\definecolor{Green}{rgb}{0.15,0.45,0.25}

\begin{document}

\title{Comparing Simulations of AGN Feedback}
\author{Mark L. A. Richardson\altaffilmark{1,2}, Evan Scannapieco\altaffilmark{2}, Julien Devriendt\altaffilmark{1,3}, 
Adrianne Slyz\altaffilmark{1}, Robert J. Thacker\altaffilmark{4}, Yohan Dubois\altaffilmark{5,6}, James Wurster\altaffilmark{4,7}, 
Joseph Silk\altaffilmark{1,5,6,8}}
\altaffiltext{1}{Sub-department of Astrophysics, University of Oxford, Keble Road, Oxford, OX1 3RH}
\altaffiltext{2}{School of Earth and Space Exploration, Arizona State University, Tempe, AZ 85287}
\altaffiltext{3}{Observatoire de Lyon, UMR 5574, 9 avenue Charles Andr\'{e}, F-69561 Saint Genis Laval, France}
\altaffiltext{4}{Department of Astronomy \& Physics, Saint Mary's University, Halifax, NS, B3L 3C3, Canada}
\altaffiltext{5}{Sorbonne Universit\'{e}s, UPMC Univ Paris 06, UMR 7095, Institut d'Astrophysique de Paris, F-75014, Paris, France}
\altaffiltext{6}{CNRS, UMR 7095, Institut d'Astrophysique de Paris, F-75014, Paris, France}
\altaffiltext{7}{Monash Centre for Astrophysics and School of Physics and Astronomy, Monash University, Victoria, 3800, Australia}
\altaffiltext{8}{Department of Physics and Astronomy, The Johns Hopkins University Homewood Campus, Baltimore, MD 21218, USA}
\setcounter{footnote}{8}

\begin{abstract}
We perform adaptive mesh refinement (AMR) and smoothed particle hydrodynamics (SPH) cosmological zoom simulations of a region around a forming galaxy cluster, comparing the ability of the methods to handle successively more complex baryonic physics.  In the simplest, non-radiative case, the two methods are in good agreement with each other, but the SPH simulations generate central cores with slightly lower entropies and virial shocks at slightly larger radii, consistent with what has been seen in previous studies.  The inclusion of radiative cooling, star formation, and stellar feedback leads to much larger differences between the two methods.  Most dramatically, at $z=5,$ rapid cooling in the AMR case moves the accretion shock well within the virial radius,  while this shock remains near the virial radius in the SPH case, due to excess heating, coupled with poorer capturing of the shock width.   On the other hand, the addition of feedback from active galactic nuclei (AGN) to the simulations results in much better agreement between the methods. In this case both simulations display halo gas entropies of 100 keV cm$^2$, similar decrements in the star-formation rate, and a drop in the halo baryon content of roughly 30\%. This is consistent with AGN growth being self-regulated, regardless of the numerical method. However, the simulations with AGN feedback continue to differ in aspects that are not self-regulated, such that in SPH a larger volume of gas is impacted by feedback, and the cluster still has a lower entropy central core.
\end{abstract}
\maketitle

\section{Introduction}

In the hierarchical cold dark matter and  dark energy ($\Lambda$CDM) model for galaxy  formation, matter condenses into small clumps that then merge to create increasingly massive objects over time. This model has provided several predictions that are in excellent agreement with observations (e.g., Spergel et al. 2007; Larson et al. 2011). For star formation to begin in $\Lambda$CDM, the temperature of gas contained within condensed dark matter `halos' must cool sufficiently to allow the formation of galaxies. Because larger galaxies have more gravitational compression, and hence a higher temperature, they might be expected to take longer to cool and form stars, with the largest galaxies only now reaching significant  star formation rates (SFRs).

Observational studies of star formation rates, on the other hand, have found several surprising trends: the cosmic star formation rate density reaches a peak at $z\simeq2$, 5-3 billion years after the Big Bang and then decreases until today (Hopkins \& Beacom, 2006; Karim et al. 2011), SFRs peaked earlier in more massive galaxies and more recently in smaller galaxies (e.g. Guzman et al. 1997; Brinchmann \& Ellis 2000), and today SFRs are lower in more massive galaxies (Heavens et al. 2004; Panter et al. 2007). These trends, known collectively as `downsizing' (e.g. Cowie et al. 1996; Bauer et al. 2005; Panter et al. 2007; Karim 2011),  are clearly at odds with the naive predictions of the $\Lambda$CDM model.

A similar such discrepancy involves the properties of galaxy clusters, as constrained by  high-resolution X-ray and radio observations such as those from the Chandra Observatory and the Very Large Array. While many clusters appear to be quiescent, about a third show strong peaks in their central X-ray surface brightness distributions, indicating that their gas is cooling  rapidly (e.g. Fabian \& Nulsen 1977; Nulsen et al. 1982; Stewart et al. 1984; Fabian 1994; Tamura et al. 2001; Cavagnolo et al. 2009).  However, this cooling is neither  accompanied by strong star formation nor a significant fraction of gas  colder than 1 keV (e.g. Peterson et al. 2001; Rafferty et al. 2006; McNamara \& Nulsen 2007). Instead galaxy formation is halted by an unknown energy source (e.g.\ Croton et al.\ 2006).

A prominent theory is that energetic feedback from active galactic nuclei (AGN) is required to explain these two discrepancies (e.g., Scannapieco \& Oh 2004; Springel et al. 2005; Thacker et al. 2006; Dunn \& Fabian 2006; Sijacki et al. 2007; Booth \& Schaye 2009; Dubois et al. 2013; Martizzi et al. 2013). AGN are among the most energetic objects in the Universe, characterized by their extremely luminous cores powered by the infall of gas from a  relativistic accretion disk (e.g., Rees 1984) onto a supermassive black hole (SMBH) with mass $M_{\rm BH} > 10^6\Msun$.  AGN are associated with two modes of  feedback into their environments. The kinematic or radio mode is associated with collimated relativistic jets, and low and radiatively inefficient accretion rates (e.g., Falcke \& Biermann, 1999; Sambruna et al. 2000; Merloni \& Heinz 2007), while the quasar or wind mode is associated with isotropic energy deposit, and high and radiatively efficient accretion (e.g., Silk \& Rees 1998). Depending on the feedback mode, it has been shown that AGN can provide the energy needed to maintain the hot ICM (e.g., Dunn \& Fabian 2006), with kinematic feedback creating the large buoyant bubbles (e.g., Dunn et al. 2006).  It is also clear that this feedback can hamper cooling of galactic halo gas, preferentially reducing the SFR first in large halos at early times, and then smaller halos at late times (e.g., Scannapieco \& Oh 2004; Scannapieco et al. 2005). AGN also act to remove, via accretion and heating, the low specific angular momentum (sAM) gas from the central region of its halo, gas that would otherwise form stars. This leads to a net increase in the galactic sAM. However, AGN can also increase galactic sAM where feedback has heated gas and reduced the of possibly high angular momentum gas into the galactic disk (e.g., Dubois et al. 2013; Genel et al. 2015; Nelson et al. 2015), . 

Unfortunately, AGN feedback is extremely difficult to simulate as its effects span several orders of magnitude, originating on sub-parsec scales, and impacting kiloparsec and even  megaparsec scales.  Thus numerical methods must implement a subgrid prescriptions for injecting the feedback model if they are resolving cluster or cosmological scales. The nature of this  feedback is also highly debated, with different studies focusing on different input mechanisms for the feedback energy, and different environments in which to study its effects.

The first models of AGN feedback were limited to smoothed particle hydrodynamics (SPH) simulations.  Springel et al. (2005) used the SPH code {\tt GADGET-2} (Springel 2005) to study individual and merging galaxies with black holes (BHs) and feedback, injected into the simulation thermally and isotropically.  They found that the feedback energy was sufficient to regulate the BH growth. 
Thacker et al. (2006b; 2009) carried out SPH simulations using the {\tt HYDRA} (Couchman et al. 1995; Thacker \& Couchman 2006) code  with an isotropic kinematic outflow model for AGN feedback. Their results reproduced the antihierarchical turnoff in the quasar luminosity function, as well as 
the spatial distribution of quasars on both small and large scales.  However, the impact of feedback was significantly less than 
 predicted by analogous semianalytic models, a difference that could be traced to in-shock cooling as occurred in they SPH simulation.
 Sijacki et al. (2007) introduced a dual mode prescription for their AGN simulations in {\tt GADGET-2}, resulting in better agreement between the simulated galaxy stellar mass density and observations than previous models. Booth \& Schaye (2009) performed SPH simulations using  the {\tt GADGET III} code with a modified version of the AGN prescription of Springel et al. (2005), and found that the BHs  greatly suppressed the star formation in high-mass galaxies, self-regulating their feedback such that they agreed with the $M_{\rm BH}$-$\sigma_{\rm v}$ relation,  a tight correlation between BH mass,  $M_{\rm BH}$ and the velocity dispersion of the host galaxy's bulge, $\sigma_{\rm v}$ (e.g., Ferrarese  \& Merritt 2000; Gebhardt et al. 2000; Tremaine et al. 2002).
Recently, Planelles et al. (2014), Le Brun et al. (2014), and Pike et al. (2014) used variants of {\tt GADGET III} and {\tt GADGET-2} to study the impact of AGN on a range of galaxy masses and clusters, and how AGN models can be tuned to produce better agreement between simulated and observed galaxy groups and clusters.  Finally, Barai et al. (2013) and Wurster \& Thacker (2013) each used SPH simulations ({\tt GADGET III} and {\tt HYDRA}, respectively) to consider the impact of  various AGN feedback models on isolated galaxies and mergers, showing some success in replicating the $M_{\rm BH}$-$\sigma_{\rm v}$ relation.

All such SPH simulations, which use Lagrangian schemes to solve the equations of fluid dynamics, are very efficient at resolving dense structures. However, SPH is not without its shortcomings.  As particles moves along with the mass, SPH is ill-equipped to resolve the low-density environment surrounding galaxies and clusters, although it is through this medium that the feedback interacts with the surrounding structure. Furthermore, traditional, also called standard, SPH has difficulties accurately modeling shocks and mixing (e.g., Morris 1996; Marri \& White 2003; Agertz et al. 2007; Hopkins 2013), which are essential when studying the impact of feedback on surrounding structure. Fortunately, there is on-going effort and success in reformulating the SPH method to overcome the mixing and shock issues discussed above (e.g., Price 2012; Hopkins 2013), although they have not yet become the standard practice.

Thus, recently adaptive mesh refinement (AMR) methods are gaining interest, as AMR is a shock-capturing method, it can set high spatial resolution in any region of the simulation volume and resolves shocks with only a few cells. Dubois et al. (2010) introduced kinematic radio-mode feedback in {\tt RAMSES} simulations, and then in Dubois et al.\ (2013) used both radio and quasar models of AGN feedback in the AMR code {\tt RAMSES} (Teyssier et al. 2002) to study the growth of a galaxy cluster in a full cosmological simulation. Their study focused on the accretion history of the cluster SMBH and the effect of feedback on the gas content and temperature. They found that only with AGN feedback were they able to greatly heat the gas and affect its ability to accrete onto the central galaxy, thus limiting the overall SFR and accretion on to the SMBH, while being in good agreement with the empirical  $M_{\rm BH}$-$\sigma_{\rm v}$ relation. Martizzi et al. (2013) also used {\tt RAMSES} to simulate an isolated cluster halo, studying how quickly cluster gas heated by AGN feedback cools back into the central region. 

Aside from using different hydrodynamical methods, these collection of studies are quite diverse in how they model the deposited feedback energy and its impact on the environment.  Thus the role of the simulation method in determining the conclusions of these studies is difficult to disentangle in the absence of a comparison that employs the same physical model across different simulation codes. The importance of  comparisons between simulation techniques has also been highlighted by the usefulness of such studies in models of structure formation without AGN feedback. For example, the Santa Barbara Cluster Comparison Project \citep{Frenk99} compared the results of twelve numerical codes using the same initial conditions to study the virialization of a massive galaxy cluster, without including feedback or radiative cooling, finding that agreement was best for properties of the dark matter and worst for the total X-ray luminosity.
Similar comparisons with higher resolution were done in Voit et al. (2005) and then Mitchell et al. (2009), focusing on the central entropy profiles of clusters, and they showed that the core entropy was lower in SPH than in AMR  due to under-mixing in SPH, and to a lesser degree, over-mixing in AMR. A detailed computational comparison between AMR and SPH methods for standard hydrodynamic turbulence was performed by \citet{Agertz07}, which demonstrated the inherent difficulties of modeling shear layers with standard SPH. This work also highlighted the effect of steep density gradients in standard SPH simulations, where an effective surface tension between the two phases would lead to limited mixing, overcooling, and angular momentum transport (Kaufmann et al. 2007).

Code comparisons continue to be important. For the Aquila project, \citet{Scannapieco12} studied the formation of a Milky Way-size galaxy, wherein 13 different numerical methods were used, starting from the same initial conditions, with no attempt to use identical sub-grid models (gas cooling and the formation and feedback of stars and AGN). They found that while the variety of different gas physics led to a large span in the physical characteristics of the final galaxy, it was inconsistent with observations of real galaxies. They also found that gas cooled more efficiently in grid codes, leading to higher star formation rates. In contrast to the Aquila project, the now underway AGORA project is combining the works of 95 scientists to use a variety of numerical codes with as identical as possible implementations of various baryonic physics to produce more observationally consistent galaxies (Kim et al. 2014). There has been a suite of work comparing galactic and cosmological simulations from the recently introduced moving-mesh code {\tt AREPO}, with the SPH code {\tt GADGET III} (e.g., Springel 2010; Bauer \& Springel 2012; Kere\v{s} et al. 2012; Sijacki et al. 2012; Torrey et al. 2012a; Vogelsberger et al. 2012; Nelson et al. 2013), which only recently included AGN feedback (Hayward et al. 2014). These have further demonstrated where shortcomings exist in the ability of SPH codes to mix merging material. In particular, Kere\v{s} et al. (2012) found that with radiative gas cooling {\tt AREPO} had either the same or lower central entropy profiles depending on the halo mass, which is opposite the results of grid codes in non-radiative simulations (e.g., Voit et al. 2005; Mitchell et al. 2009). Finally, the nIFTy Cosmology workshop has also lead to simulation comparison papers that include AMR and SPH codes, and the moving mesh code {\tt AREPO} (Sembolini et al. 2015b, 2015a). For non-radiative simulations they reproduced the radial profile results of Mitchell et al. (2009). By introducing cooling and AGN feedback, at $z=0$ they see better qualitative agreement in the halo profiles, but with larger scatter.

In this work, we wish to continue the effort to compare the results from different codes as they simulate the cosmologically consistent formation of a cluster environment, including AGN feedback. We perform two simulation suites, one with  AMR and one with standard SPH, from the same initial conditions, studying the impact of different subgrid physics, including cooling, star formation, stellar and AGN feedback models. Note that we attempt to implement nearly identical sub-grid baryonic physics models, using the same parameter values in these models to emphasize the role of the numerical method. We compare the ability of these two numerical methods to model the evolution of the cluster environment and its response to these subgrid models, including the gas temperature, SFRs, and gas content. We stress that these results are only applicable to standard SPH implementations, and future work comparing with non-standard SPH implementations is required. In a companion paper (Richardson et al. 2016 in prep) we compare the AMR and  SPH impact of AGN feedback on the characteristics of halos ranging from 10$^{11}$ to 10$^{13.5} \Msun$.

The structure of this paper is as follows. In \sect{methods} we discuss the cooling, star formation and stellar feedback, and AGN formation, accretion, merging and feedback methods in both the grid and particle codes. In \sect{results} we first give a detailed comparison of the results from our non-cooling and our fiducial simulations, where no AGN feedback is included. These constitute the backbone of our analysis, as it is only with respect to these non-AGN runs that we can determine how AGN affects the cluster growth in AMR and SPH simulations. We then present the results from our simulations with AGN feedback. We give a discussion and conclude in \sect{concl5}.

\section{Numerical Methods}\label{methods}

Simulations were conducted with either the AMR code {\tt RAMSES} \citep{Teyssier02} or the SPH code {\tt HYDRA} \citep{Couchman95}.
The initial conditions were generated using the {\tt mpgrafic} \citep{Prunet08} package which creates a realization of the density fluctuations on a grid according to the desired power spectrum, and uses the Zel'dovich approximation to calculate the corresponding particle velocities. Our initial conditions were generated at a redshift of $z=43.2$, centered on a region in which a cluster halo with virial mass $M_{\rm vir} = 2\times10^{15} \Msun$  forms by $z=0$. We assumed a $\Lambda$CDM cosmology with cosmological parameters ($\Omega_{\Delta}$, $\Omega_{\rm M}$, $\Omega_{\rm b}$, $\sigma_8$, $h$) = (0.73, 0.27, 0.044, 0.8, 0.7) from the 7-year \textit{WMAP} \citep{Komatsu11}.  Our simulations were carried out in a 100 $h^{-1}$ Mpc comoving box with periodic boundaries and were run to $z=3$. In both particle and grid simulations we assumed a zoomed-in realization of this box, with a spherical high-resolution region 25 $h^{-1}$ Mpc in diameter, an effective dark matter particle number of 1024$^3$, and a mass resolution of $8.3 \times 10^7 \Msun.$  This high resolution region was selected to contain all particles found in the halo by $z=1$, and thus constitutes a conservative estimate of the necessary high-resolution region for $z=3$.
Outside of the high resolution region, we uniformly decreased the dark matter particle resolution until reaching an effective particle number of 64$^3$, for AMR, and 128$^3$, for SPH, in the outer regions of the box.  The dark matter particle initial conditions were identical between the grid and particle simulations. 

We carried out the grid simulations with the {\tt RAMSES} code \citep{Teyssier02}, which uses an unsplit second-order Godunov scheme for evolving the Euler equations for the gas. \ramses\ variables are  cell-centered and interpolated to the cell faces for flux calculations, which are then used with a Harten-Lax-van Leer-Contact Riemann solver (van Leer 1979; Einfeldt 1988). 
For calculating the gravitational potential and accelerations, collisionless star, black hole, and dark matter particles had their mass mapped to the grid with a Cloud-in-Cell (CIC) scheme (Birdsall \& Fuss, 1997). The CIC method was also used for comparing gas and particle densities for star formation and sink particle generation, discussed further below. Gas within the high-resolution region was refined using a semi-Lagrangian technique. When more than 8 dark matter particles were in a cell, or when the baryon density in a cell was 8 times more than the cosmic average, the cell was split into 8, doubling the spatial resolution. We aimed for a fixed maximum physical resolution of $\Delta x_{\rm min} = 545$ pc, where the maximum refinement level was increased with increasing cosmic scale factor, with increments occurring at $z\simeq$ 39, 19, 9, and 4. Thus the spatial resolution at any one time varied from 435 physical pc to 870 physical pc. To avoid over-resolving the dark matter in dense gas regions, we set the maximum level to map the dark matter particles into cells at $l_{\rm max,DM}=15$. For comparison, at $z=4$, the gas was refined up to $l_{\rm max,g}=16$  levels of resolution. Thus from $z=4$ to 3 the dark matter had an effective softening length of 2 physical kpc, given by twice the width of grid cell at $l=15$.

We carried out the particle-based simulations with the  {\tt HYDRA} code (Couchman et al. 1995; Thacker \& Couchman 2006), which uses an adaptive particle-particle, particle-mesh method \citep{Couchman91} to calculate the gravitational forces and an SPH method (Gingold \& Marigold 1977; Lucy 1977) to calculate the hydrodynamic forces. This is a standard implementation of SPH without a forced conservation of entropy  \citep[e.g.][]{Springel05B}, and does not include methods for better capturing contact discontinuities (e.g. Hopkins et al. 2013). We used the same initial conditions for dark matter as for the {\tt RAMSES} run, and overlaid the gas particles onto the dark matter particle positions, which then trace the gas density field. Note, \hydra\ employs a modification of the kernel gradient to prevent the formation of the pair instability (Thomas \& Couchman 1992), and where pressure is important we confirm that gas and DM particles separate within a few time steps. {\tt HYDRA} uses the S2 gravitational softening length \citep{Hockney81}, $\epsilon$, for mediating close encounter scattering events, with the minimum smoothing length given by $h_{\rm min} = \epsilon/2$. We set  $h_{\rm min} = 2\Delta x_{\rm min} = 1090$ physical pc, giving the same softening length for dark matter particles in the two codes at late redshifts. Although the spatial resolution of a grid or particle method is not exactly equal to these two quantities, we found that the  star formation histories were sufficiently similar when relating these parameters in this way. We discuss this further in \sect{sf}. Each gas particle set its smoothing length such that it overlapped with roughly 52 neighbors, although at any given step this number could vary between 32 and 82.

Optically thin, atomic cooling from hydrogen, helium, and metals was calculated following \citet{Sutherland93} for temperatures above 10$^4$ K, and metal fine-structure cooling following \citet{Rosen95} at cooler temperatures. The gas temperature was not allowed to drop below $T_0 = 500$ K via radiative and metal line cooling,  although adiabatic cooling below this limit was permitted. After $z=8.5,$ heating from an ultraviolet background was modeled following \citet{Haardt96}. For simplicity and consistency between codes, the metallicity was set at a constant value of a third solar for the entirety of the simulation, as typically observed in the intracluster medium \citep{Loewenstein04}.  We note that a recently  identified error in the cooling prescription in \ramses\ shows that its metal cooling tables also include the contribution from hydrogen and helium, thus in regions where H and He are the main source of cooling (e.g., high temperature synchrotron tail) we were adding an extra third of cooling (since this extra term scales with the metallicity). However, for consistency, the same cooling tables were used in the \hydra\ simulation, thus this error does not account for any differences between the two methods.  
 
\subsection{Star Formation \& Stellar Feedback}\label{sf}

In both types of simulations, star formation followed a Schmidt-Kennicut law (Schmidt 1955; Kennicut 1998), with the star formation rate given as
\begin{equation}\label{eqsf}
\frac{dM_*}{dt} = c_{\rm sf} \frac{M_g}{t_{\rm ff}},
\end{equation}
where $c_{\rm sf}$ is the star forming efficiency, $M_g$ is the gas mass, and  $t_{\rm ff} \equiv \left(32 G \rho/3 \pi \right)^{1/2}$ is the gravitational free-fall time in the vicinity of the star-forming 
region, with $G$ the gravitational constant and $\rho$ the gas mass density in a given resolution element.
We set $c_{\rm sf} = 0.01,$ consistent
with observations of giant molecular clouds \citep{Krumholz07}. Star formation was implemented where the gas density was above a hydrogen number  density threshold, sufficient to overcome the local hydrodynamic pressure. For these resolutions, we set this threshold to be  $n_* = 0.05 $ hydrogen atoms cm$^{-3}$. We also incorporate a baryon overdensity threshold, comparing the local density of gas and stars to dark matter. This is necessary to prevent  star formation from proceeding in cosmic filaments at high redshift. For SPH, if the total baryon mass within two smoothing lengths of a gas particle was more than 0.25 that of the enclosed dark matter, then the particle was permitted to proceed with the star formation prescription discussed below. For AMR, we checked on a cell-by-cell case, and required a stricter threshold than SPH to ensure the gas in the smaller volume was indeed dominating the local mass.  We allowed this to vary with time since the dark matter softening transitioned to 1 kpc, and ensured agreement with the global star formation rate in the SPH simulation in the zoom-region of the box. This resulted in the gas overdensity threshold ranging from 0.25 to 20 times the dark matter mass in a given cell. We assumed that cold gas above the  density threshold  belonged to the multiphase interstellar medium (ISM), which we could not resolve. As such we employed a polytropic equation of state for such gas with $T = T_0 (n/n_*)^{\kappa -1}$,   with polytropic index $\kappa = 4/3$, and ISM temperature $T_0 = 500$ K. The density is also consistent with particle simulations of similar resolution (e.g., McCarthy et al. 2010; Scannapieco et al. 2012 and references therein; Sijacki et al. 2012; Hayward et al. 2014).

The implementation of star formation in {\tt RAMSES} is described in \citet{Rasera06}. We defined a unit stellar mass, $m_{*,R} = \Delta x_{\rm min}^3 n_* m_{\rm p} /X_{\rm H}$,  where $m_{\rm p}$ is the proton mass, and $X_{\rm H} = 0.76$ is the hydrogen mass fraction.  Thus the unit stellar mass is the mass in a cell at the threshold density, and is therefore the minimum stellar mass.  If a cell's gas density was above the star-formation threshold density and overdensity then we used \eqn{eqsf} with $M_g = \Delta x^3 \rho$ to determine the amount of stellar mass expected to be created in the next time step. Comparing this expected mass with the unit stellar mass yielded an expected number of stars to be formed, which was used as the expectation value for a random integer drawn from a Poisson distribution. A star particle was then generated with the same velocity as the cell and a mass equal to the random number times the unit stellar mass, and the cell's mass was reduced by the star particle's mass.

The implementation of star formation in {\tt HYDRA} is described in detail in \citet{Thacker00}.  Each gas particle  accumulated a stellar component following \eqn{eqsf}, where $M$ was that gas particle's mass, and $\rho$ was the local gas density at the particle. 
Once the accumulated mass was equal to the star particle mass, which we set to be half the high-resolution gas particle mass,
a star particle was formed, and the gas particle's mass was equivalently reduced. Until the time at which a star was formed, the gas dynamics used the total gas particle mass, not just the non-stellar component. The only exception to this was that the mass in \eqn{eqsf} only included the non-stellar component of the gas particle. Once a gas particle that already made one star particle had converted 80\% of its remaining mass into stars, then that entire gas particle was turned into a second star particle.

\begin{figure}[t!] 
\centering
\includegraphics[width={0.9\columnwidth}]{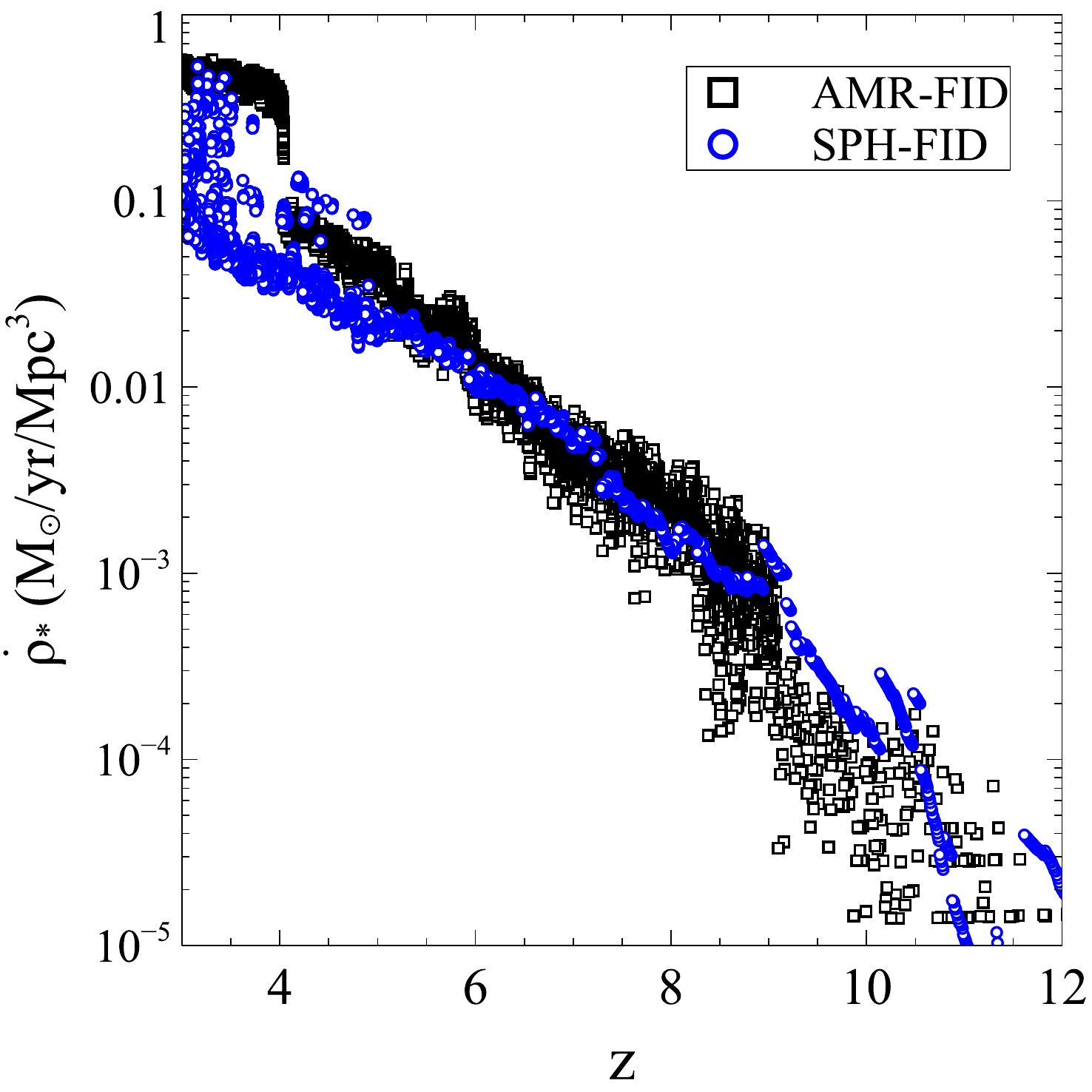}
\caption{Star formation density history up to $z=3$  in the zoom region, comparing \hydra\ and \ramses. 
Black squares show the global star formation rate density in the high-resolution zoom region for \ramses\, while blue circles show the results for \hydra. 
Note the delayed star formation for \ramses\ runs near $z=4$. The maximum refinement level increases at this redshift, followed by a strong increase in star formation as we better resolve the central
density peak of star-forming clumps. The best agreement is above $z \approx 5$ and at $z \approx 3$, when the resolutions were most comparable.}
\label{figcsf}
\end{figure}

In \fig{figcsf} we compare the history of star formation for the two codes in a 25 $h^{-1}$ comoving Mpc sphere centered on the region of interest.
The varying resolution of the \ramses\ runs is manifested by the
delay in star formation shortly before an increase in maximum refinement level, occurring at roughly $z=9$ and $z=4$. While for $h_{\rm min} = 2\Delta x$ there was fairly good agreement before using the overdensity threshold discussed above, the inclusion of this threshold results in stars forming only in galaxy cores.

We assumed $10\%$  of the stellar mass formed was contained in high-mass stars that contributed feedback through type II supernovae (SNe), with 10$^{51}$ erg of energy released for 10 $\Msun$ of high-mass stars (i.e. per 100 $\Msun$ of total stars). We thus used a total efficiency of $\epsilon_{\rm fb} = 5\times10^{15}$ erg g$^{-1}$ (Sommer-Larsen et al. 1999). This energy was deposited immediately into the vicinity of the formed star particle. Although the typical lifetime of a high-mass star was on order tens of time-steps, our single particle is representative of several giant molecular clouds, and we expect little issue to arise from injecting this energy without a delay at those kpc resolutions (see Wurster \& Thacker 2013). The implementation of the stellar feedback in {\tt RAMSES} is described in \citet{Dubois08}. This energy was deposited kinematically with a radius of a single cell around the star particle. The injected mass, momentum, and energy are consistent with a Sedov blast wave solution. The implementation of the stellar feedback in {\tt HYDRA} is described in \citet{Thacker00}. Since particle positions are not isotropic, we injected this energy thermally, using a kernel weighting for gas particles within the star particle's former smoothing 
length.  Post-feedback cooling uses a multiphase description of the gas density, which is much lower than the cold gas. In this model, pressure equilibrium is assumed, and the resulting decrease in density is determined by the net energy increase. In this way, the method is similar to other delayed cooling schemes (see for example Gerritsen \& Icke 1997). This multiphase density then gradually returns to the particle density, set by a half-life time of 1 Myr, consistent with the time for the blast wave to reach the cooling radius for densities near our star-formation threshold and feedback energy \citep{Blondin98}.

\subsection{Black Holes \& AGN Feedback}

Black holes (BHs) were modeled as sink particles in both types of simulations,  and they were formed where the local gas and stellar density were both above the star formation criteria, $n_* = 0.05 \cc$. All BHs had a seed mass of $8 \times 10^5 \Msun$,
and to ensure only one BH was made per galaxy, they were only allowed to form in locations at least 30 comoving kpc from all other BHs.  In \ramses,  this particle was given the same momentum as its cell of origin, and the cell's mass was reduced by the seed mass. In \hydra, the BHs were spawned at the same location and with the same momentum as the source gas particle, and the gas mass was reduced by the seed mass. Each BH particle in \hydra\ had a smoothing length that overlapped with roughly 60 neighboring gas particles. Black holes could merge when they were within 4 resolution units in AMR, or two smoothing lengths in SPH.  
\begin{deluxetable*}{l ccc ccc cc}[b!]
\tabletypesize{\scriptsize}
\tablecaption{Simulation Summary \label{tab_sims}}
\startdata
\multicolumn{1}{l}{\textbf{Runs}} &
\multicolumn{1}{c}{\textbf{Code}} &
\multicolumn{1}{c}{\textbf{Boxsize}$^1$} &
\multicolumn{1}{c}{$\mathbf{\Delta x}^2$} &
\multicolumn{1}{c}{$\mathbf{h_{\rm min}}^2$} &
\multicolumn{1}{c}{$\mathbf{M_{\rm DM}}^3$ } &
\multicolumn{1}{c}{$\mathbf{M_{\rm g}}^3$ } &
\multicolumn{1}{c}{$\mathbf{M_{*}}^3$} \\ \hline

{\sf AMR-} & \ramses & 100 & 550 & - & 80 & - & 0.2  \\
{\sf SPH-} & \hydra & 100 & - & 1100 & 80 & 16 & 8  \\ \hline

\multicolumn{1}{l}{\textbf{Runs}} &
\multicolumn{1}{c}{\textbf{Cool/Reion.}} &
\multicolumn{1}{c}{$\mathbf{n_{*}}$ (cm$^{-3}$)} &
\multicolumn{1}{c}{$\mathbf{c_{\rm sf}}$} &
\multicolumn{1}{c}{$\mathbf{\epsilon_{\rm fb}}$ (erg g$^{-1}$)} &
\multicolumn{1}{c}{$\mathbf{\epsilon_{\rm r}}$} &
\multicolumn{1}{c}{$\mathbf{\epsilon_{\rm f}}$} &
\multicolumn{1}{c}{$\mathbf{M_{\rm BH,s}}$ ($\Msun$)} \\ \hline

{\sf -NC}  & N & - & - & - & - & - & - \\  
{\sf -FID}  & Y & 0.05 & 0.1 & $5\times 10^{15}$  & - & - & - \\  
{\sf -QSO}  & Y & 0.05 & 0.1 & $5\times 10^{15}$  & 0.1 & 0.15 & $8\times 10^5$
\enddata
\tablenotetext{}{$^1$comoving Mpc h$^{-1}$, $^2$physical pc, $^3 10^6 \Msun$}
\end{deluxetable*}

The black holes accreted gas following the Bondi-Hoyle-Littleton rate \citep{Bondi52}, as described in \citet{Dubois13} and \citet{Wurster13}.  The accretion rate is given by
\begin{equation}\label{eqacc}
\frac{dM_{\rm BH}}{dt} = 4 \pi \alpha \frac{G^2 M^2_{\rm BH} \bar{\rho}}{(\bar{c}_{\rm s}^2 + \bar{u}^2)^{3/2}},
\end{equation} 
where $M_{\rm BH}$ is the black hole mass, $\bar{\rho}$ is the local average gas density, $\bar{c}_{\rm s}$ is the local average sound speed, $\bar{u}$ is the 
local average gas speed, and $\alpha$ is a dimensionless boost factor with $\alpha = $ max[1,$(n/n_*)^2$] \citep{Booth09}, which accounts for our inability to resolve
the cold high density ISM gas around the BH. We set the maximum accretion rate to be the Eddington accretion rate, 
\begin{equation}\label{eqedacc}
\frac{dM_{\rm Edd}}{dt} = 4 \pi \frac{G M_{\rm BH} m_{\rm p} }{\epsilon_{\rm r} \sigma_{\rm t} c},
\end{equation} 
where $\sigma_{\rm t}$ is the Thompson cross section, $c$ is the speed of light, and $\epsilon_{\rm r}$ is the radiative efficiency, set to 0.1 for the \citet{Shakura73} model of accretion onto a Schwarzschild BH.
In \ramses, each BH particle has a cloud of sensors that sample the surrounding gas to determine the average gas quantities. In \hydra\
we used a kernel-weighted average of the gas particles that overlap with the BHs smoothing length. Since mass is discretized in \hydra, we used an internal and a dynamic BH mass. The internal mass was incremented by the accreted amount, and once this mass was larger than the dynamical mass by half a gas particle mass,  the closest gas particle was accreted and the dynamical mass was increased by this amount. This dynamical mass was used to calculate the gravitational effect of the BH, while the internal mass was used to determine accretion and feedback properties.

The BH particles were advected similar to the dark matter particles. To model the effect of gas on the BHs, we included a drag force, thus avoiding spurious oscillations of the BHs about their local potential minimum. This dynamical friction is set to $F_{\rm DF} = f_{\rm gas} 4 \pi \alpha \rho (G M_{\rm BH}/\bar{c}_{\rm s})^2$, where $f_{\rm gas}$ is a factor whose value is depends on the local Mach number, with $0 < f_{\rm gas} < 2$ (Ostriker 1999; Chapon et al. 2013). In \hydra, the drag force is calculated using the smoothed gas values of the BH's neighboring particles.

In this work, we have employed the quasar-mode AGN feedback following \citet{Dubois13}. 
At every step the thermal quasar feedback mode injected energy into a sphere of radius $\Delta x$ centered on the BH, at an injection rate of $\dot{E}_{\rm AGN} = \epsilon_{\rm f} 
\epsilon_{\rm r} \dot{M}_{\rm BH} c^2$, where $\epsilon_{\rm f}$ is a free parameter set to 0.15 to reproduce the $M_{\rm BH} - M_{\rm b}$, $M_{\rm BH} - \sigma_{\rm b}$, and
BH density in the local universe (see Dubois et al. 2012). The implementation of this feedback mode in \hydra\ was built upon the existing work by \citet{Wurster13}, which heats gas particles within two smoothing lengths.
For both methods, the feedback was deposited with a uniform specific energy inside the bubble radius.

For each simulation code, we ran three simulations (see \tabl{tab_sims}). One simulation was carried out without cooling, reionization, star formation, or AGN, so as to best understand, from the ground up, how the two methods compare. This simulation is labelled \AMRNC\ and \SPHNC\ (for no cooling) in \ramses\ and \hydra, respectively. A second simulation includes cooling, reionization, star formation and SNe feedback. This simulation is labelled \AMRFID\ and \SPHFID\ (for fiducial) in \ramses\ and \hydra, respectively. 
A final simulation includes these processes, but additionally tracks the evolution of AGN and their associated feedback.  This  is labelled \AMRQSO\ and \SPHQSO\ in \ramses\ and \hydra, respectively. For the \QSO\ run, to make the two implementations as similar as possible, we used a temperature ceiling of 10$^{10}$ K,  and in the  SPH run, energy was deposited kernel-weighted to the gas particles within the BH smoothing length.  Regardless of the simulation method, if the AGN feedback energy at a given step would heat its environment above the temperature ceiling, then the excess energy was saved for the following step, and accretion was stalled until this excess energy was administered.   With this work we wish to highlight the dependence of cluster environment and galaxy gas evolution on feedback implementations and how this evolution is dependent on the numerical method.

\section{Results}\label{results}
We break up our results into four main sections, as we focus on the cluster environment and its gas, and how the cluster
responds to the addition of increasingly complex gas physics in AMR and SPH simulations. We begin by looking at the characteristics of the gas in the cluster halo and sequentially increase the physics included in the simulations.  A summary of the different simulations is given in \tabl{tab_sims}.  We first discuss the Non-Cooling runs, and then compare these simulations with our Fiducial 
runs, which include radiative cooling, star formation, and stellar feedback. We then introduce our AGN simulations and compare these with our fiducial runs. For these first three sections we focus in great detail at snapshots at $z=5$ and $z=3,$ and we end our results section by looking at the continuous evolution of the cluster gas, stellar and black hole components between these two snapshots.

\begin{deluxetable*}{l cccccc}[b!]
\tabletypesize{\scriptsize}
\tablecaption{Cluster Characteristics \label{tab_clust}}
\startdata
\multicolumn{1}{l}{$\mathbf{z}$} &
\multicolumn{1}{c}{Run} &
\multicolumn{1}{c}{$\mathbf{M_{\rm tot} (\Msun)}$} &
\multicolumn{1}{c}{$\mathbf{r_{\rm vir}} $ (kpc) } &
\multicolumn{1}{c}{$\mathbf{T_{\rm vir}}$ (K)} &
\multicolumn{1}{c}{$\mathbf{S_{\rm vir}}^{\rm a}$ (keV cm$^2$)} &
\multicolumn{1}{c}{$\mathbf{\left(\frac{M_{\rm bar}}{M_{\rm tot}}\right)/\left(\frac{\Omega_{\rm b}}{\Omega_{\rm m}}\right)}$} \\  \hline
5          	& \AMRNC	& $2.33 \times 10^{12} $	& 70.3  	& $7.6 \times 10^{6} $  	& 17 	& 0.92 \\
		& \AMRFID 	& $2.30 \times 10^{12} $	& 70.0  	& $7.5 \times 10^{6} $  	& 17 	& 1.1 \\
		& \AMRQSO	& $1.95 \times 10^{12} $	& 66.2 	& $6.7\times 10^{6} $ 	& 15 	& 0.68 \\ 
		& \SPHNC	&  $2.45 \times 10^{12} $	& 71.7  	& $7.9\times 10^{6} $ 	& 18 	& 0.92 \\
		& \SPHFID	&  $2.54 \times 10^{12} $	& 72.5  	& $8.0\times 10^{6} $ 	& 18 	& 0.92 \\
		& \SPHQSO 	&  $2.26 \times 10^{12} $	& 69.9  	& $7.4\times 10^{6} $ 	& 17 	& 0.68 \\ \hline

4		& \AMRFID 	& $7.31\times 10^{12} $     & 123.1 	& $1.36\times 10^{7} $ 	&  44 & 1.0 \\
		& \AMRQSO 	& $6.49 \times 10^{12} $	& 118.3 	& $1.25\times 10^{7} $ 	&  40 & 0.61 \\
		& \SPHFID 	&  $7.27\times 10^{12} $ 	& 123.6 	& $1.35\times 10^{7} $ 	&  44 & 0.92 \\
		& \SPHQSO 	&  $6.77 \times 10^{12} $	& 120.5  	& $1.29\times 10^{7} $ 	&  41 & 0.68 \\ \hline
		
3		& \AMRFID 	& $2.30 \times 10^{13} $    & 224.1 	& $2.33\times 10^{7} $ 	& 128 & 1.0 \\
		& \AMRQSO 	& $2.18 \times 10^{13} $	& 219.7 	& $2.25\times 10^{7} $	& 123 & 0.74 \\
		& \SPHFID 	& $2.31\times 10^{13} $	& 227.3 	& $2.30\times 10^{7} $ 	& 126 & 0.92 \\ 
		& \SPHQSO 	&  $2.19 \times 10^{13} $	& 223.2  	& $2.25\times 10^{7} $ 	& 123 & 0.68 \\ \hline
\multicolumn{7}{l}{$^{\rm a}$ Following Voit et al. (2005), but per H atom instead of electrons.}
	

\enddata
\end{deluxetable*}

\subsection{Non-cooling Runs}
We first ran the two \NC\ simulations, which did not include gas cooling, star formation, reionization, or black holes. These two simulations are crucial for ensuring the gravity solvers are in good agreement and that the artificial viscosity implemented in \hydra\ produces consisten shock heating.  They also allow us to  compare our simulations directly with other non-radiative work (e.g., Voit et al. 2005; Mitchell et al. 2009).   Unlike the more complex cases,  we only ran these simulations to $z=5$, which given their simplicity was sufficiently advanced to be able to compare with the more complex runs. Virial quantities, taken at the radius within which the average density is 200 times denser than the critical density, are presented for all simulations in \tabl{tab_clust}.

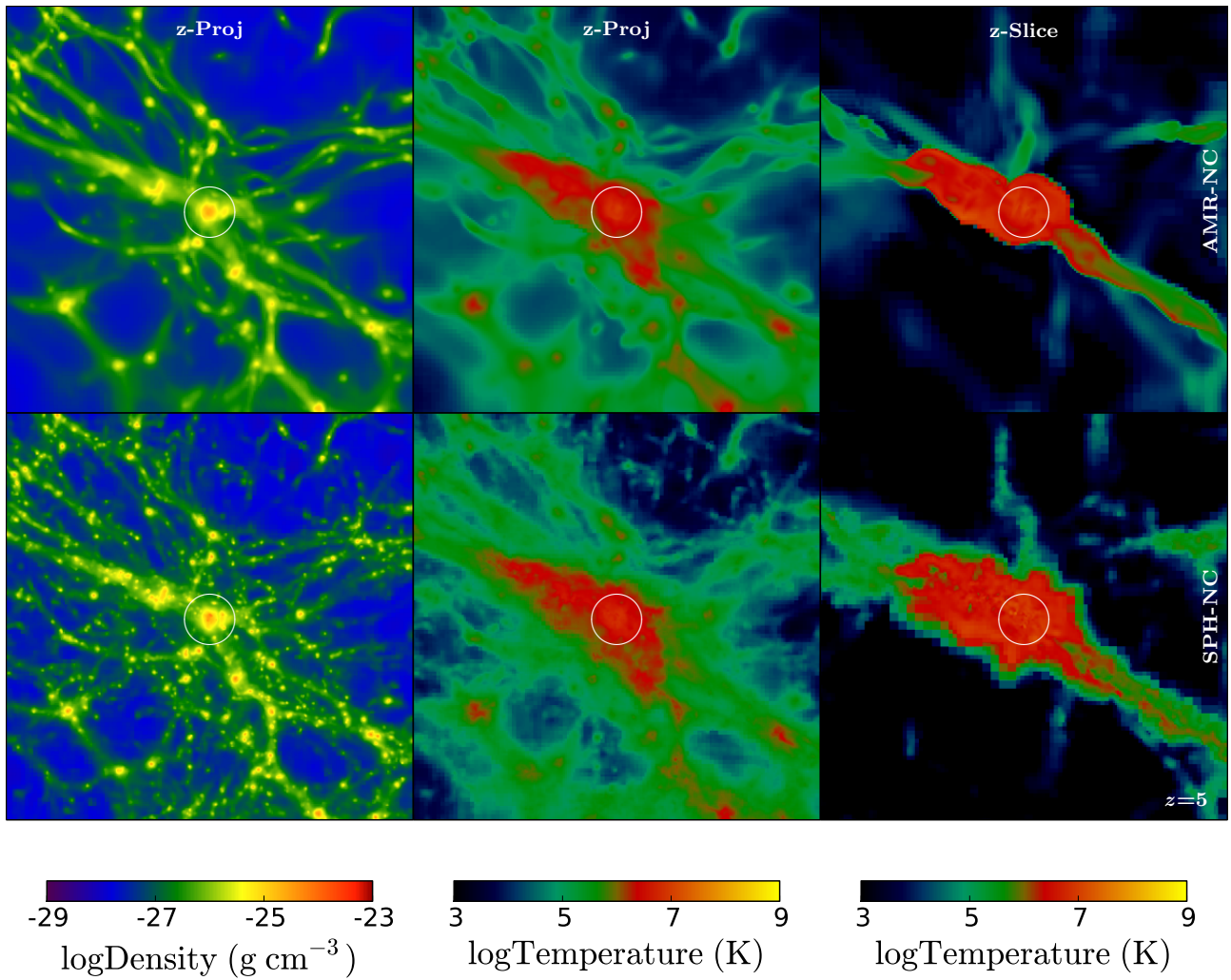
\begin{figure*}[t!] 
\centering
   \begin{overpic}[scale=0.56]{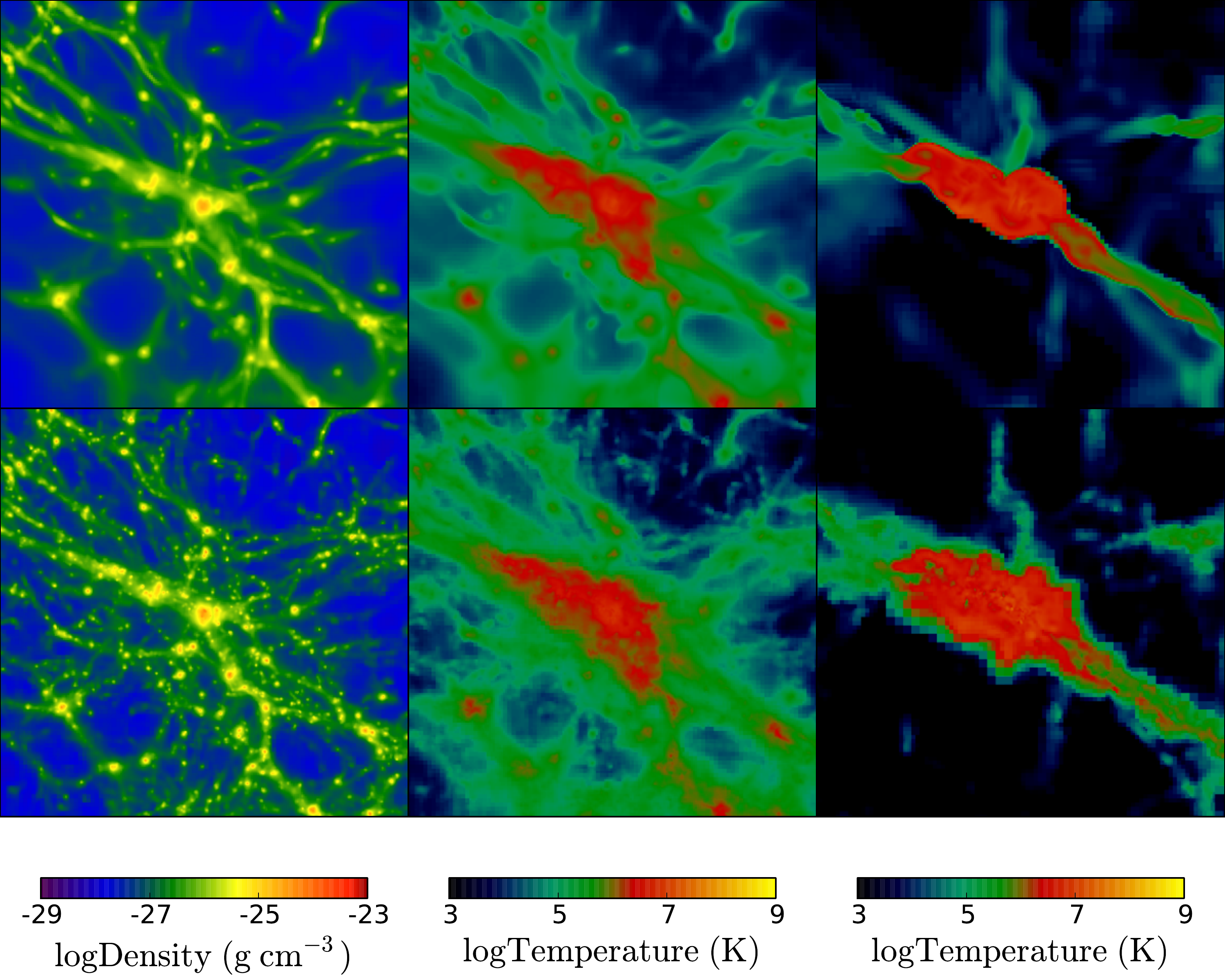}
     \put(16.6667,63.15){\color{white}\circle{4.1667}}
     \put(16.6667,29.75){\color{white}\circle{4.1667}}
     \put(50,63.15){\color{white}\circle{4.1667}}
     \put(50,29.75){\color{white}\circle{4.1667}}
     \put(83.3333,63.15){\color{white}\circle{4.1667}}
     \put(83.3333,29.75){\color{white}\circle{4.1667}}
     \put(16.6667,78){\makebox(0,0){\color{white} \textbf{z-Proj}}}
     \put(50,78){\makebox(0,0){\color{white} \textbf{z-Proj}}}
     \put(83.3333,78){\makebox(0,0){\color{white} \textbf{z-Slice}}}
     \put(96.6,15){\makebox(0,0){\color{white} \textbf{\textit{z}=5}}}
     \put(98.6,64){\makebox(0,0){\rotatebox{90}{\color{white} \textbf{AMR-NC}}}}
     \put(98.6,30){\makebox(0,0){\rotatebox{90}{\color{white} \textbf{SPH-NC}}}}
  \end{overpic}
\caption{Comparisons of density and temperature for \AMRNC\ (top) and \SPHNC\ (bottom)
out to 8 virial radii ($r_{\rm vir}=70$ physical kpc, indicated by a white circle) at $z=5$. Each image is thus 1.1 physical Mpc across centered on the halo. 
Left (Middle): Projections along the $z$-axis showing the gas density (density-weighted temperature). Right: Slices of the same region for temperature highlighting the position of the virial shock.}
\label{fig_adiab_images}
\end{figure*}

The left column of Figure \ref{fig_adiab_images} compares the density of the gas out to 8 virial radii surrounding the cluster at $z=5$ from the two simulations (top: AMR, bottom: SPH).  Unsurprisingly, there is significant correspondence between the \ramses\ and the \hydra\ plots. However we note significantly more dense gas  present in the SPH run. We attribute this to a combination of the different gravitational calculations on small scales between the two methods, and numerical dissipation of entropy in dense regions in SPH. On small scales \hydra\ uses a particle-particle gravity solver, while \ramses\ uses the standard adaptive particle mesh. Particle-particle methods have better short-range force resolution, which gives more clumps (e.g., O'Shea et al. 2005). Additionally, \hydra\ uses an energy-conserving, rather than entropy-conserving, implementation of SPH. Therefore, in dense regions some entropy is dissipated numerically, further differentiating these clumps in SPH. The larger number of dense knots in classic SPH is well established, and has been discussed in detail (e.g., Frenk et al. 1999; Kaufmann et al. 2006; Power et al. 2014).

The middle and right columns of \fig{fig_adiab_images} show the temperature in projection and slice, respectively, of the gas for the same region. 
The pixelation seen in the SPH slice is a result of the mapping used to produce the image with the same visualization software as for the AMR images (Turk et al. 2011, http://yt-project.org/).
 In projection we see better agreement between the two simulations, while the slices highlight the different extent of the virial shocks.
The SPH run has a larger shock radius that is spread over a wider spatial extent, consistent with the artificial viscosity injecting entropy at a somewhat larger radii than in AMR. This earlier shock heating causes more high-entropy gas in SPH since the densities are very similar at the virial radius. This increased entropy may also be related to the poor ability of SPH to model subsonic turbulence, as described in \citet{Bauer12}. Gas accretion into clusters is dependent on subsonic gas, which in AMR correctly cascades to smaller scales but in SPH thermalizes near the driving scale. It is this, combined with heating by the artificial viscosity, that leads to an expanded virial shock. 

\begin{figure*}[t!] 
\centering
\includegraphics[width={1.6\columnwidth}]{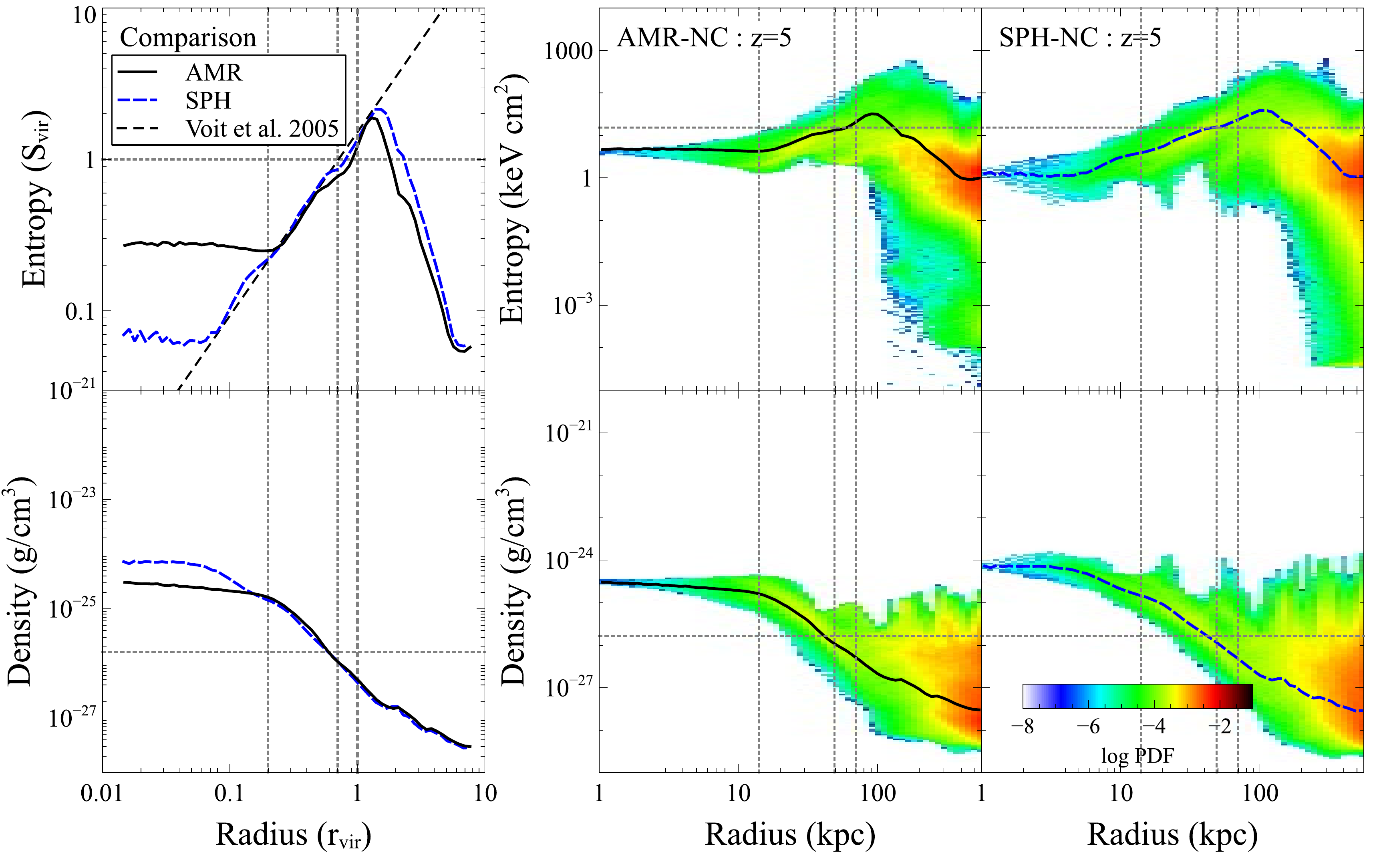}
\caption{Radial profile plots at $z=5$ showing the gas entropy (top) and gas density (bottom) vs radius for \AMRNC\ (middle) and \SPHNC\  (right) out to 8 virial radii from the cluster. 0.2, 0.7, and 1.0 virial radii, and the virial density and entropy are indicated by the grey dotted lines (see \tabl{tab_clust}). The virial
entropy is calculated following Voit et al. 2005, except we use the hydrogen density instead of the electron density. Color corresponds to log gas mass probability distribution fraction, where the integral of this quantity over a given plot is unity. The solid black (dashed blue) lines in the middle (right) plots demark the volume-weighted average density and entropy (taken as the average temperature over average density to the 2/3 power) with radius for all gas in the AMR (SPH) simulation.
The left plots compare the average values of AMR vs SPH, using the same color and line scheme as the middle and right plots, scaled by the virial quantities so that they are directly comparable.\vspace{4pt}}
\label{fig_adiab_prof}
\end{figure*}

\fig{fig_adiab_prof} presents radial profiles of the gas density and entropy, here expressed as $S \equiv k_{\rm B}T/n_{\rm H}^{2/3}$, where $k_{\rm B}$ is the Boltzmann constant, and $n_{\rm H}$ is the hydrogen number density, for all gas, out to eight times the virial radius ($r_{\rm vir} = 70$ physical kpc). Virial quantities are annotated with dotted grey lines, to aid in comparisons with works  that look at multiple halos, switching to scale-free units normalized by these virial quantities (e.g, Voit et al. 2005; Mitchell et al. 2009; Kere\v{s} et al. 2012).  Also included in the left plots of this figure are volume-averaged density and entropy (taken as the volume-averaged temperature divided by the volume-averaged density to the 2/3 power), which for direct comparison between AMR and SPH, we normalize by the virial quantities. 

For 0.2$r_{\rm vir} < r < r_{\rm vir},$ the average entropy and density profiles agree very well, and they are consistent with previous comparisons looking at a range of mass scales  (Frenk et al. 1999; Voit et al. 2005; Mitchell et al. 2009; Richardson et al. 2013; Sembolini et al. 2015).  In particular, the entropy profile matches the observations of Voit et al. (2015), where $S \propto r^{1.2}$. The average AMR gas is slightly more dense through the virial shock, which may be due to SPH smoothing out the shock width. Outside of the virial radius, the average density is again in agreement between the two methods, although the entropy is slightly increased in SPH near the virial radius. Sembolini et al. (2015) saw a similar small increase in the entropy at the virial radius in their \hydra\ simulation compared to their AMR comparison run. As we discussed above, this is due to higher temperature at the virial radius. 

Within 0.2$r_{\rm vir},$ on the other hand, the SPH gas density is higher than in AMR,  and the entropy reaches a lower value than AMR before plateauing. The higher, more centrally peaked density in SPH is seen in \fig{fig_adiab_images} in the center of the cluster. This is also consistent with previous comparisons (Frenk et al. 1999; Voit et al. 2005; Mitchell et al. 2009; Richardson et al. 2013; Sembolini et al. 2015), and Mitchell et al. (2009) demonstrated that the lower entropy in the central region of SPH simulations is due to a reduced amount of mixing, with the low-entropy particles instead sinking to the center. It is unclear, however, just how much over-mixing occurs in AMR simulations, thus the true central profile is expected to be slightly lower than in AMR. However, the density discrepancy is not sufficient to explain the entropy difference, thus the SPH gas is slightly cooler in the center. 

Looking at the full distribution of density beyond the virial radius, we see that the density in \hydra\ extends to higher values than in \ramses, consistent with satellite objects also collapsing to slightly higher densities in SPH, and visible in \fig{fig_adiab_images}. The entropy distribution has a wider distribution of entropy in the center of the SPH simulations, consistent with less mixing of high and low entropy particles.

\begin{figure}[t!] 
\centering
\includegraphics[width={0.9\columnwidth}]{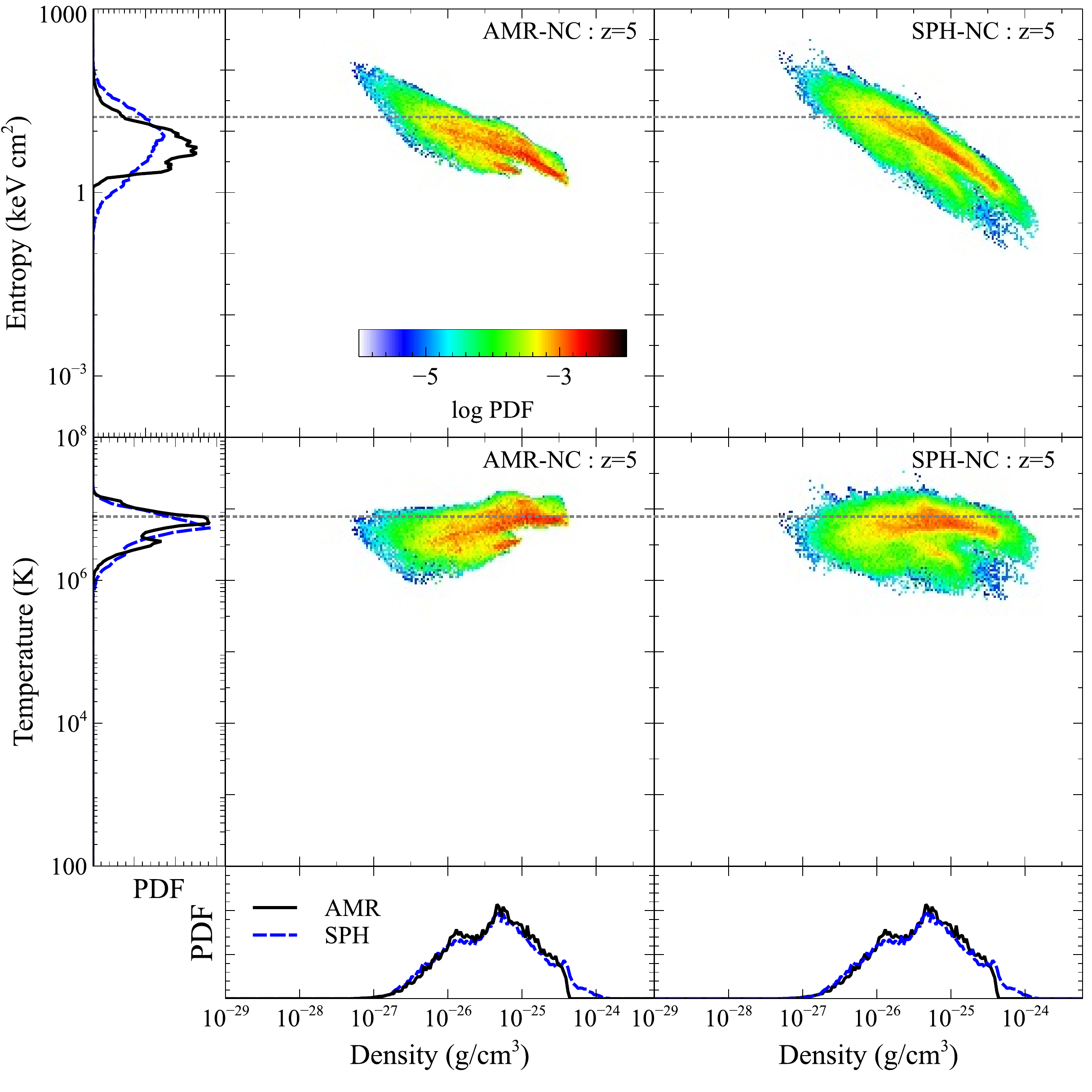}
\caption{Phase plots at $z=5$ showing the gas entropy (top) and temperature (bottom) vs density for \AMRNC\ (left) and \SPHNC\ 
(right) within the virial radius. Color corresponds to log gas mass fraction. The horizontal dotted line demarks the virial entropy and temperature of the cluster. 
On the left edge in linear units are one-dimensional
probability distributions of the log entropy (top) and log temperature (bottom) comparing the relative distribution of gas mass for both AMR (black solid line) and SPH (blue dashed line), while the bottom edge shows in linear units the one-dimensional probability
distribution of log gas density, shown twice to facilitate comparison with the above phase plots.}
\label{fig_adiab_phase}
\end{figure}

\fig{fig_adiab_phase} shows two-dimensional entropy-density and temperature-density distributions functions for gas
within the virial radius, along with one-dimensional distribution functions for each of these quantity. In the SPH simulation, high-density gas is found at lower entropy and even higher densities than in the AMR case (compare with \fig{fig_adiab_prof}). This gas is at the center of the  cluster, where the entropy is numerically dissipated, and not accurately reinjected through mixing (Mitchell et al. 2009) or from large scales (Bauer \& Springel 2011). This larger fraction of high-density, low-entropy gas in SPH is also seen for halo substructure in the nIFTy comparison \citep{Sembolini15A}. A larger fraction of gas is found at very high entropy in SPH, corresponding to low density, hotter gas at the virial radius, as we discussed above. Besides the very high density gas in SPH, the distribution of density and temperature in SPH and AMR are very consistent. 

In general, the subtle differences that appear between the simulations are not surprising due to the difficulty in implementing an \textit{ad hoc} artificial viscosity in SPH, the energy-conserving implementation of the fluid equations in SPH, and the tendency for SPH to undermix and AMR to overmix. We proceed, aware that these small differences may compound when including cooling, star formation, and feedback.

\subsection{Fiducial Runs}\label{sect_fid}
\begin{figure*}[t!] 
\centering
   \begin{overpic}[scale=0.56]{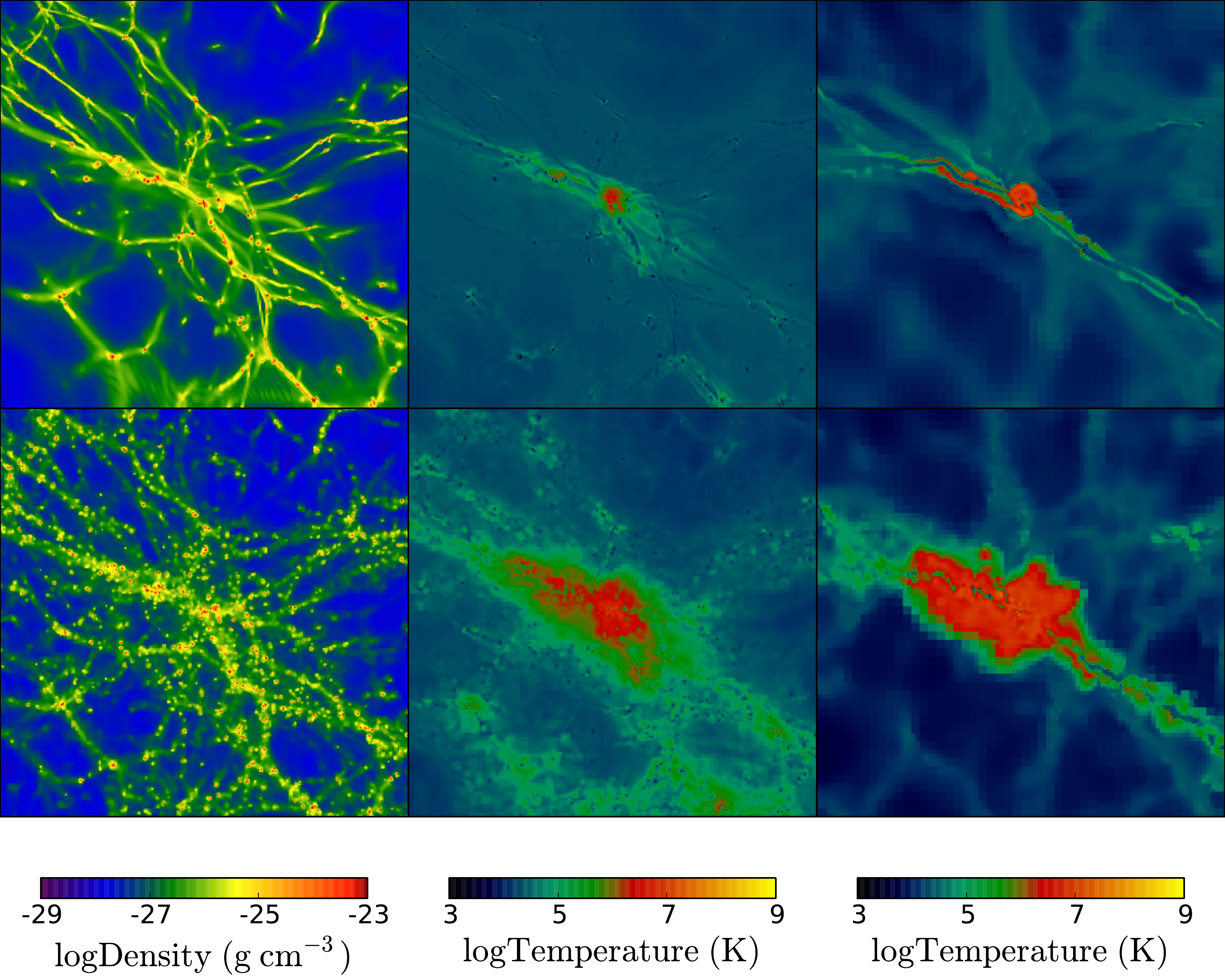}
     \put(16.6667,63.15){\color{white}\circle{4.1667}}
     \put(16.6667,29.75){\color{white}\circle{4.1667}}
     \put(50,63.15){\color{white}\circle{4.1667}}
     \put(50,29.75){\color{white}\circle{4.1667}}
     \put(83.3333,63.15){\color{white}\circle{4.1667}}
     \put(83.3333,29.75){\color{white}\circle{4.1667}}
     \put(16.7,78){\makebox(0,0){\color{white} \textbf{z-Proj}}}
     \put(50.6,78){\makebox(0,0){\color{white} \textbf{z-Proj}}}
     \put(84.6,78){\makebox(0,0){\color{white} \textbf{z-Slice}}}
     \put(96.6,15){\makebox(0,0){\color{white} \textbf{\textit{z}=5}}}
     \put(33.4,46.7){\includegraphics[scale=0.1, trim=  175 132 185.0 48, clip=true]{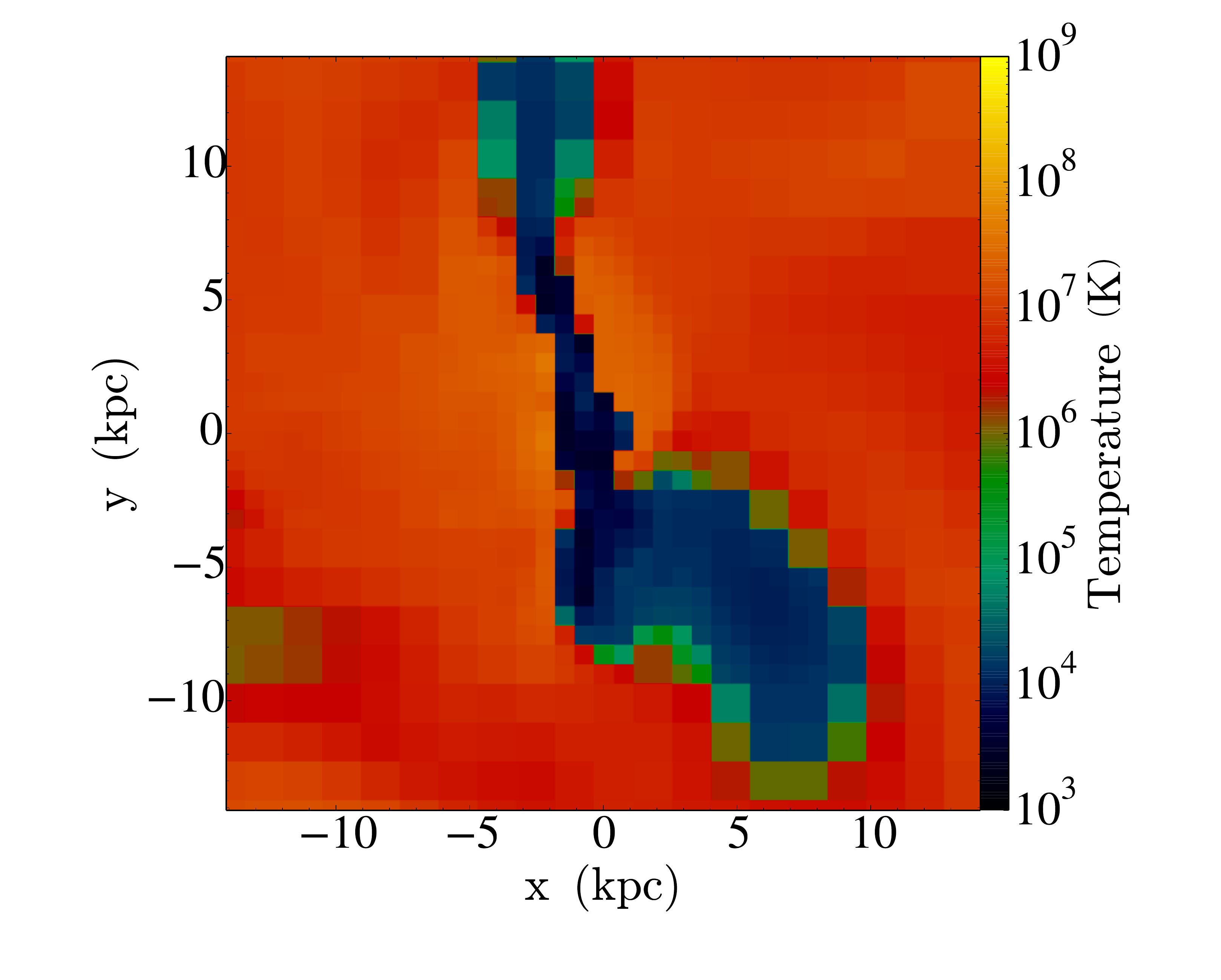}}
     \put(0.07,46.7){\includegraphics[scale=0.1, trim=  116 86 144.0 30, clip=true]{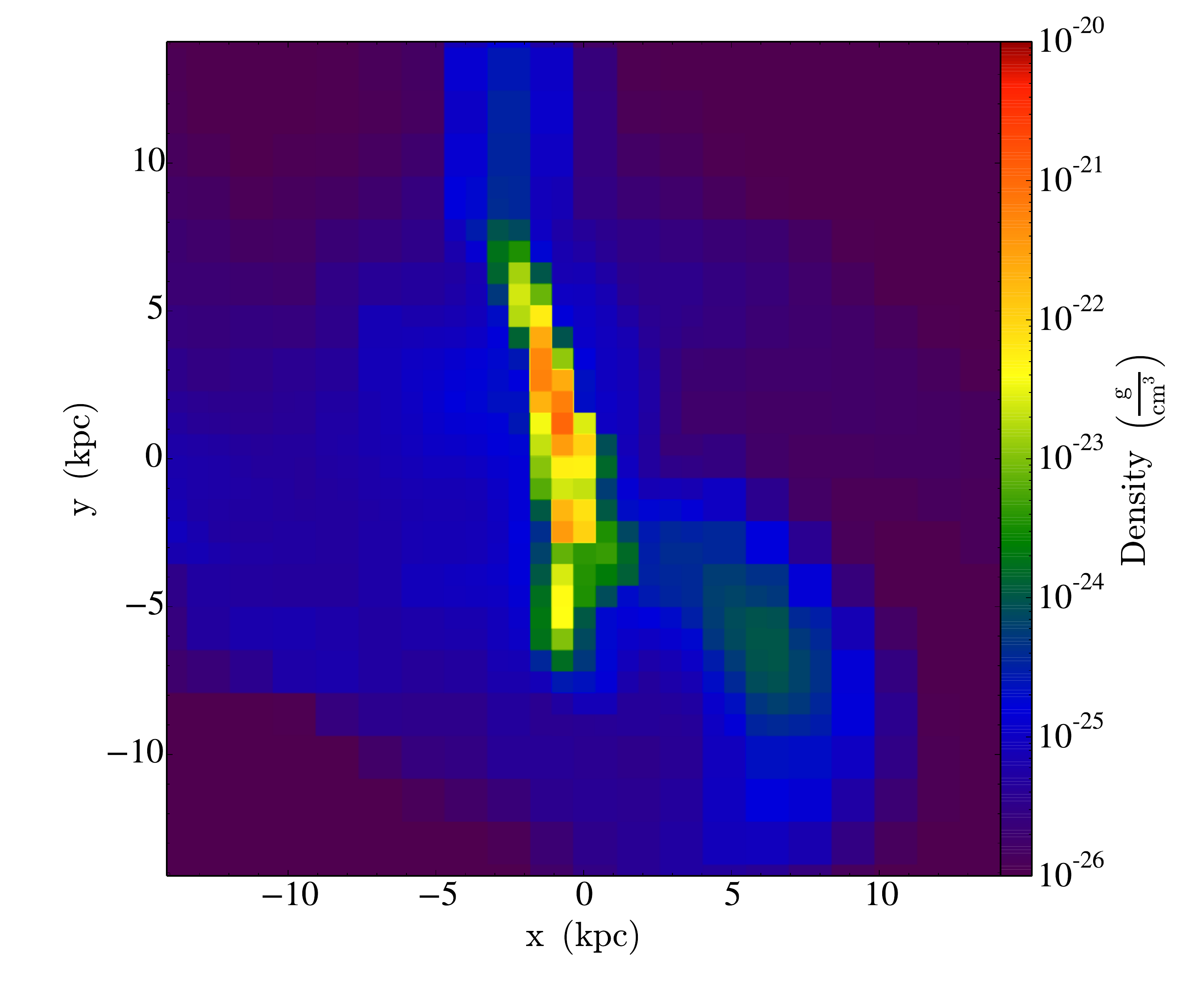}}
     \put(33.4,13.4){\includegraphics[scale=0.1, trim=  175 132 185.0 48, clip=true]{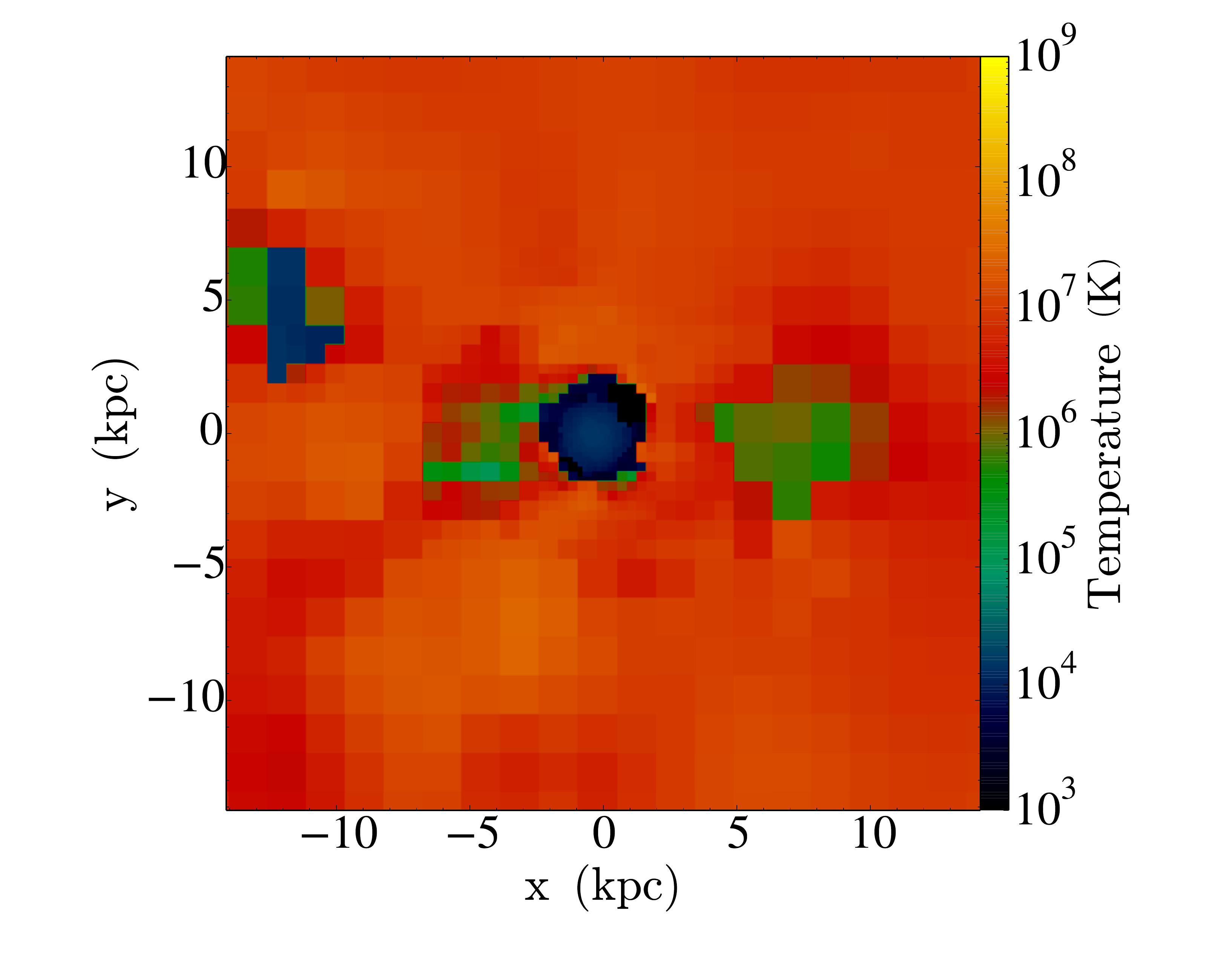}}
     \put(0.07,13.4){\includegraphics[scale=0.1, trim=  116 86 144.0 30, clip=true]{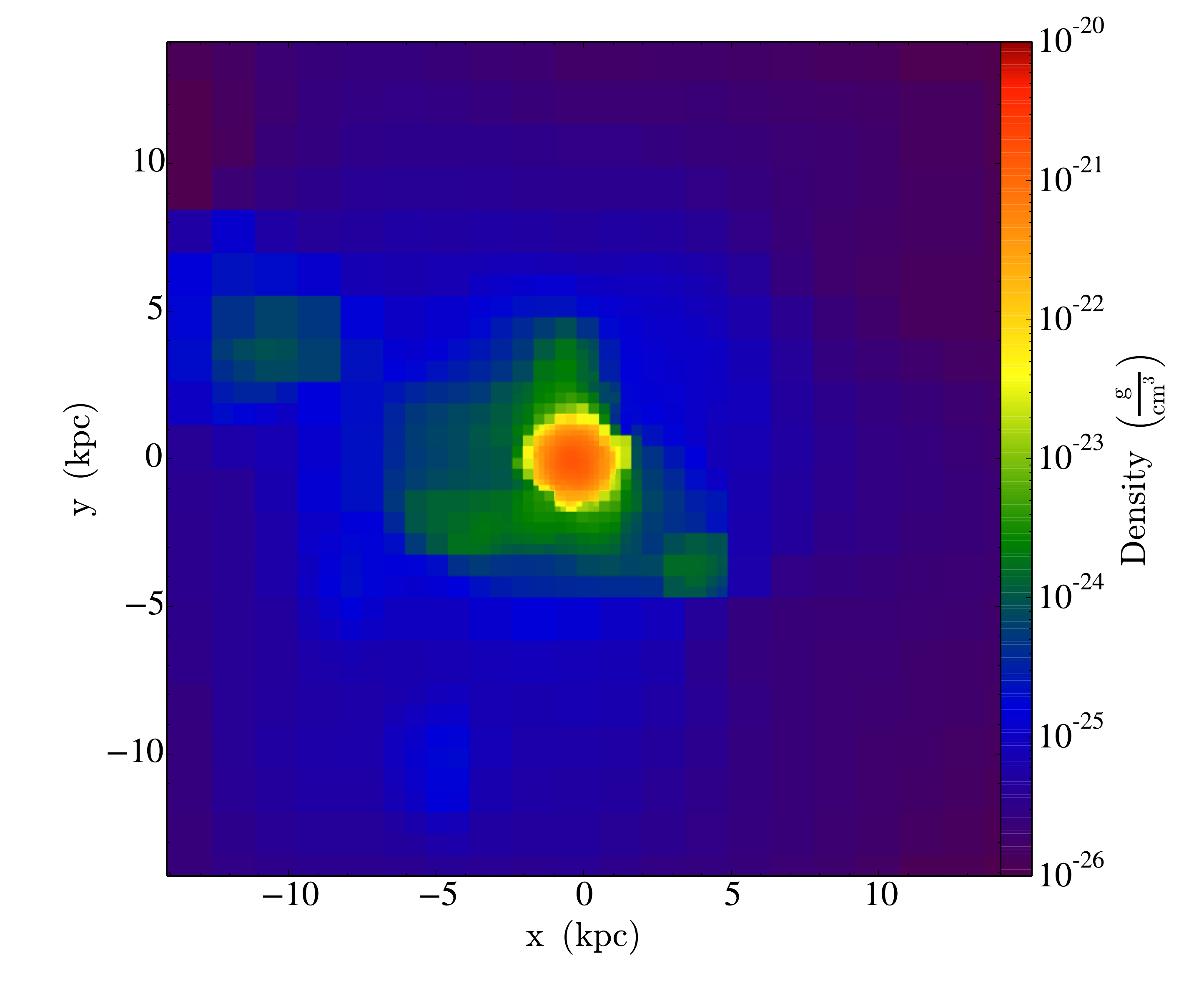}}
     \put(98.6,64){\makebox(0,0){\rotatebox{90}{\color{white} \textbf{AMR-FID}}}}
     \put(98.6,30){\makebox(0,0){\rotatebox{90}{\color{white} \textbf{SPH-FID}}}}
  \end{overpic}
\caption{Same as \fig{fig_adiab_images} but for the \FID\ simulations, measuring 1.1 physical Mpc across. The inserts in the projection
plots are slices of density (left) and temperature (middle) focused on the galaxy scale, measuring 28 physical kpc across, and reach 
out to 0.2 $r_{\rm vir}$.} 
\label{fig_fid_images5}
\end{figure*}
Next we look at the \FID\ simulations, which compare AMR and SPH with the inclusion of cooling, reionization, star formation, and stellar feedback. We first begin by comparing these results with the $z=5$ \NC\ results, and then compare the FID results in more detail at $z=3$. The halo virial quantities  are listed in \tabl{tab_clust}, which shows that the inclusion of cooling has led to a higher gas fraction than in the \AMRNC\ simulations, with the \AMRFID\ values now slightly exceeding the cosmic average.

In \fig{fig_fid_images5} we show projections of the gas density for the \AMRFID\ and \SPHFID\  simulations at $z=5.$ In the \FID\ simulations the gas can cool, resulting in more condensed structures. Thus the filaments are thinner, and the galaxies are collapsed to thin disks. On these scales in density we see little impact of star formation or feedback. In comparison, we see an amplification of the differences seen in \fig{fig_adiab_images}, with more clumps in SPH than in AMR.  In AMR, the gas filaments are
smooth ribbons of near uniform density gas with large galaxies residing within their nodes. In SPH, these filaments are much more inhomogeneous, housing many more  small clumps that are less dense than AMR galaxies, but more dense than the surrounding filament gas. 
The clumpiness in \SPHNC\ is now  compounded in \SPHFID\ by the fact that this denser, lower-entropy gas has  shorter cooling times, leading to quicker fragmentation times for the filament as a whole.  Inside the virial radius the dense gas is completely fragmented into individual parcels. The larger clumps agree between AMR and SPH, and are cospatial with clumps in dark matter. However, the additional clumps found in SPH, which are seen in other studies of classic SPH (e.g., Frenk et al. 1999; Kaufmann et al. 2006; Power et al. 2014), are not cospatial with dark matter clumps and are due
to artificial dissipation. We stress however that as the densest clumps fragment along the polytrope, they are forced to increase in temperature. Now that the filling factor of dense gas has decreased, we expect that the addition of AGN feedback will be more efficient in SPH and better able to blow away the more tenuous ambient gas, as we discuss in detail in \sect{gas_agn}.

In the fiducial runs the virial shocks in the SPH case are at a significantly larger radii than in the AMR case. 
These differences are visibile in the temperature projections and slices shown in the middle and right panels of \fig{fig_fid_images5}, and are 
much more apparent than in the NC comparison runs shown in \fig{fig_adiab_images}. 
Thus the increased post-shock temperature and slightly wider shock  in SPH (e.g., Hubber et al. 2013) leads to a lower cooling rate. We discuss the expected cooling times below.

\begin{figure*}[t!] 
\centering
\includegraphics[width={1.6\columnwidth}]{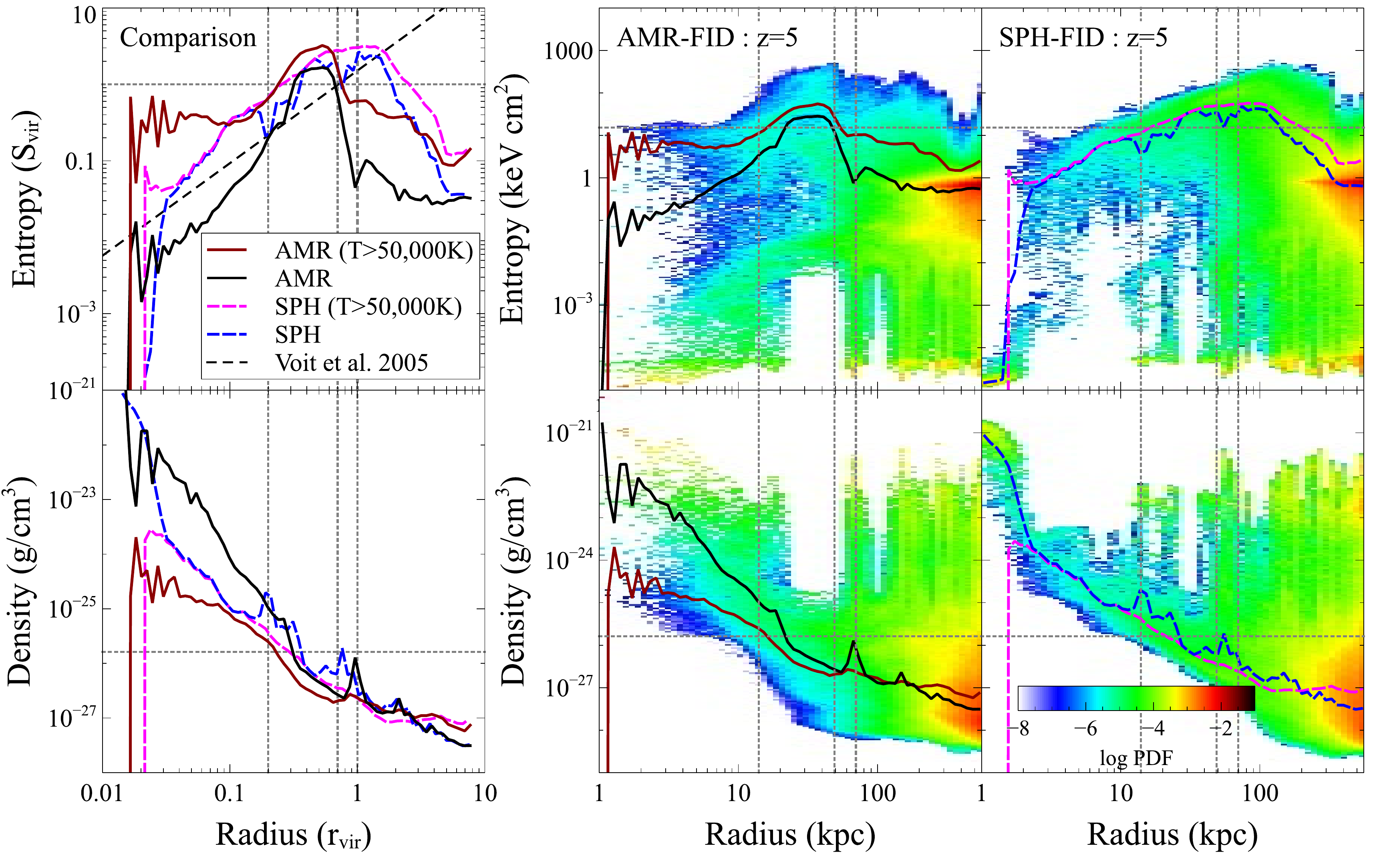}
\caption{Same as \fig{fig_adiab_prof}, but for the \FID\ simulations. The red (magenta) lines in the middle (right) plots demark the volume weighted average density with radius for 
gas hotter than 50,000 K in the AMR (SPH) simulation, respectively.
The left plots compare the average values, scaled by the virial values, of AMR vs SPH, using the same color and line scheme as the middle and right plots.}
\label{fig_fid_prof5}
\end{figure*}

\fig{fig_fid_prof5} presents the profile diagram of the gas at $z=5$ out to 8 virial radii, to compare with \fig{fig_adiab_prof}. Here we have split the average quantities into a hot  component above 50,000 K, and a total component. The hot gas is more directly relatable to other work (e.g., Kere\v{s} et al. 2012) and with our \NC\ runs, and less susceptible to the different clumping behavior. Cooling in the fiducial case also leads to a large population of gas at very low entropies and high densities, which is seen in both the AMR and SPH simulations. 
However in the AMR run there is a two-phase medium, with cold and dense material found at the same radial distances as the hot, tenuous gas. In the SPH run, on the other hand,  the hot and cold phases are segregated, with cold material found almost exclusively in the center, surrounded by a hot diffuse region.  Thus, the addition of cooling has amplified the ability of low entropy gas in SPH to sink to the center of the halo, and the two methods yield very different results within 0.2 $r_{\rm vir}$. 

Beyond this radius, out to 0.7 $r_{\rm vir}$, there is better agreement of the hot gas between the two methods, while AMR is better able to model cold streams with entropy of roughly $10^{-2}$ keV cm$^2$. Beyond 0.7 $r_{\rm vir}$ and out to roughly the virial radius, the entropy differences are even stronger than in the \NC\ runs. Here the post-shock gas can efficiently cool in the AMR run, leading to the shock radius lying within the virial radius. In the SPH run, on the other hand, the shock radius is at the virial radius.  This is not what is expected from theory. At $z=5$, this system has a  dynamical time of roughly 0.11 $H^{-1}$ while the cooling time is approximately 0.03 $H^{-1}$. Thus the post-shock gas should cool quicker than the typical growth time 
of the halo, and we expect the virial shock to lie within the virial radius (e.g.\ White \& Rees 1978; Birnboim \& Dekel 2003). In our \hydra\ run, no star particles have been made within the virial radius of this cluster by $z=5$, thus this larger virial shock and smaller filling factor of cold gas is not due to a difference in the SPH SNe feedback. Instead, this appears to be due to the excess heating near the virial radius seen in the \SPHNC\ run, coupled to a reduced cooling rate as the gas shock is broadened in SPH. This lower cooling at the virial radius in SPH then leads to an inflated post-shock region. This is similar to the comparison of the projected temperature of a galaxy halo taken from the SPH code {\tt GADGET} with the moving mesh code {\tt AREPO} reported in Nelson et al.\ (2013), in which the SPH virial shock was located at larger radii than in the moving-mesh case.

\begin{figure}[h!] 
\centering
\includegraphics[width={0.9\columnwidth}]{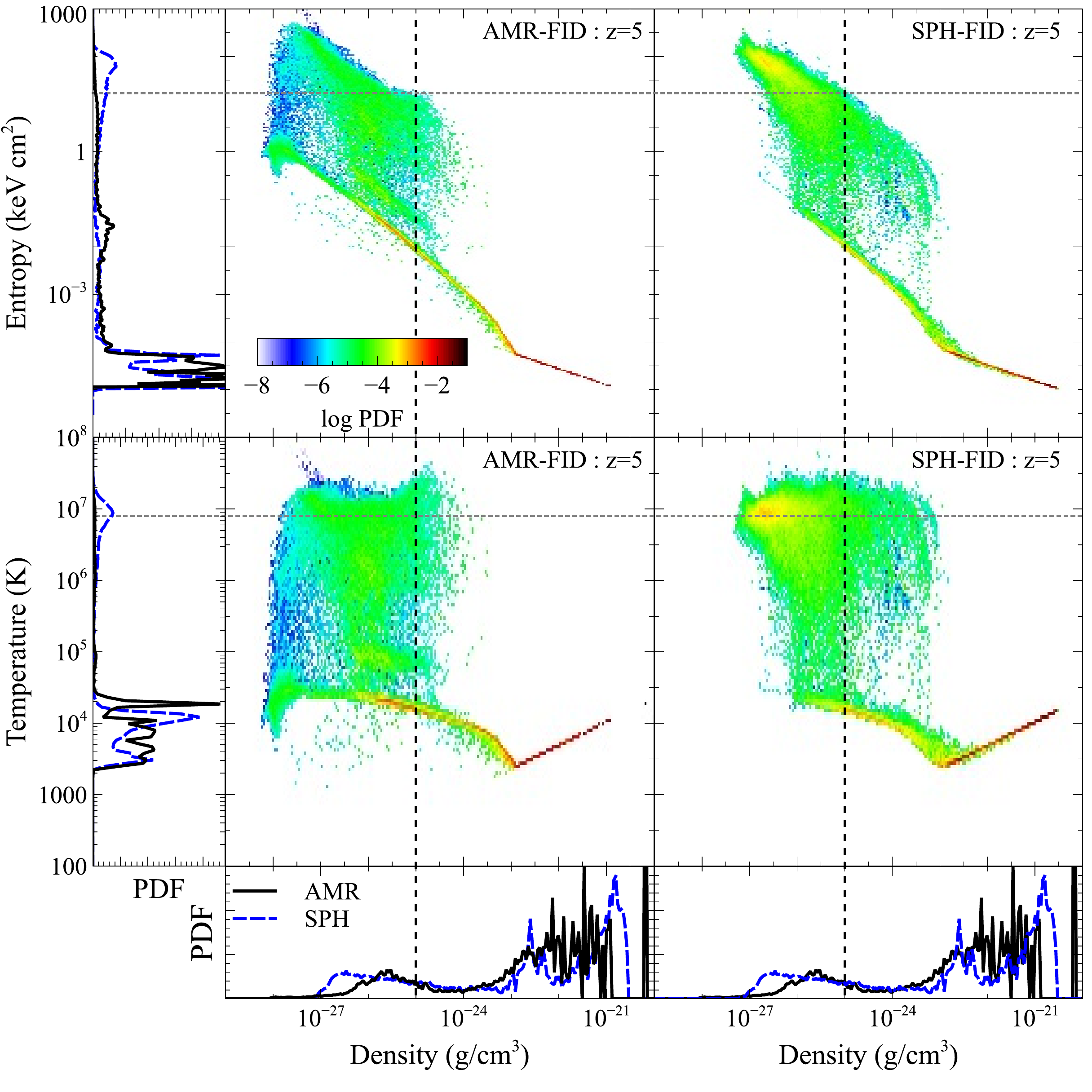}
\caption{Same as in \fig{fig_adiab_phase}, but for the \FID\ simulations. The vertical dotted line marks the star-formation density threshold at roughly $10^{-25} \gcc$.}
\label{fig_fid_phase5}
\end{figure}

Phase diagrams of the gas at $z=5$ out to the virial radii are given in \fig{fig_fid_phase5}.  These plots show many new features not seen in the the \NC\ results shown in Figure \ref{fig_adiab_phase}.  The medium is now found mostly at 10,000 - 20,000 K, where the cooling function has a local minimum. As gas slowly cools through this regime, it becomes denser and therefore has lower entropy. At densities above $3\times 10^{-24} \gcc$ the gas cools more efficiently, until cooling to the enforced polytope, where the temperature is forced to scale with $\rho^{1/3}$. Finally, post-shock gas is heated to just above the virial temperature, and cools inefficiently, except at high densities. The \SPHNC\ gas at high temperatures that extends to a density of $\rho \simeq 10^{-24} \gcc$ is able to cool in the \SPHFID\ simulation, forming lines extending down to the cooler, $T\simeq 10^4$ K regime. Finally, given that gas was found at slightly higher densities in \SPHNC\ than in \AMRNC, we naively expected the cooling rate for the \SPHFID\ gas to be faster. However, since the polytrope gas extends to higher densities and temperatures in AMR, SPH appears to prevent gas in the polytrope from moving to higher densities. This may be due to undermixing, as discussed before. Additionally, AMR creates star particles of smaller mass than in SPH, thus AMR is more quickly removing pressure support from the densest regions, possibly leading to the buildup of more high-density polytrope gas. A future study of star particle mass and star formation rate is needed to better understand this effect.

\begin{figure*}[t!] 
\centering
   \begin{overpic}[scale=0.56]{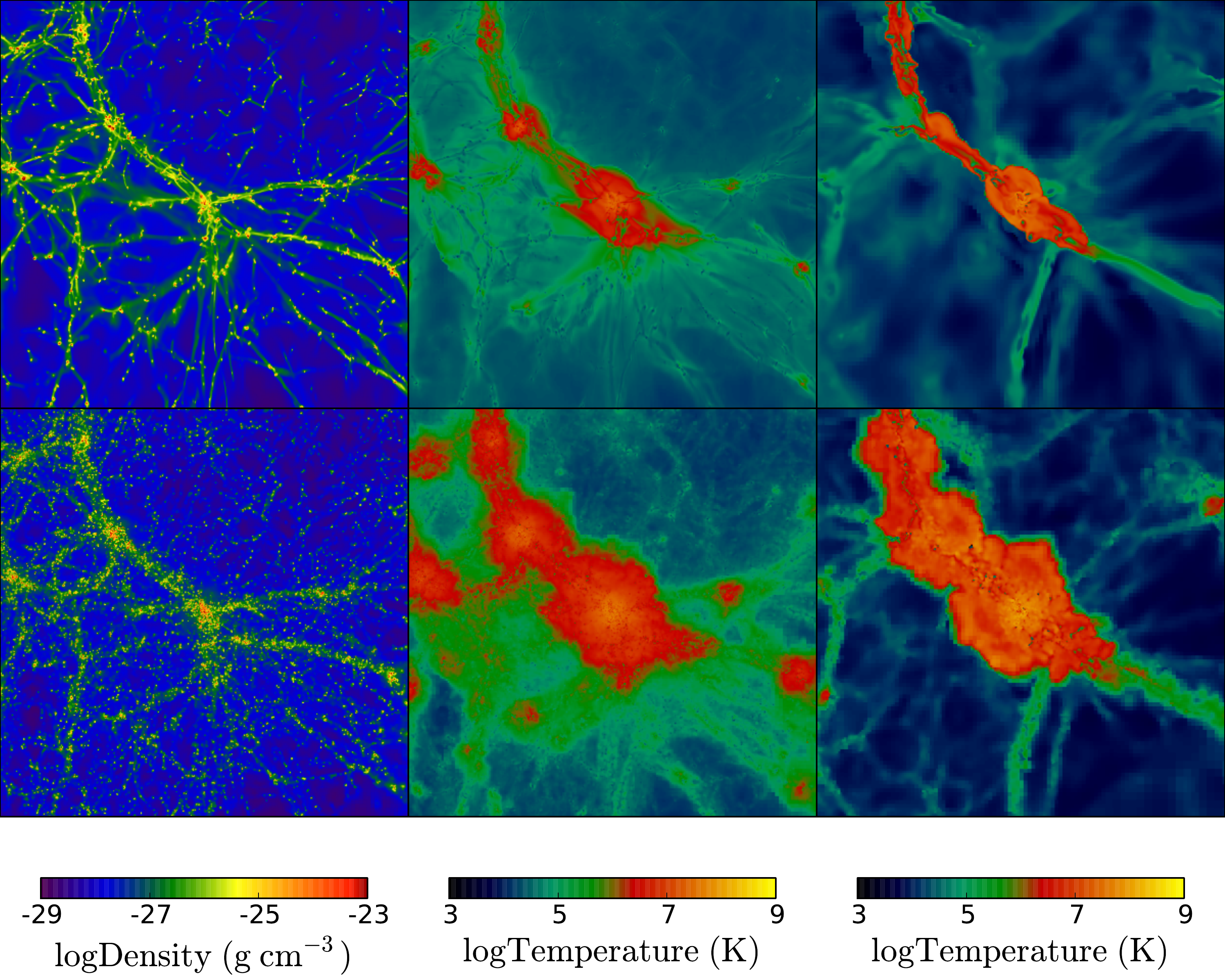}
     \put(16.6667,63.15){\color{white}\circle{4.1667}}
     \put(16.6667,29.75){\color{white}\circle{4.1667}}
     \put(50,63.15){\color{white}\circle{4.1667}}
     \put(50,29.75){\color{white}\circle{4.1667}}
     \put(83.3333,63.15){\color{white}\circle{4.1667}}
     \put(83.3333,29.75){\color{white}\circle{4.1667}}
     \put(16.7,78){\makebox(0,0){\color{white} \textbf{z-Proj}}}
     \put(50.6,78){\makebox(0,0){\color{white} \textbf{z-Proj}}}
     \put(84.6,78){\makebox(0,0){\color{white} \textbf{z-Slice}}}
     \put(96.6,15){\makebox(0,0){\color{white} \textbf{\textit{z}=3}}}
     \put(33.4,46.7){\includegraphics[scale=0.1, trim=  175 132 185.0 48, clip=true]{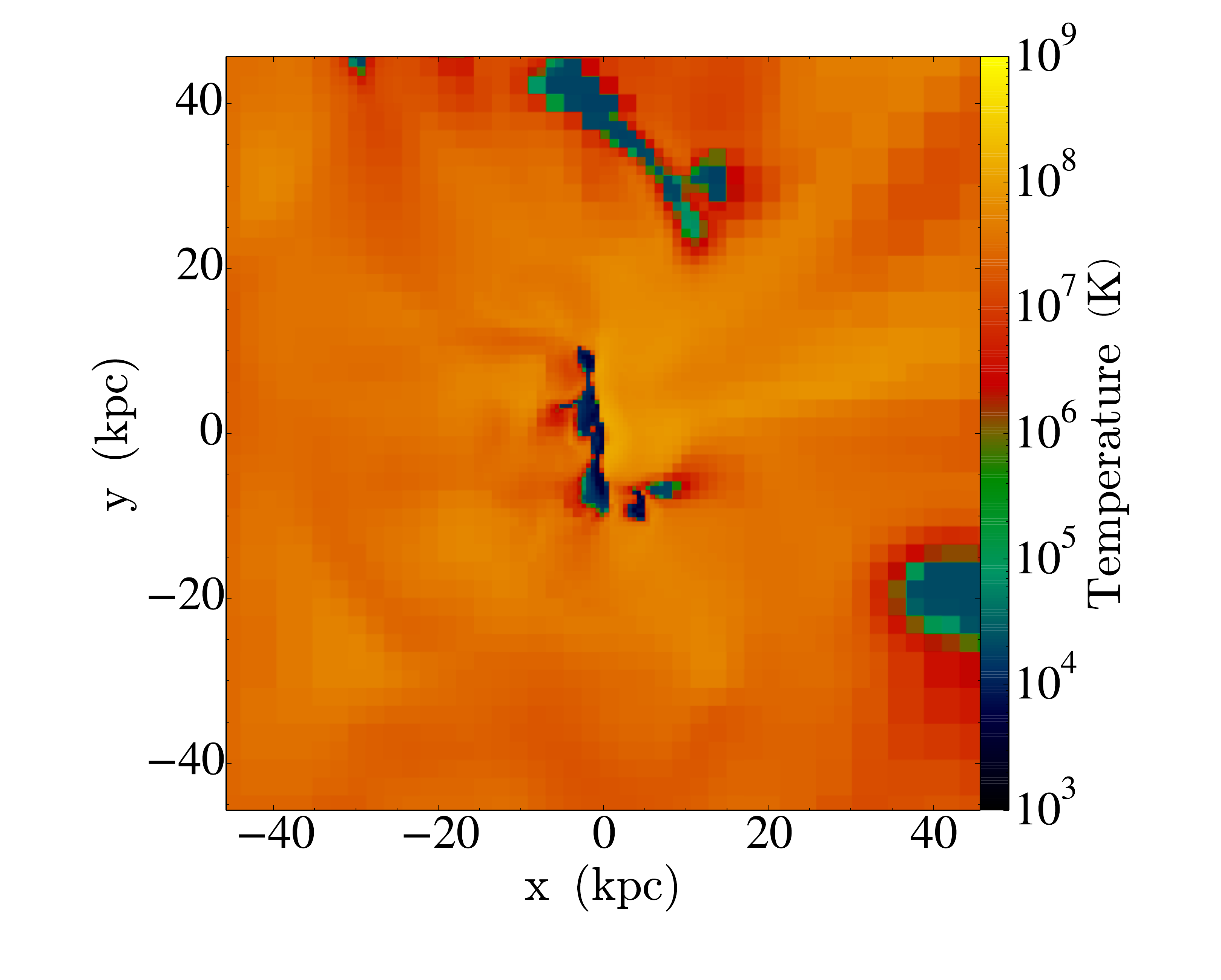}}
     \put(0.07,46.7){\includegraphics[scale=0.1, trim=  116 86 144.0 30, clip=true]{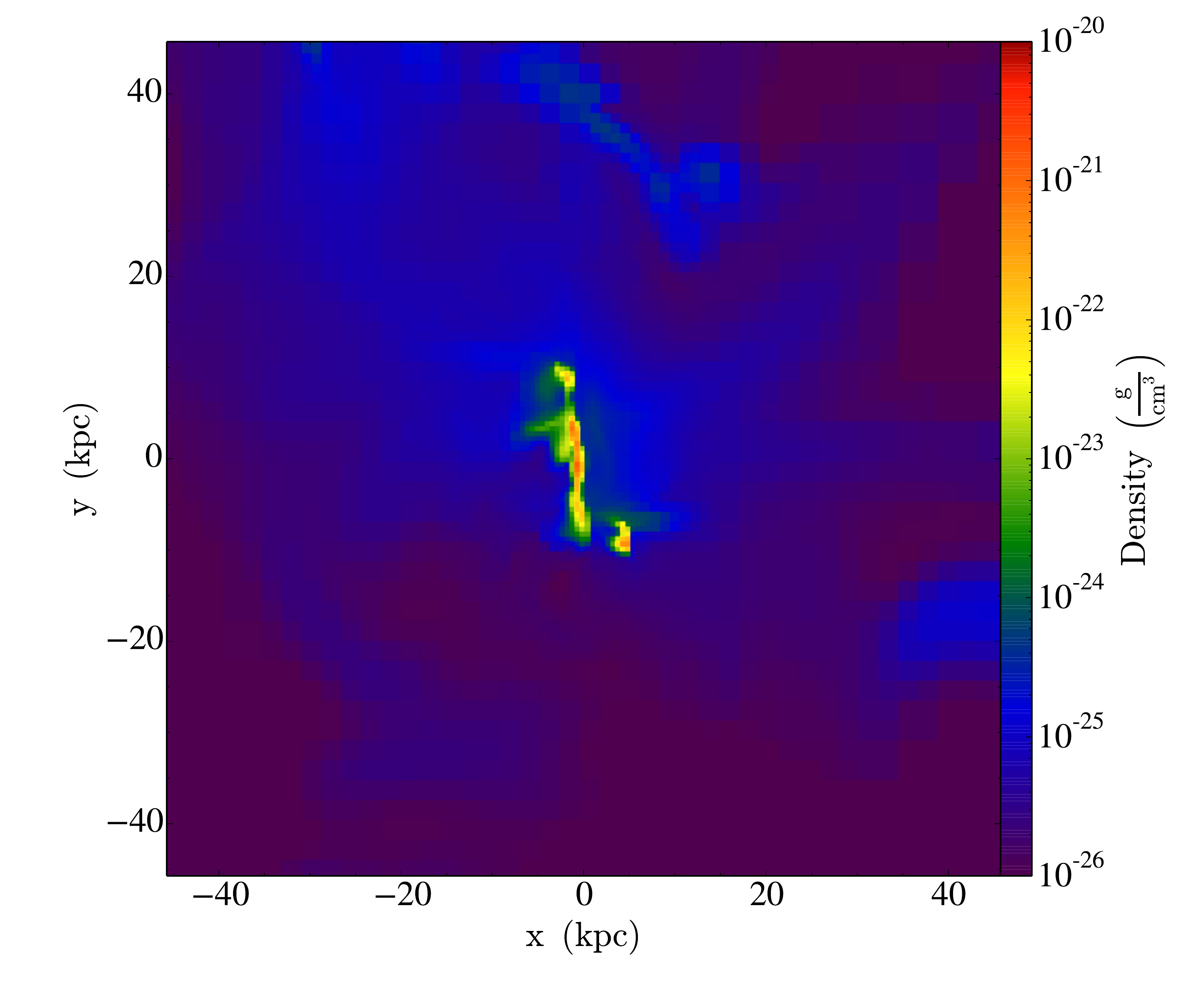}}
     \put(33.4,13.4){\includegraphics[scale=0.1, trim=  175 132 185.0 48, clip=true]{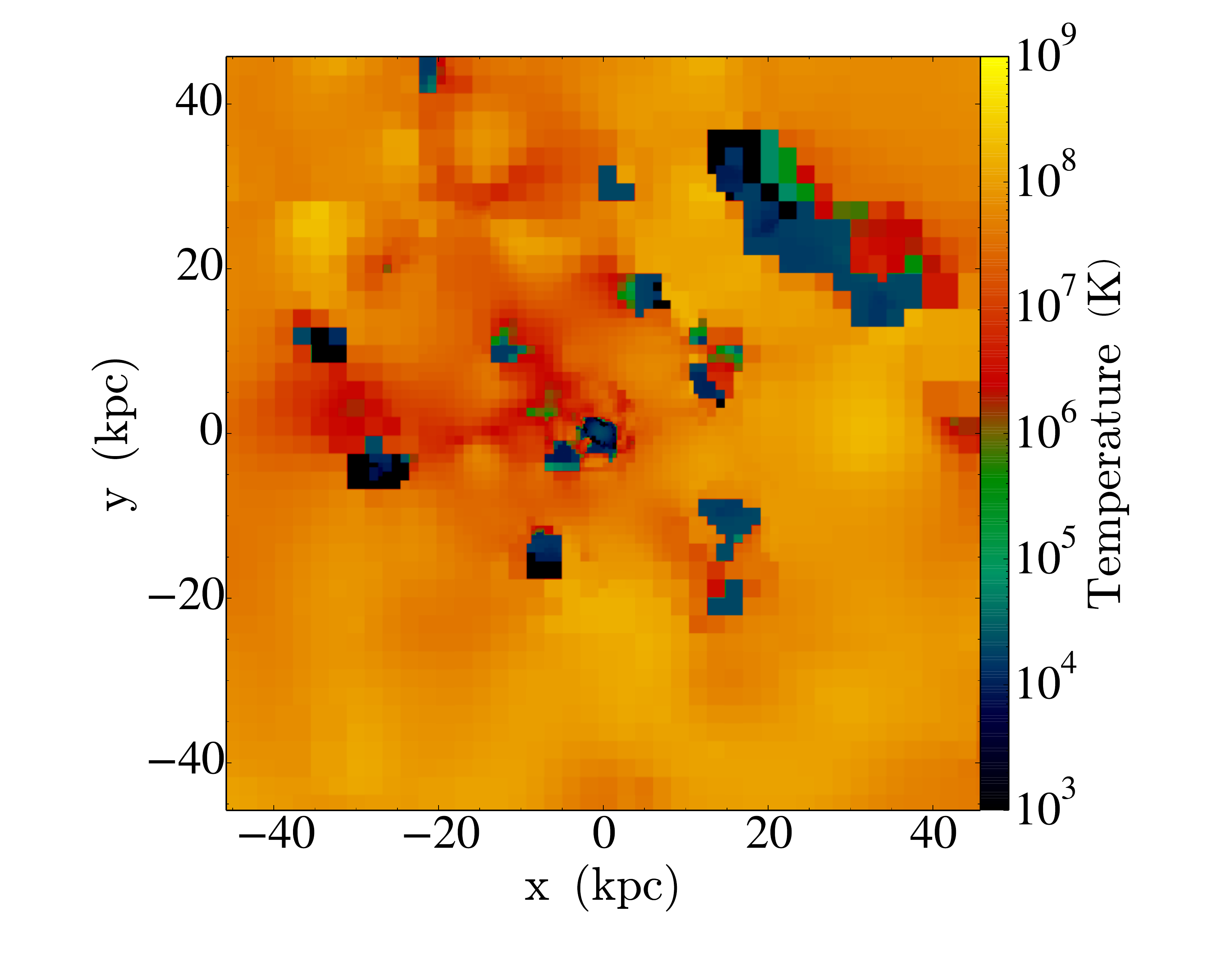}}
     \put(0.07,13.4){\includegraphics[scale=0.1, trim=  116 86 144.0 30, clip=true]{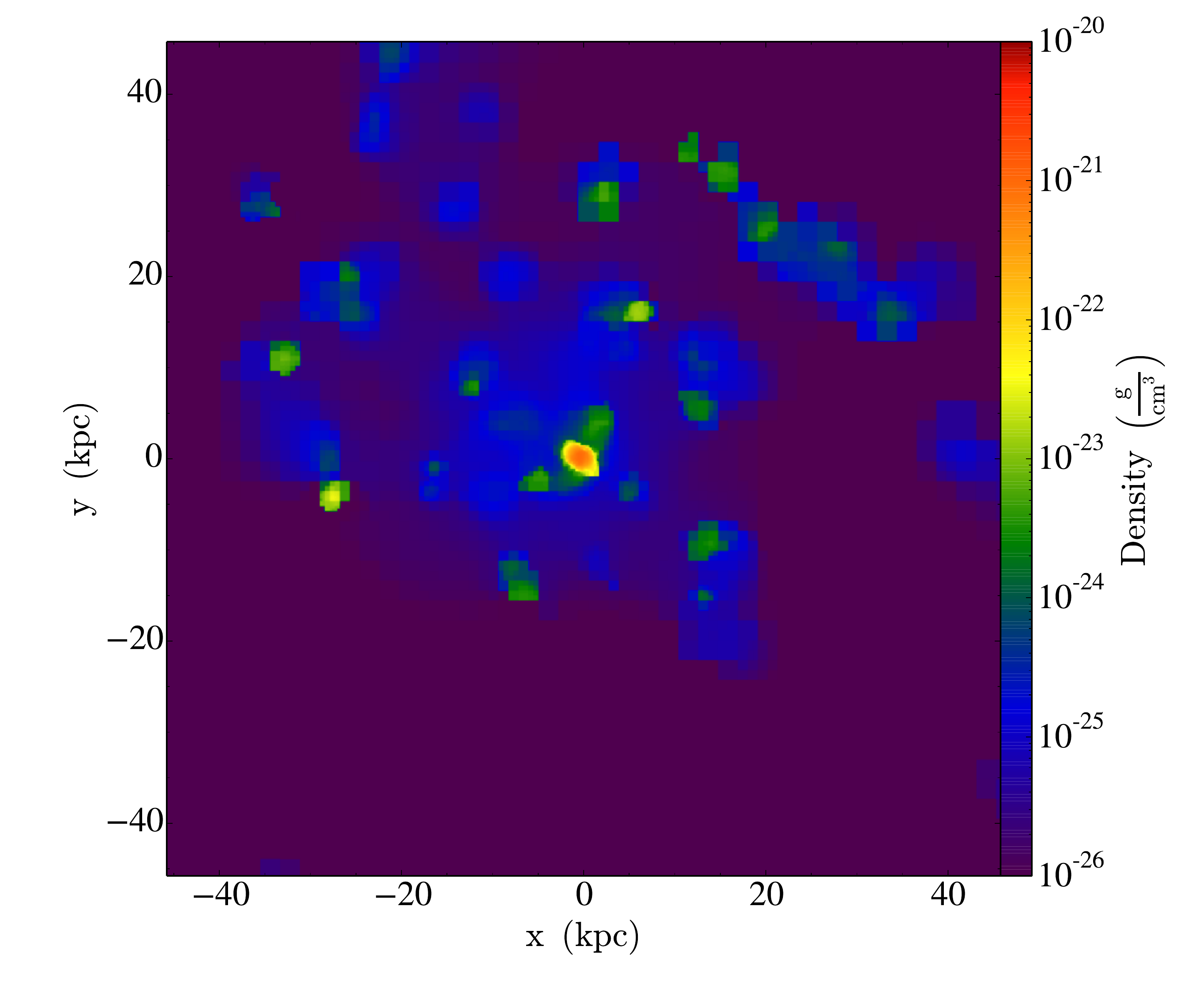}}
     \put(98.6,64){\makebox(0,0){\rotatebox{90}{\color{white} \textbf{AMR-FID}}}}
     \put(98.6,30){\makebox(0,0){\rotatebox{90}{\color{white} \textbf{SPH-FID}}}}
  \end{overpic}
\caption{Same as \fig{fig_fid_images5} but for $z=3$, where now $r_{\rm vir}=220$ physical kpc. The images thus measure about 3.5 Mpc across centered on the halo.}
\label{fig_fid_images3}
\end{figure*}
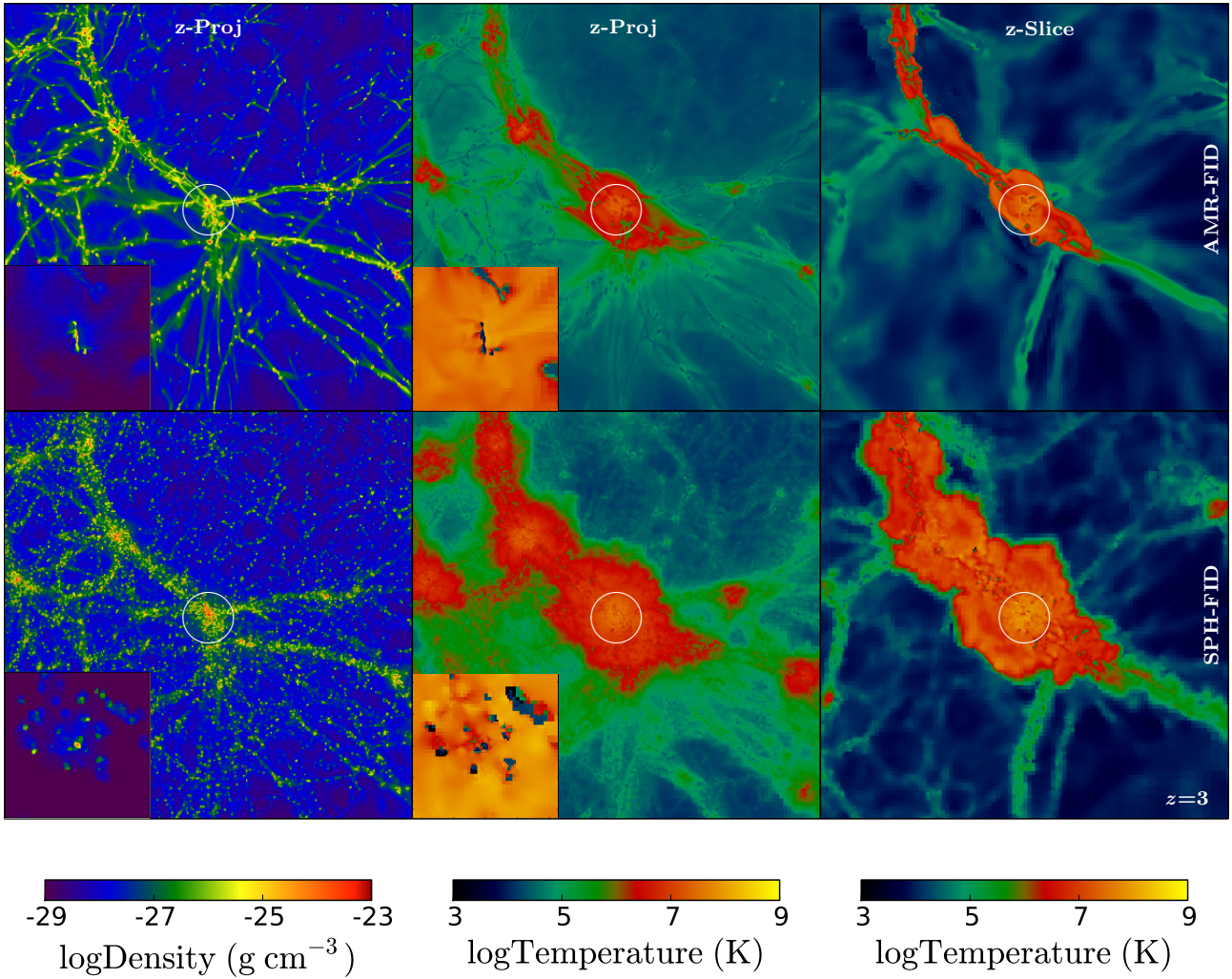
Next we carry out the same analysis at $z=3,$ which we will also use to compare with the \QSO\ runs, in the redshift regime where AGN feedback becomes more important. Density and temperature projections of the gas at this redshift are given in \fig{fig_fid_images3}. In density, the various accretion filaments have coalesced into a more condensed cosmic web that is less volume filling, with denser gas in the cosmic nodes. The temperature projections also show an increased virial temperature, that extends to a larger radius, and the filaments are also encased within post-shock heated gas. The virial shock radius in SPH and AMR are both now at or beyondthe virial radius at this redshift, since both the cooling time and the dynamical time are roughly 0.11 $H^{-1}$.  In AMR the filaments clearly remain relatively cold, while this is much more difficult to see in SPH. Looking at the temperature slices (right), indeed the filaments inside the shock-heated gas are cool in SPH, but they are comprised of smaller, more fragmented, gas clumps than in AMR, while the post shock gas is much more extended. The discrepant behavior, that is the overheating and undercooling at the shock, that explained the hotter gas in SPH at $z=5$ at the virial radius has had a runaway effect, and by $z=3$ its impact is even more extreme. In the galaxy-scale inset, it is clear that there is cold gas in both AMR and SPH, within which stars form, but this gas is more ordered in AMR, consistent with a higher specific angular momentum.

\begin{figure*}[t!] 
\centering
\includegraphics[width={1.6\columnwidth}]{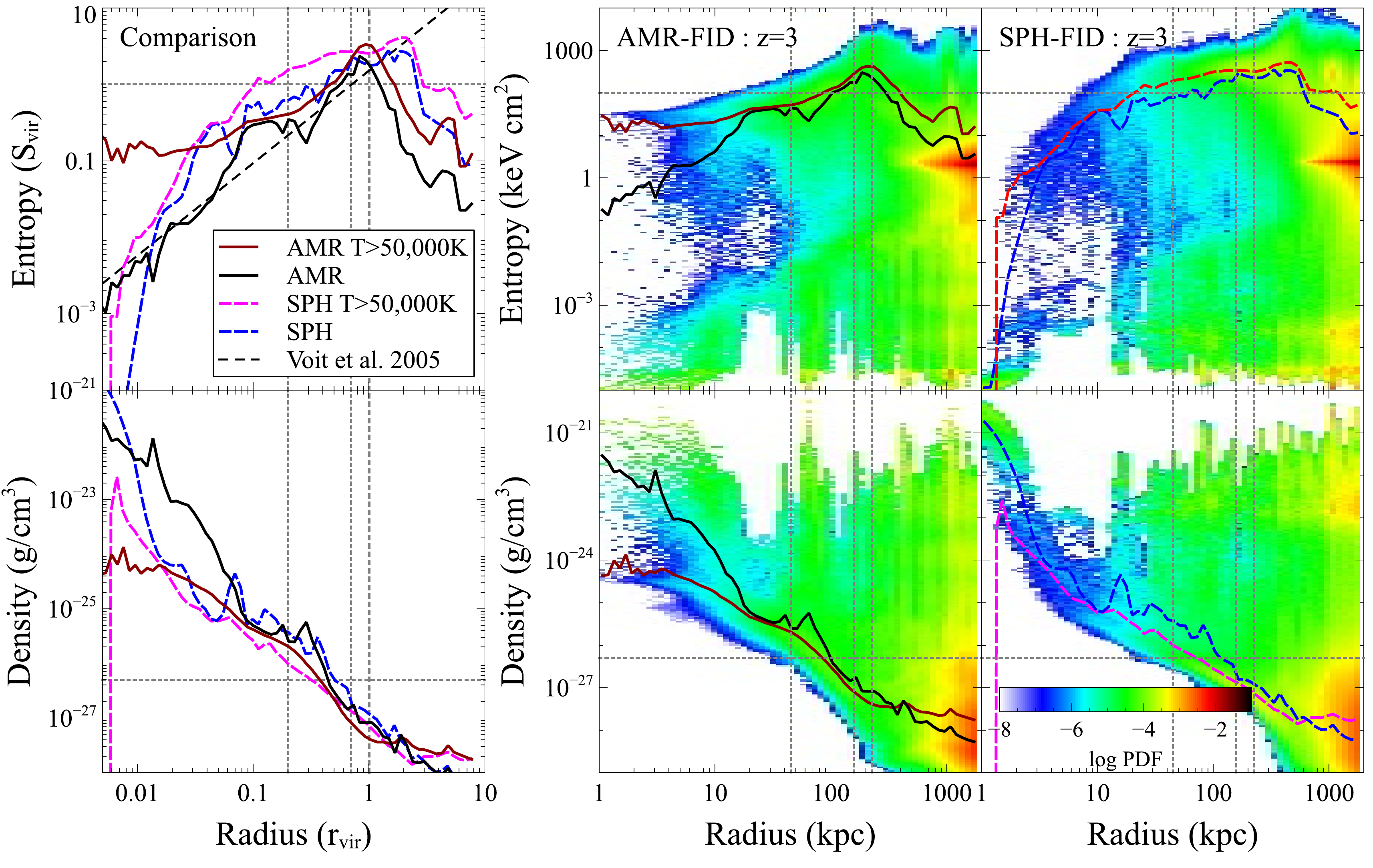}
\caption{Same as \fig{fig_fid_prof5}, but for $z=3$.\vspace{4pt}}
\label{fig_fid_prof3}
\end{figure*}

In the profile plots, presented in \fig{fig_fid_prof3}, we see the same qualitative behavior in entropy and density for AMR and SPH as at $z=5$. A similar trend in the entropy of the hot gas is visible in the volume-average profiles, in particular at small radii ($r<0.05r_{\rm vir}$) the SPH entropy is lower than in AMR, at intermediate radii ($0.05r_{\rm vir} < r < r_{\rm vir}$) the SPH entropy is larger than in AMR, at $r\simeq r_{\rm vir}$ the entropies are in good agreement, and out to $2r_{\rm vir}$ the SPH entropy is again higher, with temperatures near the virial temperature. However, the ratio of AMR and SPH entropies are larger than their values at $z=5,$ thus the behavior of the two codes is even less consistent. The densities, on the other hand, show the same level of consistency between the two codes at $z=3$ as at $z=5$.

Thus at both $z=5$ and $z=3$ with radiative cooling, star formation, and reionization, the overall trend is for standard SPH to have lower entropy cluster cores and larger entropy at larger radii  compared with AMR. This is consistent with other comparisons between standard SPH and AMR for non-radiative simulations (e.g., Voit et al. 2005; Mitchell et al. 2009), but oddly, this is not what is seen in comparisons between standard SPH and the moving mesh code {\tt AREPO}. In fact, \citet{Keres12} saw that for intermediate mass halos similar to our halo that the moving mesh simulations have even lower entropy values in the core and higher entropy values near the virial radius than the SPH runs. The authors argued that the moving mesh was capturing a cooling flow, and thus SPH was not capturing sufficient cooling in the gas. While this may be the case, the authors also discuss the mixing of low entropy gas at the center of the cluster in {\tt AREPO} which is not captured in {\tt GADGET}. Yet Mitchell et al. (2009) compared SPH to Eulerian simulations to show that such mixing  injects heat into the central region, leading to a higher, not lower entropy value. While Mitchell et al. (2009) did not include cooling, we still find the same behavior in our AMR simulations with cooling,  suggesting that dissipative heating is in fact sufficient to offset the cooling in the center. This is even with our assumed constant metallicity of one third solar, and an overestimate of the cooling from H and He (see \sect{methods}).  
\begin{figure}[t!] 
\centering
\includegraphics[width={0.9\columnwidth}]{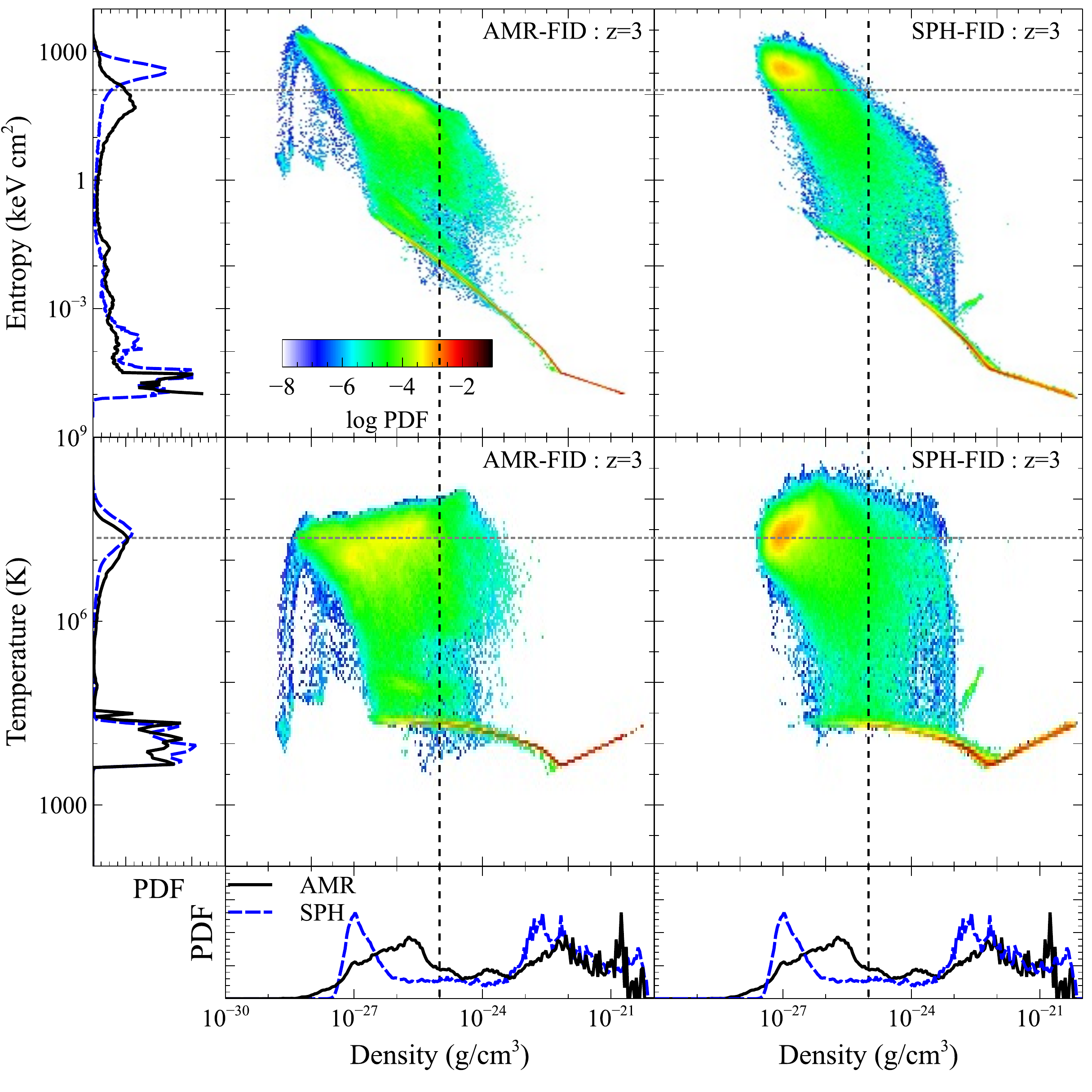}
\caption{Same as in \fig{fig_fid_phase5}, but for $z=3$.\vspace{4pt}}
\label{fig_fid_phase3}
\end{figure}

The tendency for SPH to have more diffuse, high entropy gas is made clearer in \fig{fig_fid_phase3}. The extra heating occurring at the virial radius has led to higher entropy gas in the SPH run, and this material cannot cool on a Hubble time ($S$ is above 100 keV cm$^2$; Oh \& Benson 2003). In AMR, this gas instead only reaches $S=10$ keV cm$^2$. The temperature profiles again are in good agreement, although the post-virial shock gas does still extended to slightly hotter temperatures. Thus SPH locks up more gas in the diffuse, high-entropy phase, which will have an impact on the amount of star-forming gas in the cluster. Finally, in SPH there is a small feature extending from the star-forming ISM to hotter, denser gas. This is a post-SNe feedback region, where gas is instantly moved to higher temperatures in the phase diagram, and then has its cooling artificially delayed. This results in the heated gas first expanding adiabatically, dropping in density and temperature, until as its cooling ramps up it cools more quickly and its entropy drops.

\subsection{AGN Feedback Runs}\label{gas_agn}

We now look at the \QSO\ simulations to see how the inclusion of AGN feedback impacts the halo and galaxy gas, and how AMR and SPH compare in this context. We first begin by comparing these results with the $z=5$ \NC\ and \FID\ results, followed by the \FID\ results at  $z=3.$  

The halo virial quantities for these simulations are listed in \tabl{tab_clust}. The inclusion of AGN feedback has led to a much lower gas fraction than in the \FID\ simulations, which is now consistent between the AMR and SPH simulations. As we show below, with AGN feedback the virial shock is now beyond the virial radius in both AMR and SPH, and it is for this reason that the baryonic fraction has dropped, as gas spends a longer time in the outer halo before cooling and falling within the virial radius.
\begin{figure*}[t!] 
\centering
   \begin{overpic}[scale=0.56]{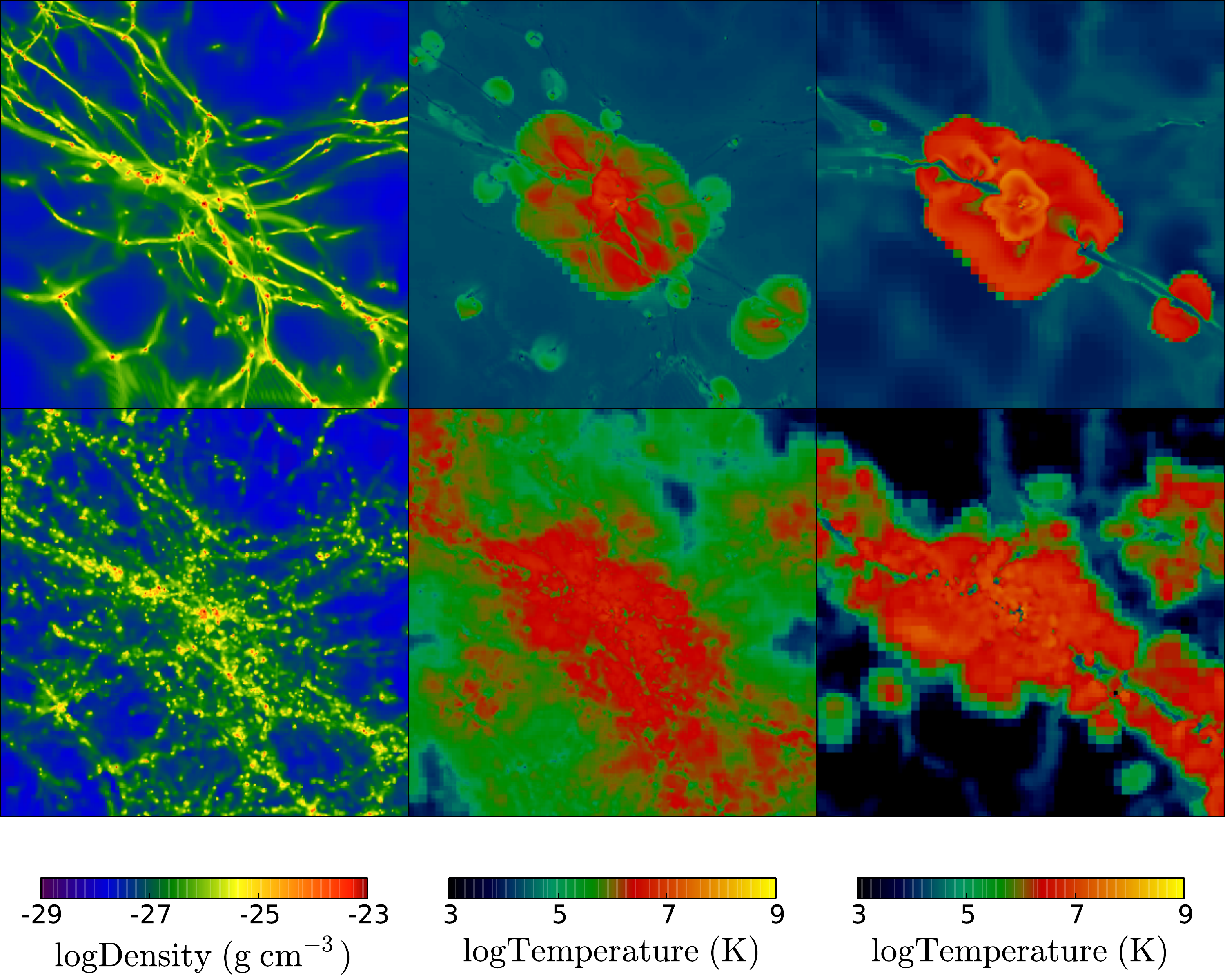}
     \put(16.6667,63.15){\color{white}\circle{4.1667}}
     \put(16.6667,29.75){\color{white}\circle{4.1667}}
     \put(50,63.15){\color{white}\circle{4.1667}}
     \put(50,29.75){\color{white}\circle{4.1667}}
     \put(83.3333,63.15){\color{white}\circle{4.1667}}
     \put(83.3333,29.75){\color{white}\circle{4.1667}}
     \put(16.7,78){\makebox(0,0){\color{white} \textbf{z-Proj}}}
     \put(50.6,78){\makebox(0,0){\color{white} \textbf{z-Proj}}}
     \put(84.6,78){\makebox(0,0){\color{white} \textbf{z-Slice}}}
     \put(96.6,15){\makebox(0,0){\color{white} \textbf{\textit{z}=5}}}
     \put(33.4,46.7){\includegraphics[scale=0.1, trim=  175 132 185.0 48, clip=true]{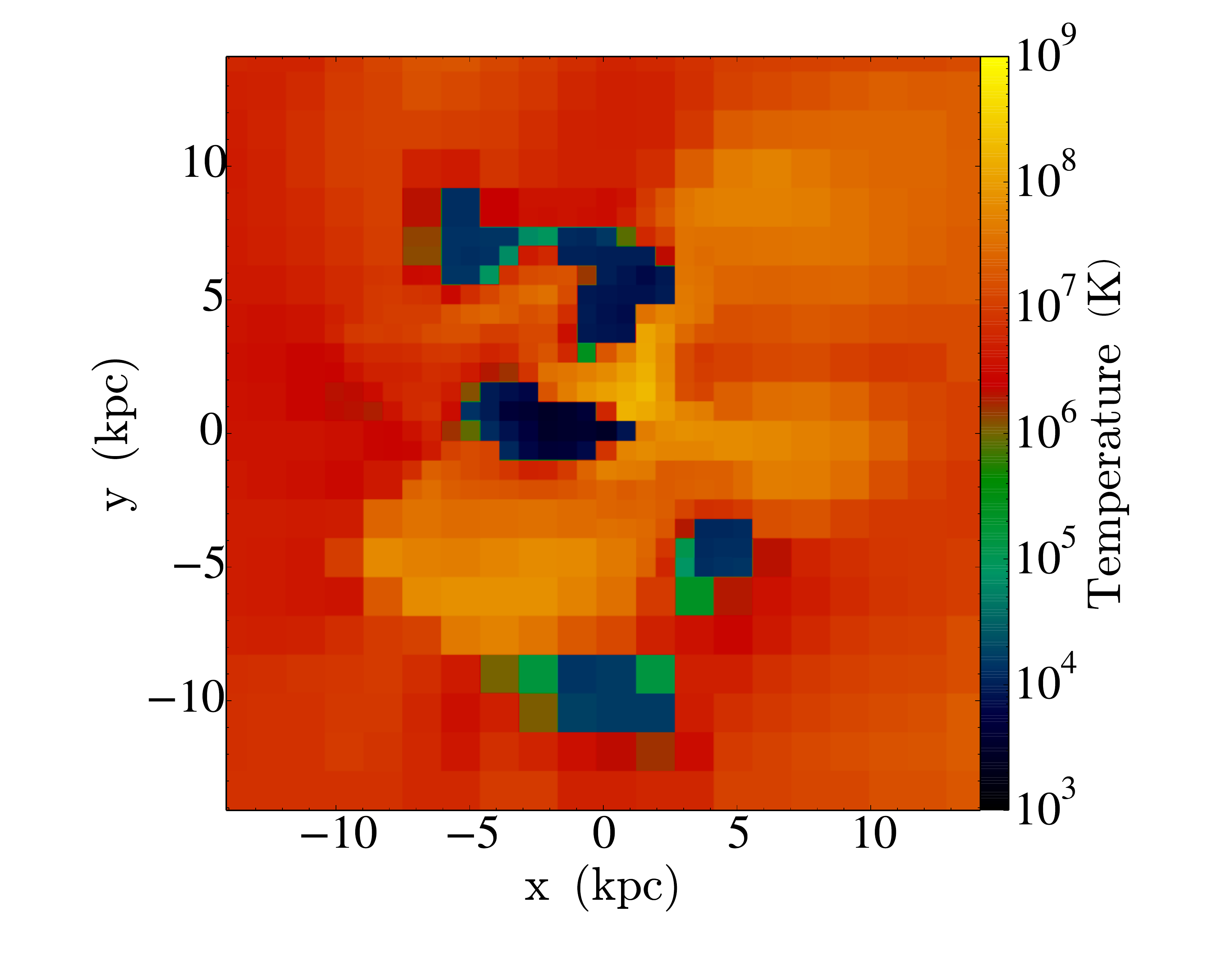}}
     \put(0.07,46.7){\includegraphics[scale=0.1, trim=  116 86 144.0 30, clip=true]{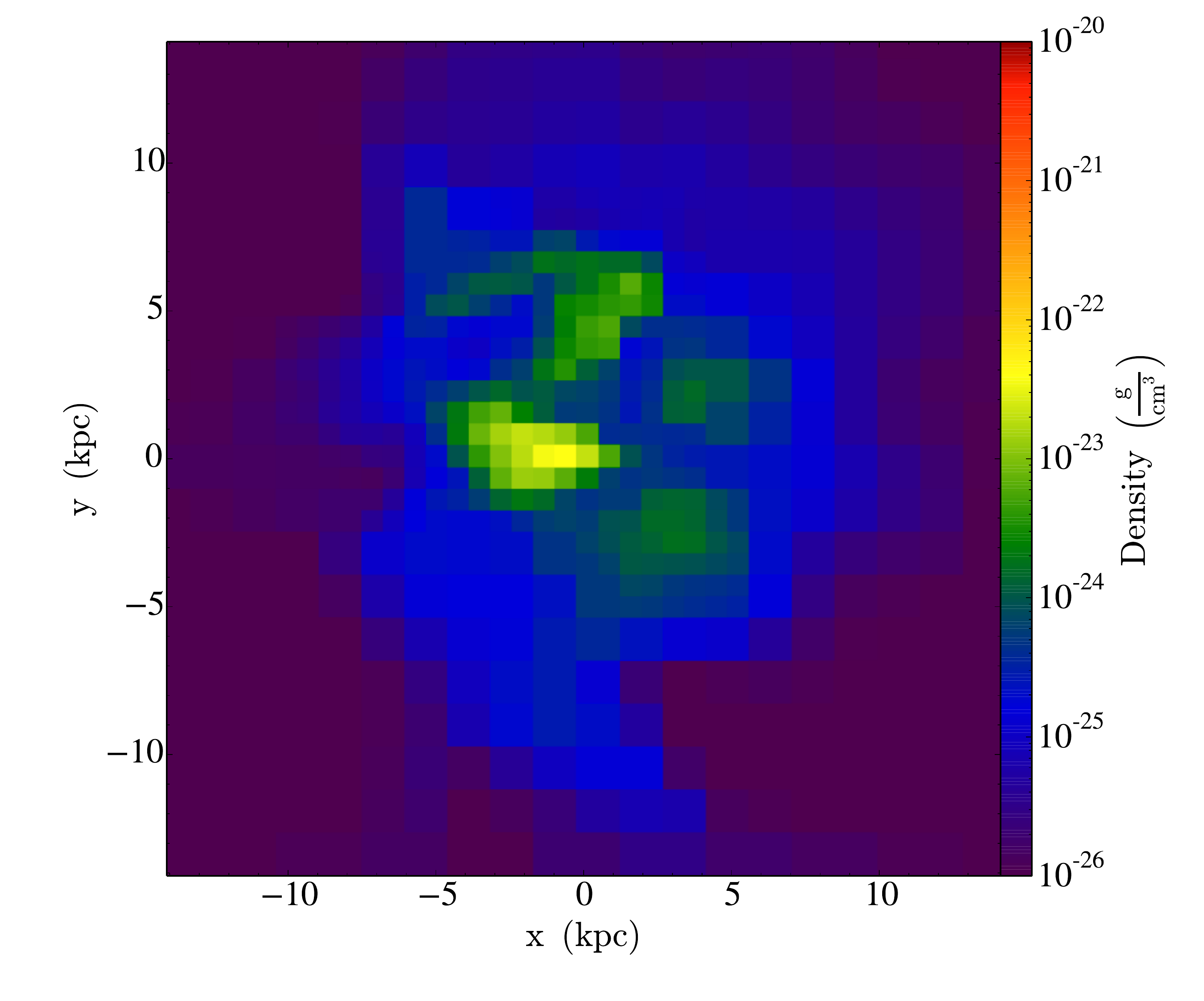}}
     \put(33.4,13.4){\includegraphics[scale=0.1, trim=  175 132 185.0 48, clip=true]{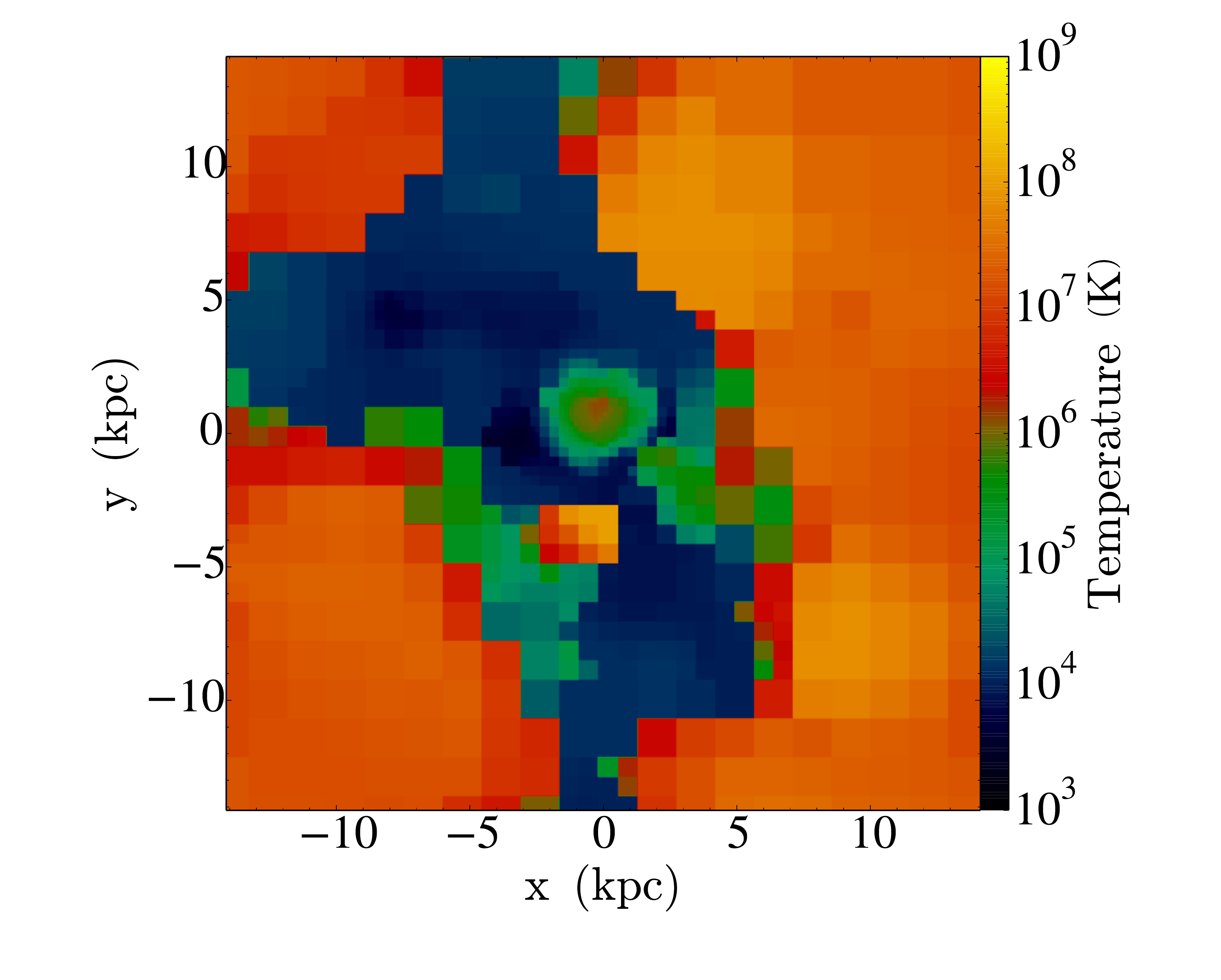}}
     \put(0.07,13.4){\includegraphics[scale=0.1, trim=  116 86 144.0 30, clip=true]{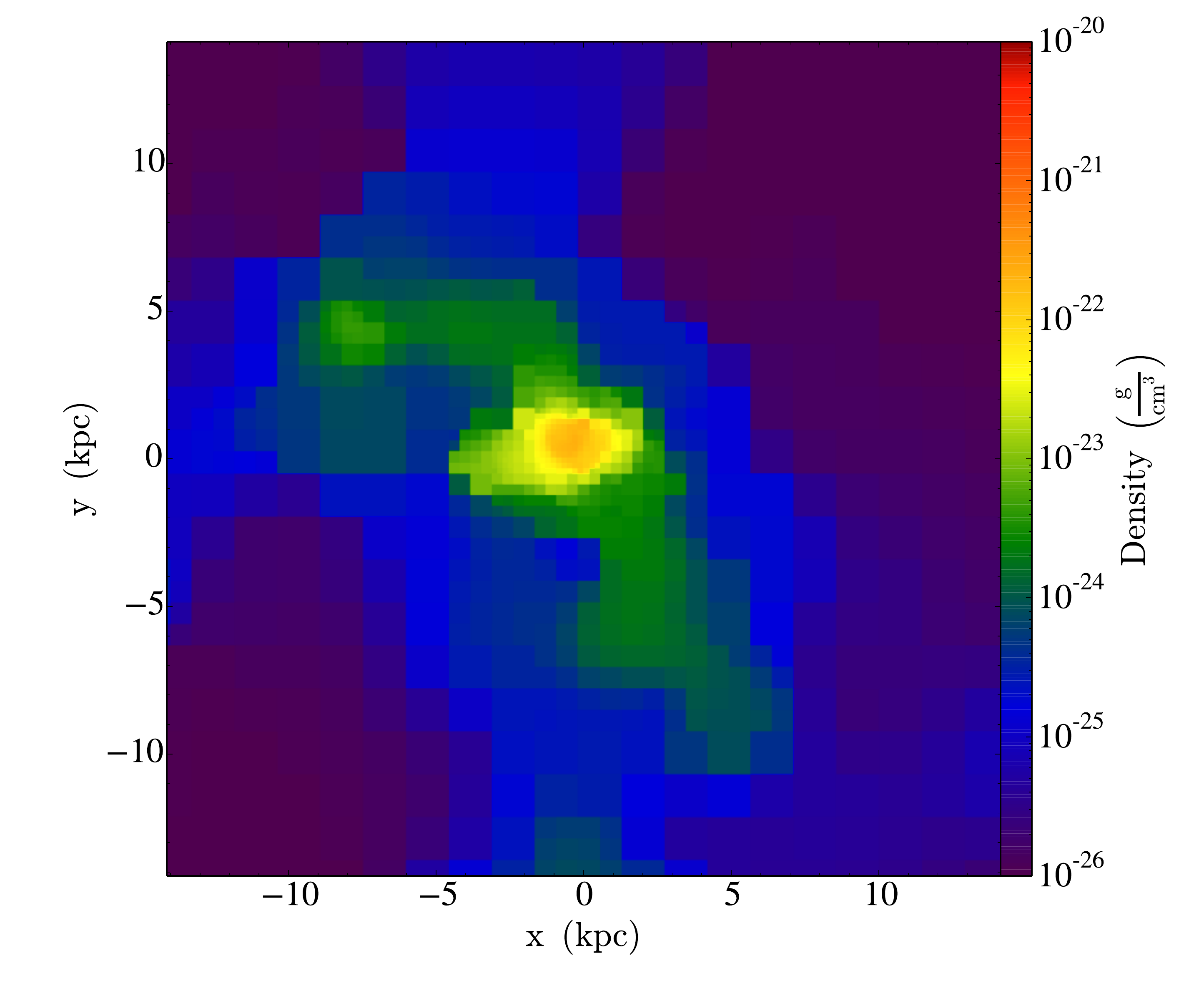}}
     \put(98.6,64){\makebox(0,0){\rotatebox{90}{\color{white} \textbf{AMR-QSO}}}}
     \put(98.6,30){\makebox(0,0){\rotatebox{90}{\color{white} \textbf{SPH-QSO}}}}
  \end{overpic}
\caption{Same as \fig{fig_fid_images5} but for the \QSO\ simulations.}
\label{fig_qso_images5}
\end{figure*}

In the left column of \fig{fig_qso_images5}, we show large-scale (out to 8 $r_{\rm vir}$) projections of the gas density for the \AMRQSO\ (top) and \SPHQSO\ (bottom) simulations at $z=5,$ to compare with Figures \ref{fig_adiab_images} and \ref{fig_fid_images5}. The introduction of AGN feedback results in very
little apparent impact on the density distribution at this redshift. This is mostly due to only a 30\% drop in overall gas fraction, which is difficult to see with the
scale spanning 6 orders of magnitude. However, under close comparison with \fig{fig_fid_images5}, we see that the densest regions in the \QSO\ runs have the largest decrement in density compared with the \FID\ runs, while the filaments inside the cluster are somewhat denser. In AMR we see that the filaments are slightly pushed off-center, and are thinner. In SPH, there is a lower number of collapsed clumps, suggesting that the feedback is offsetting the numerical dissipation. Finally, the diffuse halo medium inside the virial radius is slightly denser with AGN. Thus, while by $z=5$ AGN feedback is not effective in moving large amounts of matter out of the cluster environment, it does reduce the central densest peaks and delay gas accretion into the halo.   

\begin{figure*}[t!] 
\centering
\includegraphics[width={1.6\columnwidth}]{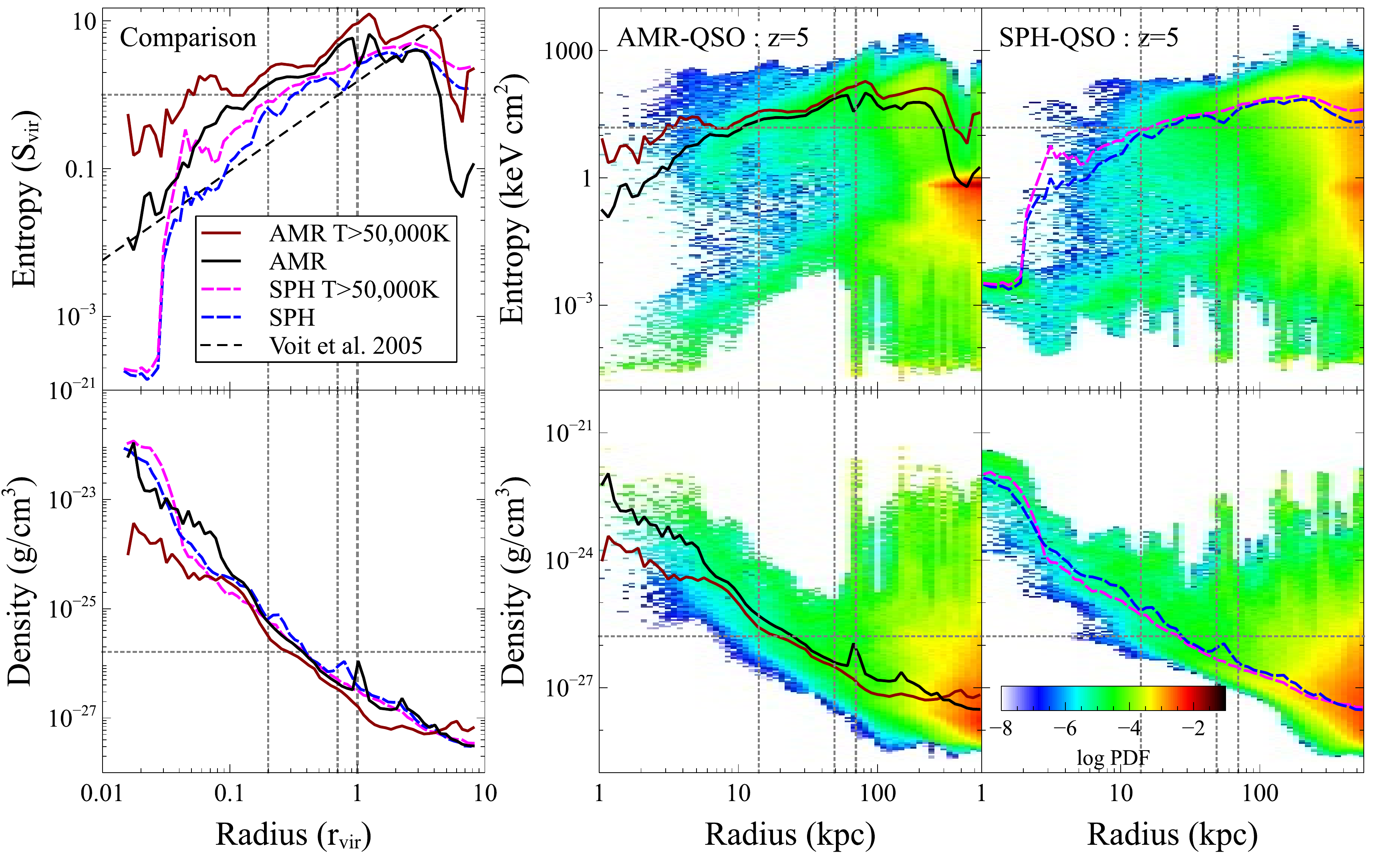}
\caption{Same as \fig{fig_fid_prof5} but for the \QSO\ simulations.\vspace{4pt}}
\label{fig_qso_prof5}
\end{figure*}

The middle and right columns of \fig{fig_qso_images5} present projections and slices of the gas temperature, respectively. While AGN feedback does not affect the gas density, it clearly affects the gas temperature. In both simulation types, the AGN feedback results in a larger volume of hot gas than in the \FID\ runs, with an increase in the hot gas volume-filling factor of roughly 10, and a virial shock radius well beyond the virial radius.  However, note that in both \FID\ and \QSO\ simulations the amount of heated gas is consistently greater for the SPH simulations. Given that more gas is heated in SPH, it is thus surprising that the temperature slices reveal that the two methods give a consistent picture of the halo gas (within the white circle).  The halo virial radii are 66 physical kpc and 70 physical kpc for \AMRQSO\ and \SPHQSO, respectively, and within the halo the gas is nearly uniform at the virial temperature of $8\times10^6$ K.  Although the AMR filaments are more continuous than in SPH, similar to what is seen in the \FID\ simulations, the cold filaments in \AMRQSO\ truncate  at a larger radius than in \AMRFID\ and point slightly away from the center of the cluster. The filaments' truncation radius is signified by a second shock located at the virial radius.  Thus, although it is not visible in the large-scale projections, the AGN is impacting filament gas on small scales in the AMR simulation, which is also seen in higher resolution runs in Dubois et al. (2013). We do not see a similar impact on the filaments in SPH however, and this is consistent with what is seen in Di Matteo et al. (2012).

We interpret the similarity of the AMR and SPH halo gas as follows: although the injection scale of the AGN feedback for the two methods is not identical, the AGN eventually heats the halo gas sufficiently to self-regulate its accretion, and this self-regulating gas configuration occurs at a ``critical entropy''  (Oh \& Benson 2003; Scannapieco \& Oh 2004). We find that this critical entropy is largely code independent. However, it is clear that SPH causes more collateral gas heating, such that the gas at very large radii is heated to $10^5$ K in the process of increasing the halo gas to the same critical entropy. This becomes much clearer at $z=3,$ as discussed below.

Beyond the halo, we see that the intergalactic medium in the SPH simulation is significantly cooler than in  the AMR simulation.   This difference is unfortunately caused by  an erroneous switch that turned off reionization heating for gas below 500 K in our \SPHQSO\ simulation.  However, the shock increases the temperature in AMR by three orders of magnitude, and therefore the fact that the IGM gas is artificially cooled results in little impact on the halo itself. 
  
The impact of AGN feedback on the gas profiles of the cluster environment at $z=5$ are presented in \fig{fig_qso_prof5}.
Even though the temperature projections show a strong signature of AGN feedback in both simulations, the average profiles in this figure are largely similar to the fiducial cases shown in \fig{fig_fid_prof5}.   However, in the \QSO\ simulations, the distribution of gas at very high entropies has increased, extending roughly an order of magnitude higher than in \fig{fig_fid_prof5}, with the  largest increases occurring at very small radii ($r < 20$ physical kpc) and very large radii ($r > 100$ physical kpc). Additionally, the inclusion of AGN has resulted in less low entropy gas at 10-20 kpc in the AMR simulation, while in the SPH simulation there is less low-entropy gas ($S < $ 0.001 \kevcs) at all radii, but more mid-entropy gas ($S \simeq $ 0.1 \kevcs) at 5-20 physical kpc.   Finally, by comparing the \QSO\ results with the non-radiative profile given in Voit et al. (2005), we see that the inclusion of AGN feedback has resulted in a higher average entropy value at all radii for AMR, and a higher average entropy for the central gas in the SPH case.

AGN feedback acts to diminish the SPH and AMR central density values and average values, bringing the two methods into better agreement. However, the hot SPH gas extends to higher densities, indicative of a very recent feedback event in which the gas has not yet had an opportunity to expand. This central, hot, feature is visible in the insets  shown in \fig{fig_qso_images5}, which also show the presence of more cool gas in SPH,  explaining the lower entropy profile.  Overall there is much better agreement between the two methods than in the \FID\ runs, where the AGN heating results in similarly increased entropy profiles  necessary for self-regulating their accretion. 

\begin{figure}[t!] 
\centering
\includegraphics[width={0.9\columnwidth}]{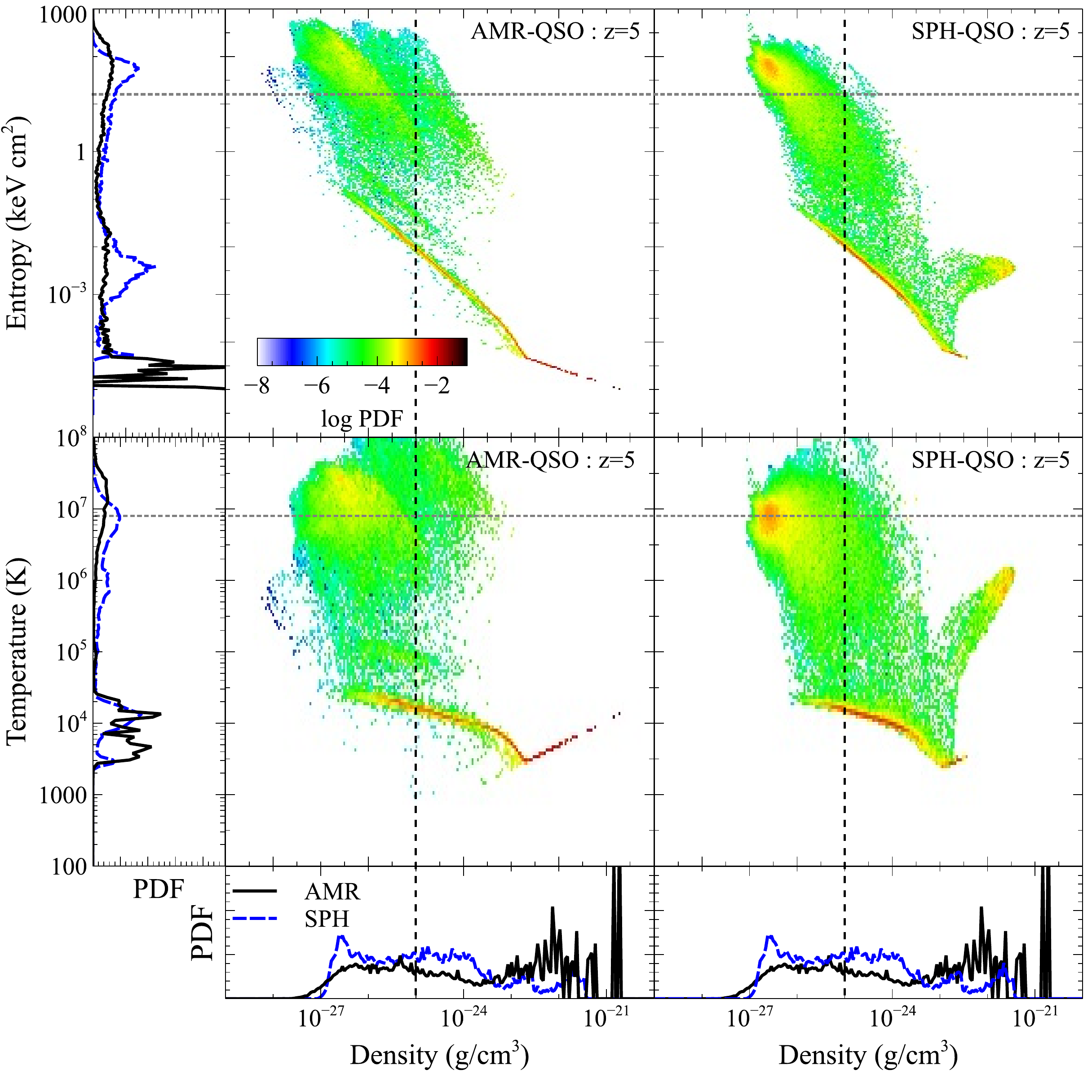}
\caption{Same as in \fig{fig_fid_phase5}, but for the \QSO\ simulations.}
\label{fig_qso_phase5}
\end{figure}

\begin{figure*}[t!] 
\centering
   \begin{overpic}[scale=0.56]{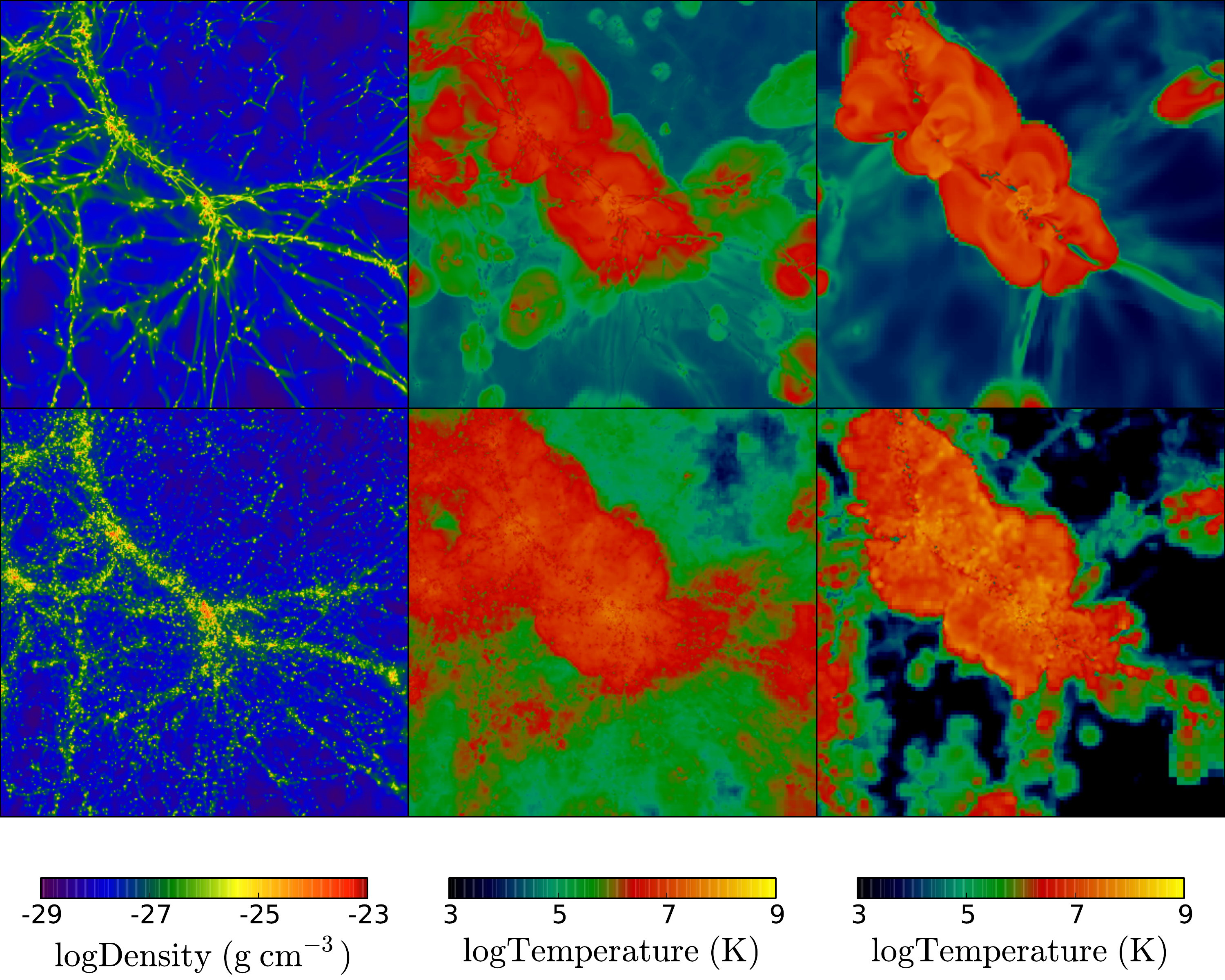}
     \put(16.6667,63.15){\color{white}\circle{4.1667}}
     \put(16.6667,29.75){\color{white}\circle{4.1667}}
     \put(50,63.15){\color{white}\circle{4.1667}}
     \put(50,29.75){\color{white}\circle{4.1667}}
     \put(83.3333,63.15){\color{white}\circle{4.1667}}
     \put(83.3333,29.75){\color{white}\circle{4.1667}}
     \put(16.7,78){\makebox(0,0){\color{white} \textbf{z-Proj}}}
     \put(50.6,78){\makebox(0,0){\color{white} \textbf{z-Proj}}}
     \put(84.6,78){\makebox(0,0){\color{white} \textbf{z-Slice}}}
     \put(96.6,15){\makebox(0,0){\color{white} \textbf{\textit{z}=3}}}
     \put(33.4,46.7){\includegraphics[scale=0.1, trim=  175 132 185.0 48, clip=true]{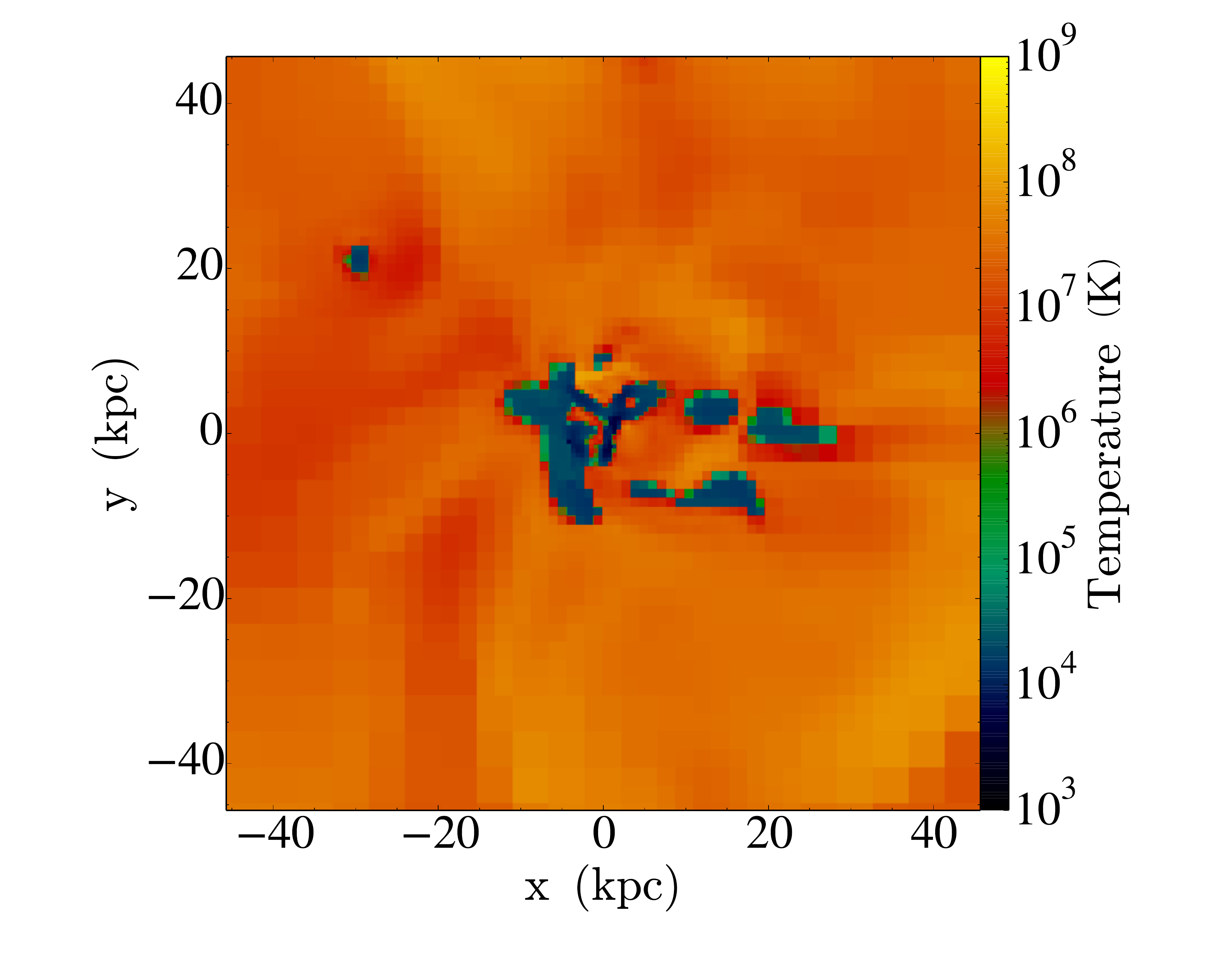}}
     \put(0.07,46.7){\includegraphics[scale=0.1, trim=  116 86 144.0 30, clip=true]{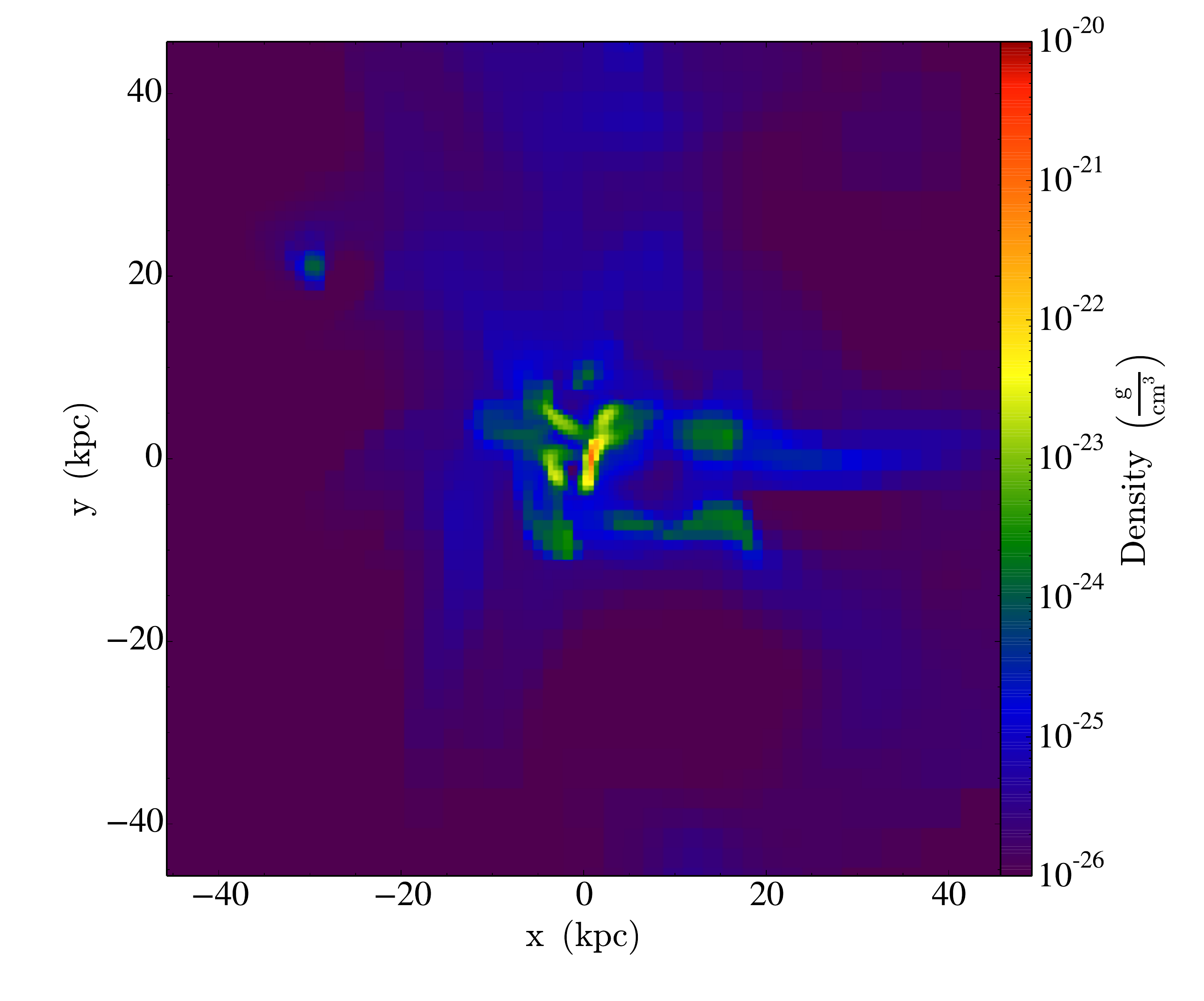}}
     \put(33.4,13.4){\includegraphics[scale=0.1, trim=  175 132 185.0 48, clip=true]{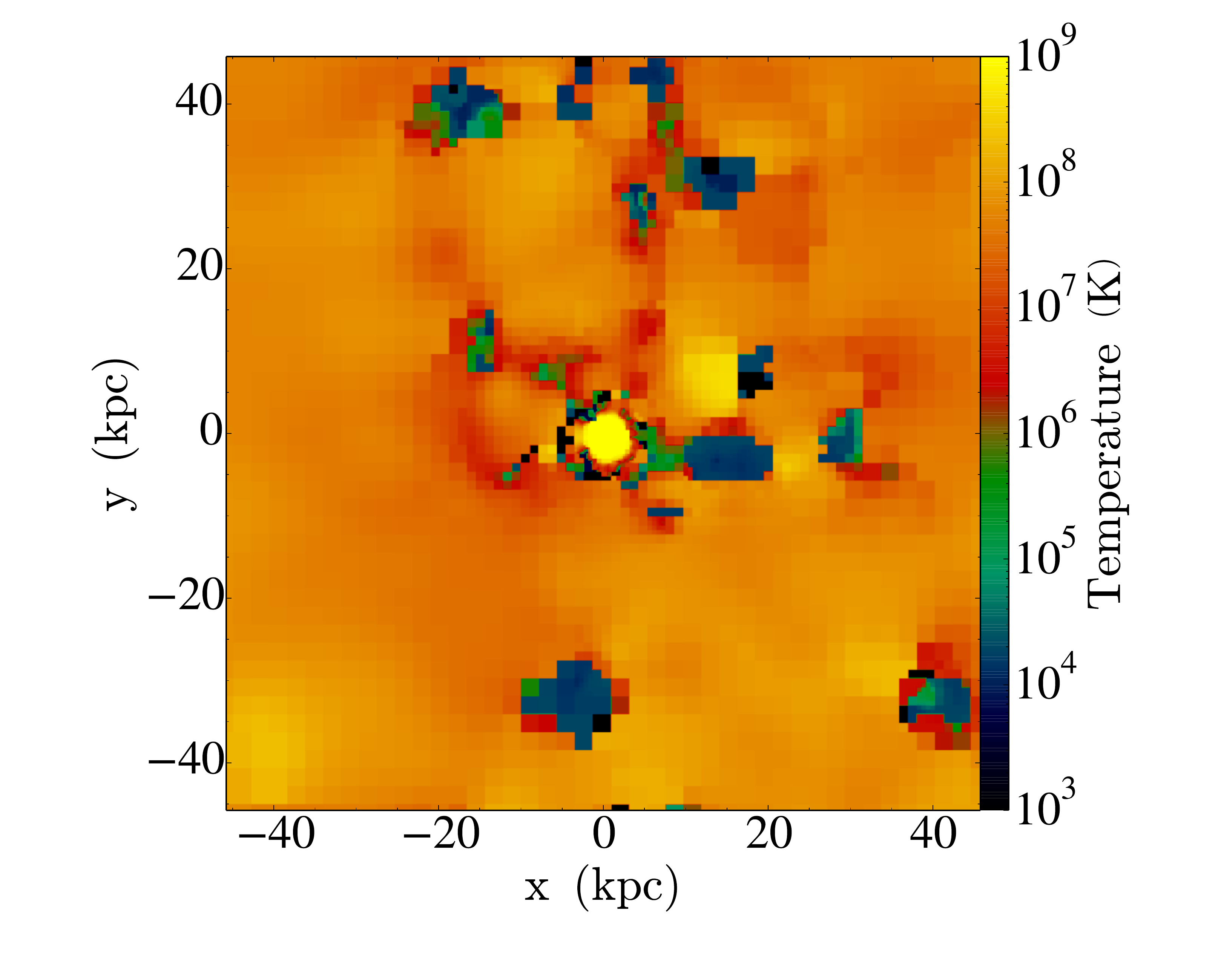}}
     \put(0.07,13.4){\includegraphics[scale=0.1, trim=  116 86 144.0 30, clip=true]{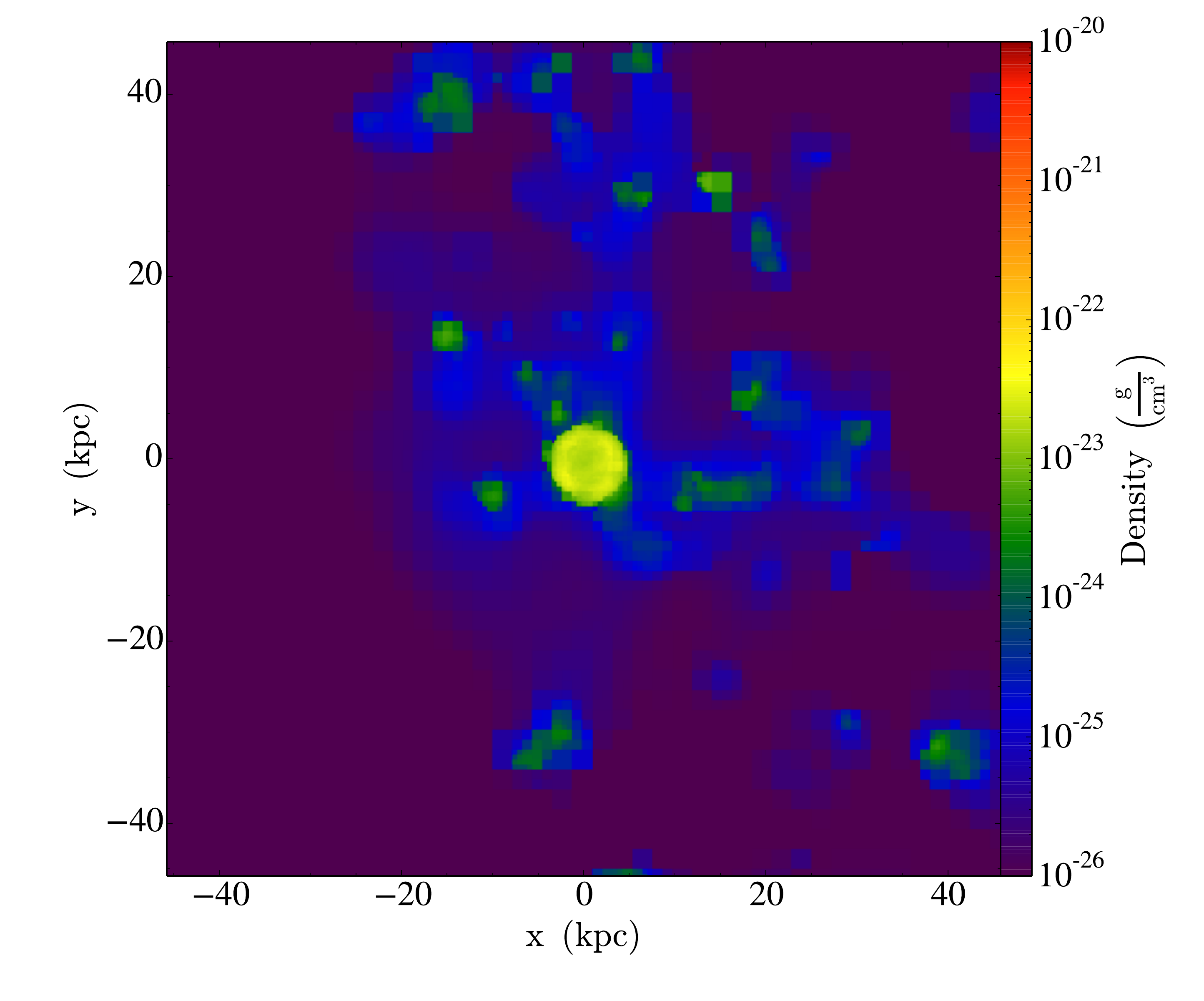}}
     \put(98.6,64){\makebox(0,0){\rotatebox{90}{\color{white} \textbf{AMR-QSO}}}}
     \put(98.6,30){\makebox(0,0){\rotatebox{90}{\color{white} \textbf{SPH-QSO}}}}
  \end{overpic}
\caption{Same as \fig{fig_qso_images5} but for $z=3$.}
\label{fig_qso_images3}
\end{figure*}

In \fig{fig_qso_phase5} we show the phase diagrams of the $z=5$ gas within the virial radius. 
The one-dimensional PDF plots of the entropy make it much clearer that the amount of high-entropy gas has increased by including AGN feedback.  In \AMRQSO, much more gas has $S \ge 100$ \kevcs\ than in \AMRFID, and this high entropy gas is found over a range of densities, from  $\rho = 3\times10^{-27} \gcc$ to the star-formation threshold density of $10^{-25} \gcc$. In SPH, there is a smaller increase in gas with $S \ge 100$ \kevcs, and most of this is  at low densities $\approx 3\times10^{-27} \gcc$. Thus, within the halo AGN are heating gas at a range of densities in AMR, preventing it from cooling to very high densities, while in SPH,  AGN are more efficient  at heating the surrounding diffuse gas. This increase in high entropy, hot gas is also  shown in the one-dimensional PDF plots of temperature, which illustrate that  in the AMR case the AGN heat less gas but this gas is heated to
higher entropies and temperatures than in the SPH case. This is partially due to the way the feedback energy is injected into the simulation, with AMR distributing the energy to the neighboring cells of the BH particle, and SPH distributing the energy over the BH particle's smoothing length.  

Finally, as shown in the profile plots, in SPH the AGN is causing significant heating of dense gas in the polytrope. This gas occupies the high-density region of the SPH plots,  where the very dense but also hot gas is unphysical. Yet it appears to be a long-lived feature. This occurs when a sink particle resides in a  cold, compact, dense clump (such as seen in the inset of \fig{fig_fid_images5}) but
it is not sufficiently massive  for Eddington-limited feedback to overcome the clump's binding energy. Instead, the AGN must grow while keeping the clump at a hot temperature, until finally dumping sufficient energy into the clump to overcome gravity.

\begin{figure*}[t!] 
\centering
\includegraphics[width={1.6\columnwidth}]{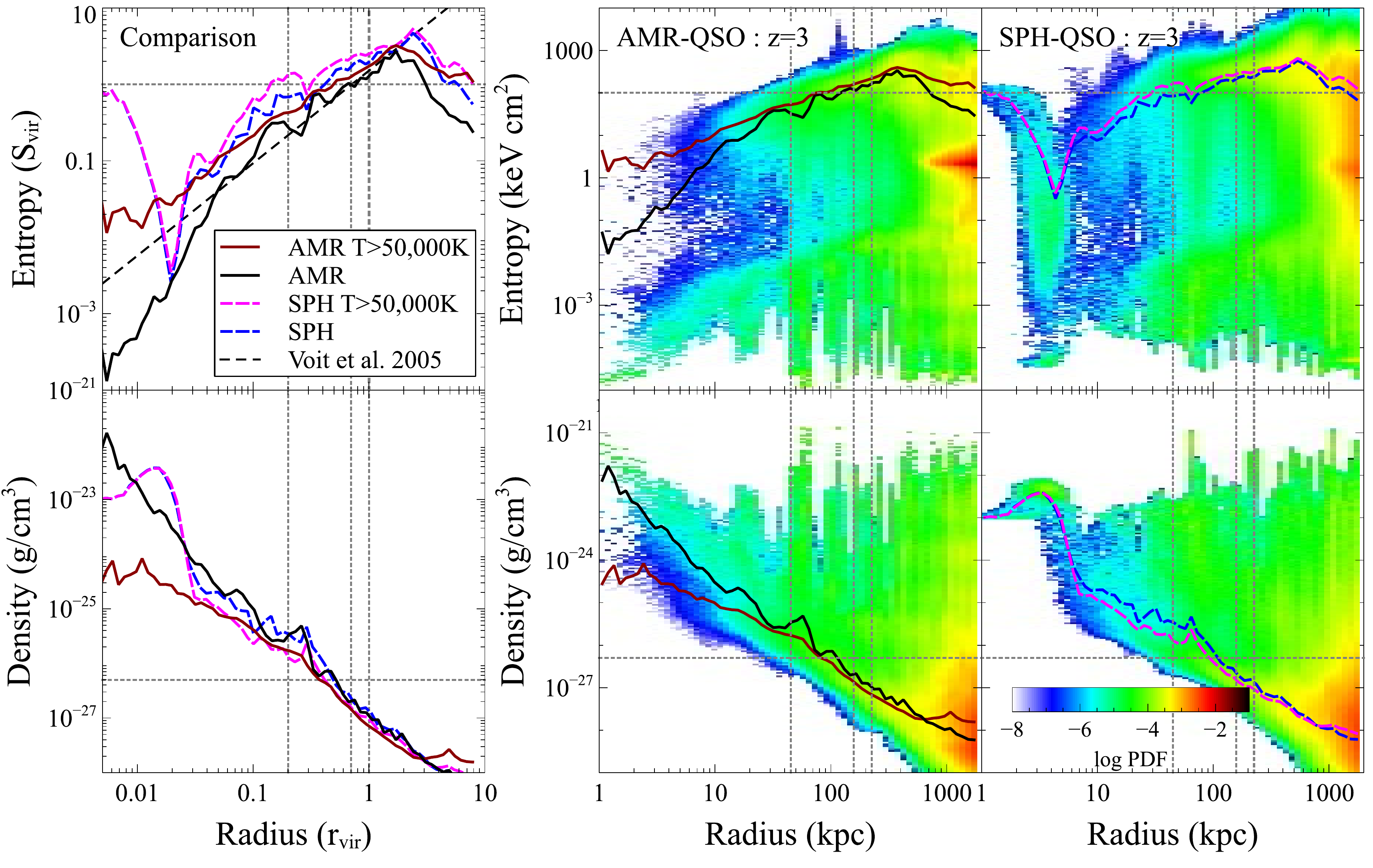}
\caption{Same as \fig{fig_qso_prof5}, but at $z=3$.}
\label{fig_qso_prof3}
\end{figure*}

Moving to $z=3$, projections of gas density and projections and slices of gas temperature out to 8 $r_{\rm vir}$ are shown in \fig{fig_qso_images3}. We
see the same small effects as at $z=5$. Thus, even at later times when AGN have a larger impact, thermal feedback does not lead to large-scale redistribution of gas.  The temperature projections now present significantly more extended hot gas, which, if we consider all gas heated above the typical IGM temperature of $10^4$ K, is more volume filling in SPH.  However, if  we consider the region with projected temperature above $10^{6.5}$ K, then we again see that there is fairly good agreement between the two methods. This region in SPH is only slightly larger and  reaches temperatures only slightly hotter than in AMR.  Thus we find that the methods produce a halo gas temperature distribution that is much more consistent than in the $z=3$ \FID\ runs and the \QSO\ runs at $z=5$. By $z=3$ the AGN has heated the gas to  roughly the same entropy in both methods, the entropy required for self-regulation.

This agreement on intermediate and large-scales aside, the inserts, showing the gas density and temperature in the inner 0.2$r_{\rm vir}$, do show a few differences. Thus while the  AGN feedback acts to make the bulk of the halo material consistent thermodynamically, on smaller scales the gas is impacted differently. In SPH there is a more centrally collapsed, hot, gas clump, which is broader in density than in temperature, and whose outskirts are denser than its interior. This region is in the process  of a feedback-driven expansion, leading to a shock front with higher density and therefore increased cooling. This expansion phase is made clearer in the gas profiles shown in \fig{fig_qso_prof3}. In SPH the gas density peaks outside of the center at roughly 4 physical kpc, where entropy drops. We have confirmed that this is not due to an improper choice of halo center.

On intermediate and large scales in \fig{fig_qso_prof3} there is again better agreement between the two methods. Outside of the central clump, the AGN has led to a much larger amount of gas at high entropy at large radii, compared with the \FID\ simulations in \fig{fig_fid_prof3}. Within the virial radius, the average entropy and density profiles are even more consistent than in the \FID\ and $z=5$ \QSO\ runs. Given that in SPH there is a central clump with higher density, we would expect to see lower entropy in the core of the SPH halo. However, the gas has finally been heated sufficiently to drive an outflow, giving a central peak in the SPH entropy. We have looked at $z=3.1$, and found at this earlier time that the central SPH entropy is indeed lower than in AMR. Thus the inversion at $z=3$ is a recent event, and in general lower entropy cores in SPH are more common.

\begin{figure}[t!] 
\centering
\includegraphics[width={0.9\columnwidth}]{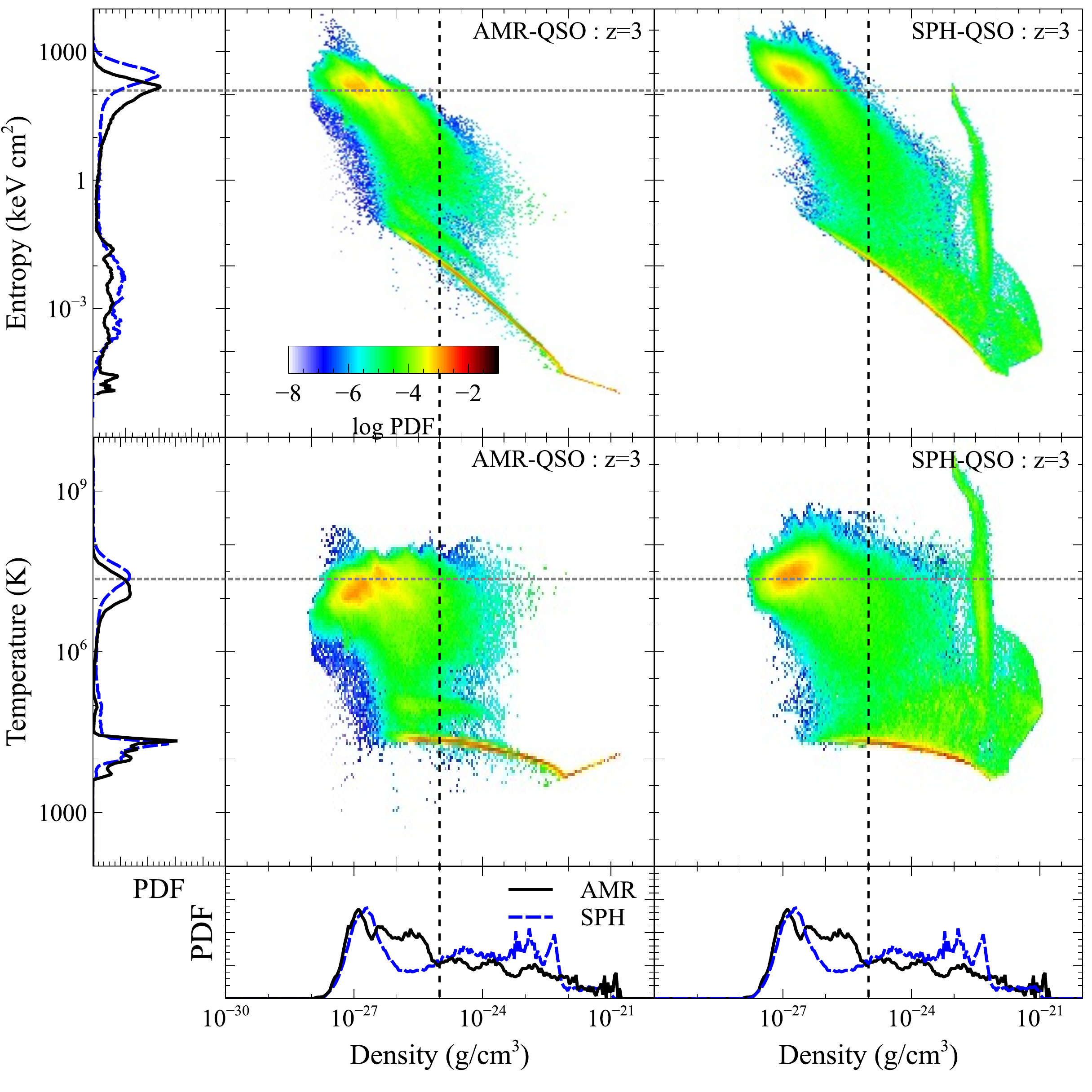}
\caption{Same as \fig{fig_qso_phase5}, but at $z=3$. }
\label{fig_qso_phase3}
\end{figure}
Finally, in \fig{fig_qso_phase3} we compare the phase plots at $z=3$ for the cluster. We can see the expanding central clump in the SPH diagram, where a large 
plume of isopycnic gas has continuously been heated via AGN feedback until reaching a sufficiently high temperature that the pressure can unbind the clump.
However, note that although this unphysical region appears quite large, it contains only a small fraction of the gas in the halo. Instead, a 
greater amount of gas is found near 10$^{-27} \gcc$ in both methods. The AGN feedback has led to similar amounts
of gas in the hot diffuse halo, although in AMR there is more gas held at $10^4$ K just below the star formation density threshold, while this gas is found above this threshold in SPH. This
is due to the clumping nature of the cold gas in SPH, where hot gas is less able to disrupt these clumps through instabilities. In AMR, on the other hand, the dense gas forms
streams instead of clumps, with larger surface area that is more susceptible to turbulence. Thus in AMR the AGN feedback is better able to dissolve and/or physically move the very inner filamentary gas into the warmer, diffuse medium, consistent with what has been seen in previous work (e.g., Dubois et al. 2013; Nelson et al. 2015).

\begin{figure*}[t]
\centering
\includegraphics[scale=0.54]{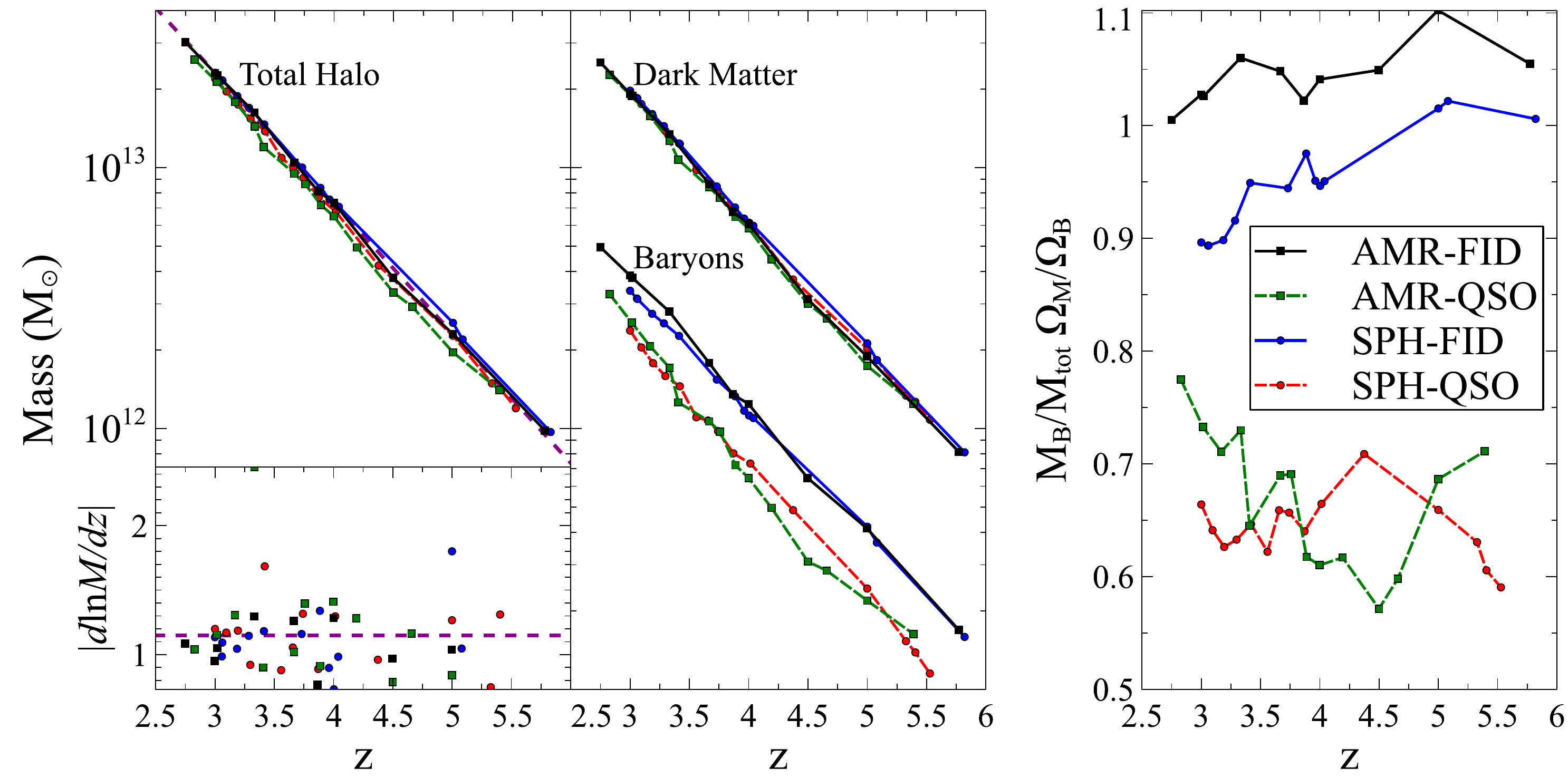}
\caption{Top Left: Evolution of the total halo mass of the cluster. The \AMRFID\ (\SPHFID) simulation  is presented with a solid black (blue) line where we mark the measured values in an 
output by squares (circles). The \AMRQSO\ (\SPHQSO) simulation is presented with a dashed green (red) line. The magenta dotted line gives an exponential mass
law, $M=M_{\rm f}e^{-\alpha (z-z_{\rm f})}=2.3\times10^{13}e^{-1.15(z-3)}$, following Wechsler et al. (2002). Bottom Left: Evolution of the halo mass accretion histories, given by $d$ln$M/dz$. Line and point styles match the above plot, 
with the magenta line giving a constant value of 1.15, consistent with the exponential parameterization for $M(z)$. Middle: Evolution of the dark matter and baryon components. Right: Evolution
of the gas fraction normalized by the cosmic mean baryon fraction.}
\label{fig_halo_evo}
\end{figure*}

In summary, the inclusion of AGN has lead to a more consistent halo environment for the two methods. This is suggestive of a self-regulation scenario, which becomes stronger at later times. The result is that the AGN heats the halo environment to sufficiently high entropies so that it turns off further gas accretion. It is very interesting that, even though
the two methods cannot probe precisely the same spatial scales, the resulting entropy is very similar. On large scales in SPH, however, the result of AGN feedback heating the halo gas to this self-regulatory temperature is that a much larger volume of gas is also impacted. This would have interesting implications for surveys of the thermal Sunyaev-Zel'dovich effect (e.g., Spacek et al 2016).

\subsection{Evolution of Gas, Stars and Black Holes}\label{evo}
\subsubsection{Halo Gas}\label{haloevo}

Finally, we turn to a more detailed study of the halo's evolution. In \fig{fig_halo_evo} we summarize the mass history of the halo, taken from 
a range of outputs from roughly $z\simeq5-3$.  The total halo mass evolution is consistent with pure exponential growth, similar to that seen for a range of halos in \citet{Wechsler02}. Using an offset at $z=3$, so that 
the leading constant is roughly the mass of the halo at the end of our simulations, we get an equation of the form 
\begin{equation}\label{massfit}
M_{\rm tot}(z) = M(z=3) e^{-\alpha (z-3)},
\end{equation}
where we find $M(z=3) = 2.3\times10^{13} \Msun$, and $\alpha=1.15$.  This $\alpha$ value is consistent with the most massive objects considered in \citet{Wechsler02}. By breaking up the mass into components, we see the dark matter mass is more consistent between the AMR and SPH runs than the baryons. 
The \AMRFID\ run has a higher baryon fraction than the cosmic mean at all times, but this fraction drops with time. This is due to the virial temperature increasing as the cluster grows,  leading to longer cooling times at further radii, slowly depleting the amount of baryons that fall inside the halo. In the \SPHFID\ run we see similar behavior, with the gas fraction dropping in time. 
However, the gas fraction starts out at roughly the cosmic mean, and then drops more quickly than in AMR. This difference is consistent with a higher post-shock temperature in SPH that is at  a larger radius than in AMR.  The inclusion of AGN feedback has little impact on the parameterization of the halo's growth. By looking at the effective slope, we can see there is a bit more spread in the \QSO\ runs, with a slightly shallower $\alpha$ at large redshifts that becomes steeper at lower redshifts.

\begin{figure}[t!] 
\centering
\includegraphics[scale=0.4, trim=0 230 0 0, clip=true]{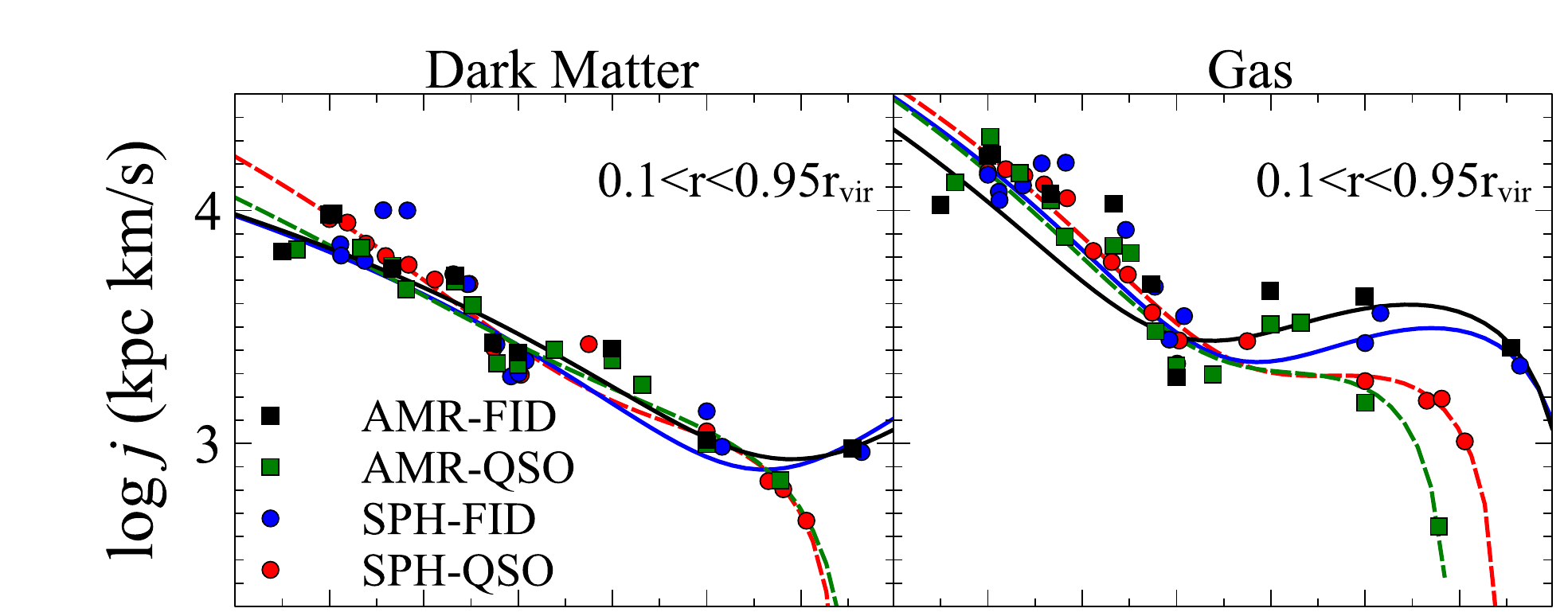}
\includegraphics[scale=0.4, trim=0 230 0 32, clip=true]{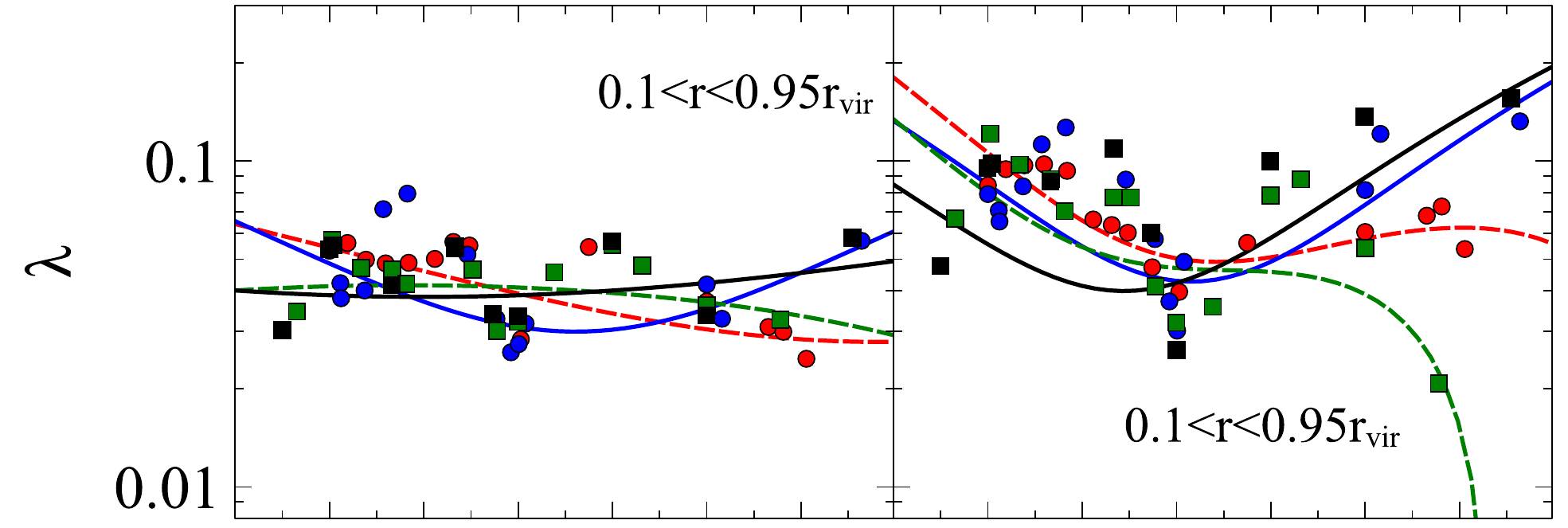}
\includegraphics[scale=0.4, trim=0 45 0 212, clip=true]{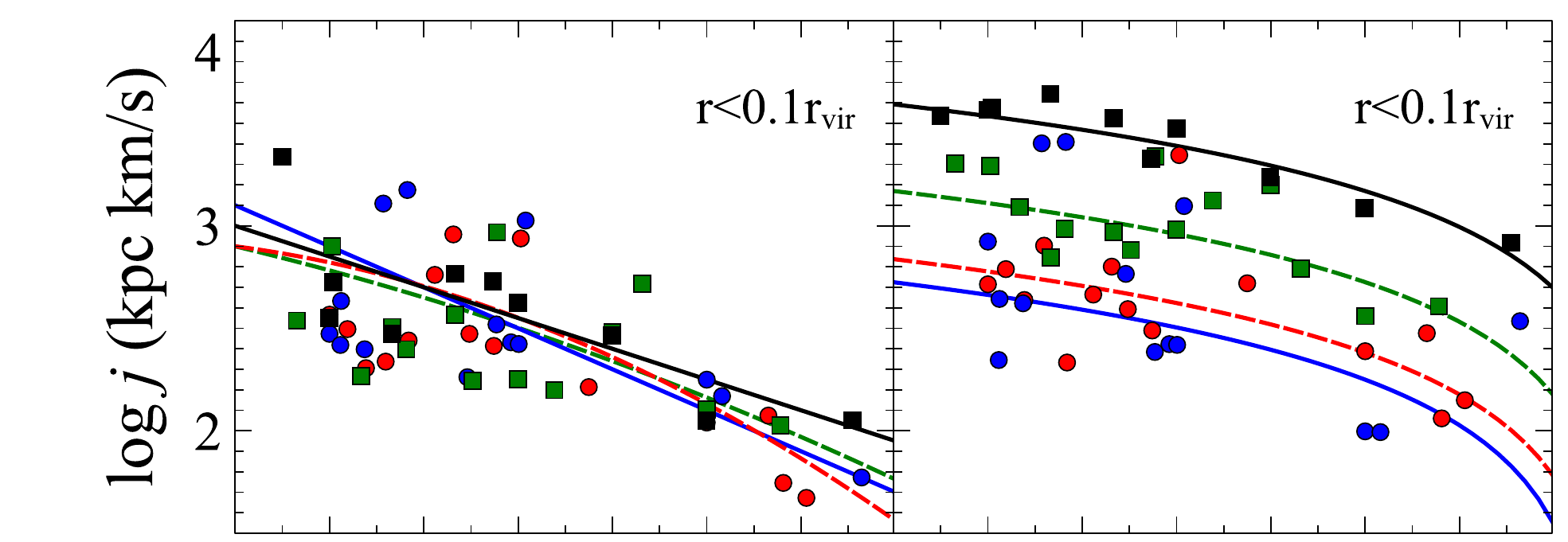}
\includegraphics[scale=0.4, trim=0 0 0 218, clip=true]{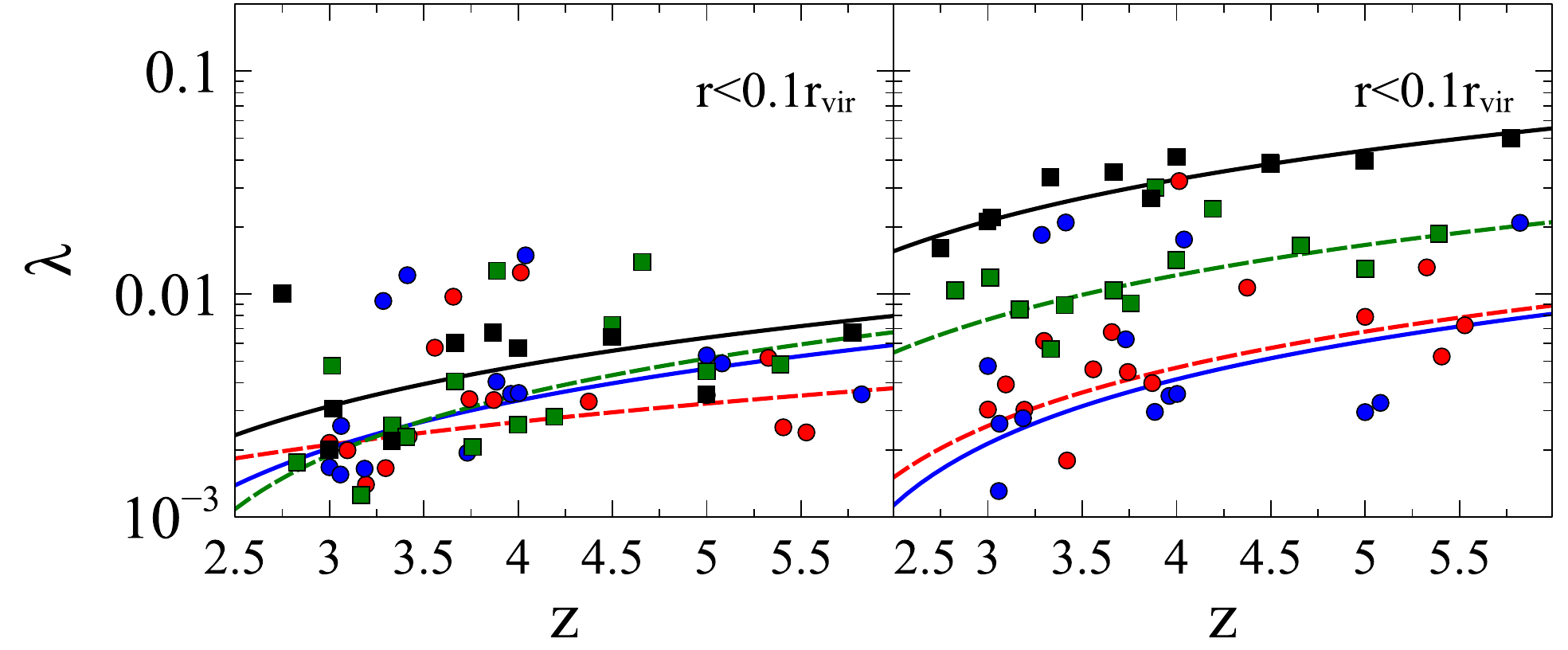}
\caption{Top two rows: Evolution of the specific angular momentum (top row) and spin parameter (second row) of the cluster dark matter (left) and gas (right) outside 0.1 $r_{\rm vir}$. Bottom two rows: Evolution of the specific angular momentum (third row) and spin parameter (bottom row) of the cluster dark matter (left) and gas (right) within 0.1 $r_{\rm vir}$. An approximate fit to the \AMRFID\ (\SPHFID) simulation is presented with solid black (blue) lines where we mark the measured values in an 
output by squares (circles). An approximate fit to the \AMRQSO\ (\SPHQSO) simulation is presented with dashed green (red) lines.}
\label{fig_halo_evo_ang}
\end{figure}

In the evolution of the halo mass, only the baryons appear affected by AGN feedback,  resulting in a drop of roughly 30\% in the amount of baryons in the cluster. With AGN feedback, there is more scatter in the evolution of the baryon fraction, but the AMR and SPH results are more consistent than in the \FID\ runs. As we have seen, the AGN leads to roughly the same halo gas entropy,  thus better agreement in the baryon gas fraction is expected. 

The specific angular momentum (sAM), $j$, and spin parameter, $\lambda = j/(\sqrt{2}r_{\rm vir}v_{\rm vir})$ (e.g., Bullock et al. 2001), are presented in \fig{fig_halo_evo_ang}, where $v_{\rm vir}$ is the virial velocity. The sAM of dark matter in the outer halo ($r>0.1r_{\rm vir}$) grows at a smooth rate over the course of the simulation. Except at very early times, the growth of sAM is unaffected by AGN feedback. By normalizing by the virial radius and velocity it is clear that $\lambda$ for the outer halo is roughly constant in time with a value near 0.034, consistent with tidal torque theory (e.g., White 1984; Bett et al. 2007). Accreting dark matter slowly exchanges its sAM with the halo, leading to a slower rate of growth of sAM in the central $0.1r_{\rm vir}$, corresponding to the galactic scale. Even on small scales the growth of sAM is not impacted by AGN feedback. As seen in Danovich et al. (2015), the dark matter sAM on small scales is roughly an order of magnitude below the accreting dark matter value.

The gas sAM behaves quite differently. Before $z=4$ the sAM is roughly constant in the outer halo, until at $z=4$ it has roughly the same sAM as dark matter, at which point it grows faster than the dark matter sAM. By $z=3$ the gas sAM is a factor of 3 above that of dark matter. At very early times, the lower gas fractions seen previously has led to lower sAM in the \QSO\ simulations. This suggests that at early time much of the angular momentum generation is via accretion of high sAM gas. However, by $z=4$ the outer gas sAM has the same value in all simulations, which suggests that a significant amount of the torque generating the angular momentum is gravitational in nature, consistent with tidal torque theory. The spin parameter is also roughly constant for gas in the outer halo, staying a factor of roughly two above the dark matter, except at $z=4$. Considering we are presenting the angular momentum of all gas, and not just cold gas, it is encouraging to see the agreement with Danovich et al. (2015), who see that cold gas in the outer halo has spin parameters of roughly three times that of dark matter. A more detailed study of the angular momentum of the different gas phases is beyond the scope of this work.

At small radii, however, there are significant differences between \SPHFID\ and \AMRFID, where the SPH sAM is consistently lower than in AMR. While SPH explicitly conserves angular momentum in the absence of artificial viscosity, Okamoto et al. (2003) demonstrated that due to its difficulty in modeling the layer between different fluids, standard SPH can transfer angular momentum from the dense disk to the hot diffuse halo. Additionally, Kaufmann et al. (2007) showed that disks in low-resolution SPH simulation, of comparable resolution to this and other cosmological simulations, unphysically transport angular momentum to the outer medium. However, AMR does not explicitly conserve angular momentum, its grid has a preferred direction, and it suffers from advection errors. Thus AMR also suffers from spurious angular momentum dissipation. Yet, we see here that the central 0.1$r_{\rm vir}$ has an order of magnitude more angular momentum in AMR than in SPH. Thus, the SPH resolution issues may be playing the dominant role. Additionally, it is possibly that the more collimated filament streams in AMR incur more sAM from tidal torques since it has a high quadrupole moment (e.g., Danovich 2015), which is also weakly supported at high $z$ and large radii.

The inclusion of AGN feedback also leads to markedly different behavior in the inner region for the two methods. In SPH, the inclusion of feedback has led to a combination of accretion of low angular momentum gas, and removal of this low AM gas from the central region. This leads to a net increase in the central SPH sAM. However, in AMR the sAM drops when we include AGN feedback. The two mechanisms leading to an increased sAM in SPH are also operating in AMR. Thus the difference is that gas heated via AGN feedback moves the cold stream filaments outwards, inhibiting gas accretion from these outer regions and transfering the filamentary gas' angular momentum through shocks into the outer halo. While the sAM is increasing for all simulations, the spin parameters are dropping. Thus the angular momentum buildup in the central region is not keeping pace with the growth of the cluster, regardless of feedback processes.

\subsubsection{Star formation}
\begin{figure}[t!] 
\centering
\includegraphics[scale=0.54]{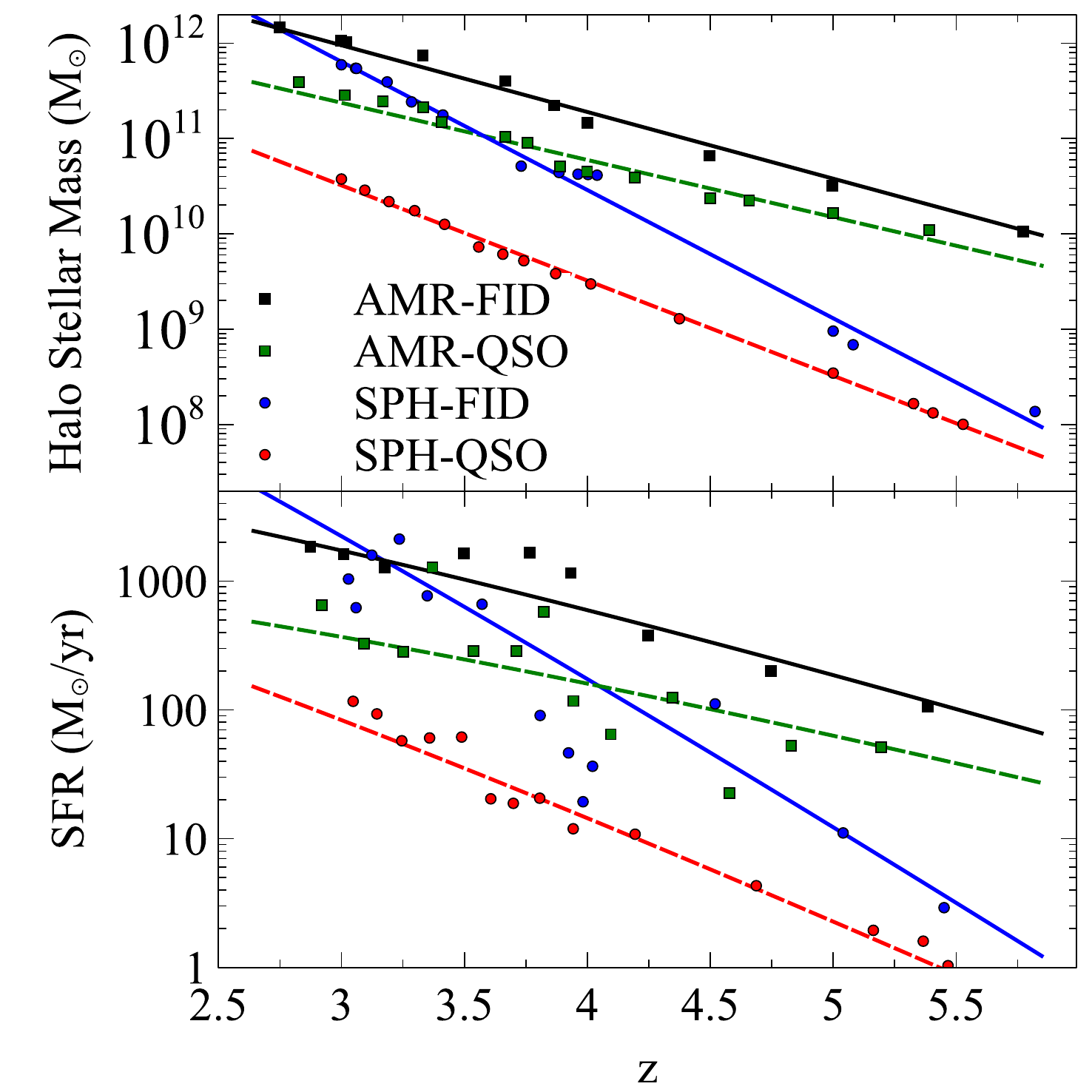}
\caption{Top: Evolution of the halo stellar mass of the cluster. The \AMRFID\ (\SPHFID) simulation  is presented with a solid black (blue) line where we mark the measured values in a 
output by squares (circles). The \AMRQSO\ (\SPHQSO) simulation is presented with a dashed green (red) line. Lines are fitted exponential laws of the same form as \eqn{massfit}. 
Bottom: Evolution of the halo star formation rate, taken as the change in halo 
stellar mass between adjacent outputs. Lines, given by the derivative of the exponential fits in the top plot, follow the same color scheme.}
\label{fig_sfr}
\end{figure}

In \fig{fig_sfr} we present the halo stellar mass and star formation rates for the four simulations containing stars. We fit the halo stellar mass functions with exponential power laws, similar to the total halo mass fits in \sect{haloevo}. The values for the fit are given in \tabl{tab_sfr_fit}. Up to $z=3$, it is always the case that the AMR simulations have more stars than in the SPH simulations. This is mostly due the gas reaching higher densities in AMR, as shown in \fig{fig_fid_phase5}. Since the star formation rate scales with $\rho^{1.5}$, the higher density gas in AMR leads to a much larger stellar masses. The inclusion of AGN feedback leads to a significantly reduced amount of stars, due to removal of low angular momentum gas, and heating of surrounding halo gas. Unsurprisingly, AGN have the biggest impact in SPH, where there is an average increase in the central angular momentum and more large-scale heating, which reduces the star formation rate in surrounding structures outside the virial radius, and thus a lower amount of stellar mass is accreted.

\begin{center}
\begin{longtable}{l cc}
\caption[Stellar Mass Fitting parameters]{Halo stellar mass fitting parameters} \label{tab_sfr_fit} \\
\hline \hline
\multicolumn{1}{l}{\textbf{Run}} &
\multicolumn{1}{c}{$\mathbf{M_{3*}}$} &
\multicolumn{1}{c}{$\mathbf{\alpha_*}$} \\ \hline
\endfirsthead
AMR-FID 	        & $9.9\times10^{11}\Msun$      	& 1.6		\\
AMR-QSO 	& $2.2\times10^{11}\Msun$      	& 1.4 	\\
SPH-FID   	& $6.4\times10^{11}\Msun$      	& 3.1 	\\ 
SPH-QSO   	& $3.2\times10^{10}\Msun$      	& 2.3 \\ \hline
\end{longtable}
\end{center}
\vspace{-20pt}

The average SFR is provided in \fig{fig_sfr} as well, taken as the change in halo stellar mass divided by the change in time between snapshots. We include the parametric fits, equivalent to the derivative 
of the exponential fits, given by \vspace{10pt}
\begin{eqnarray}
\frac{dM_*}{dt} &=& \alpha_* M_{*} H_0 (1+z)\sqrt{\Omega_M(1+z)^3 + \Omega_\Lambda}, \nonumber \\
&=&14.3 \Msun \mbox{ yr}^{-1} \left(\frac{\alpha_*}{2}\right)\left(\frac{M_{*}}{10^{11} \Msun}\right)\left(\frac{h}{0.7}\right) \times \nonumber \\
& & (1+z)\sqrt{\Omega_M(1+z)^3+\Omega_\Lambda},
\end{eqnarray}
where $\alpha_*$ is the exponential slope parameter, and $M_*$ is the stellar mass at a particle redshift, which follows an exponential growth fit equivalent to \eqn{massfit}, 
whose parameters are given in \tabl{tab_sfr_fit}. The star formation rate is consistently lower in SPH, but grows faster than in AMR. Since the rate is lower, there is more gas
present at later times from which to form stars in SPH. For the \FID\ runs, this is consistent with a longer cooling time in SPH. In the \QSO\ runs, the SPH feedback has a larger impact on large scales than in AMR,
thus the SFR is lower  at high redshifts. However, at late times the halo gas becomes consistent between the two methods, which results in the SFRs approaching 
better agreement. \vspace{10pt}

 \begin{figure}[t!] 
\centering
\includegraphics[scale=0.54]{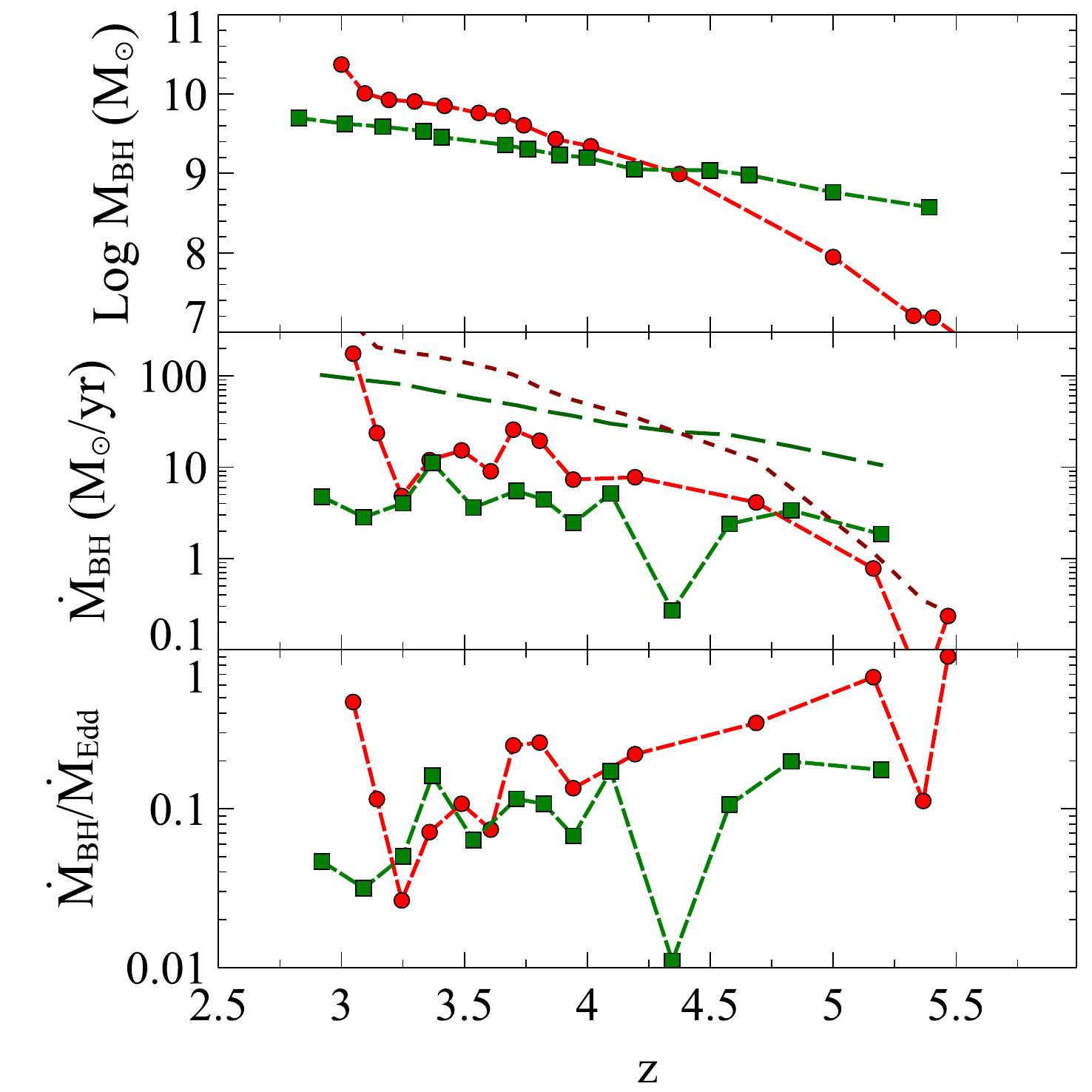}
\caption{Top: Evolution of the sum of the mass of the halo's three most massive black holes. The \AMRQSO\ (\SPHQSO) simulation is presented with a dashed green (red) line, where we mark the measured values in an output by squares (circles). Middle: Evolution of the total growth rate of the three most massive halo black holes, taken as the change in mass between adjacent outputs, and thus includes mass growth due to merging. Lines follow the same color scheme as above. The dark green long-dashed (dark red dotted) line indicates the sum of the Eddington accretion rates for these 3 \AMRQSO\ (\SPHQSO) black holes. Bottom: Same as middle, but normalized by the Eddington accretion rate sum.}
\label{fig_bh}
\end{figure}

\subsubsection{Cluster black hole growth}
In Figure \ref{fig_bh}, we show the evolution of the sum of the mass of the halo's three most massive black holes, and their net growth rate, both in absolute terms and normalized by the sum of their Eddington accretion rate, each given in \eqn{eqedacc}.  Since the star formation rate is lower in the SPH runs, it takes longer before the stellar density reaches the threshold to trigger the formation of a BH. Thus, the SPH BHs form later and are less massive at very high redshifts. Since BHs are generated at the limit of density resolution, this could be a large source of discrepancy between the two codes. It is thus very interesting that the halo environment is so similar at late times. Due to the delayed formation, the SPH BHs total growth rate is near Eddington, until the BH masses approaches that of their AMR counterparts, at which point the growth rates are in good agreement, consistent within a factor of two in absolute terms, or roughly the same value relative to their Eddington rate. This is consistent with the AGNs entering the self regulation phase. At $z<3.2$ a medium-sized SPH black hole falls into a clump (see \fig{fig_qso_images3}), where its accretion increases very quickly. This accretion will remain high until the pressure is increased enough to turn off accretion, and this is what occurs at $z=3$ in \fig{fig_qso_prof3}. During the period of best agreement between the two methods, $3.2 < z < 4$, we note that the baryon fractions are also in very good agreement (see \fig{fig_halo_evo}), but we see almost no correlation between the SFR during this time. Unfortunately, without sampling the star formation over smaller intervals, we are unable to determine whether fluctuations in star formation correlate with fluctuations in black hole accretion, suggested in other works (e.g., Dubois et al. 2013). Thus, for a self-regulation scenario, constraining AGN feedback with simulations will not come from the characteristics of the halo environment, but by the characteristics of the black holes themselves, their agreement with observed empirical relations, and the impact of AGN on the IGM and surrounding structure.
 
\section{Conclusions}\label{concl5}
To better understand the impact of AGN feedback on the formation and evolution of a large cluster and how this evolution can be biased by numerical effects, we have simulated the growth of a cluster from identical initial conditions in two different numerical methods. Using the AMR code \ramses\
and the SPH code \hydra\ we have attempted to match their radiative cooling, star formation, stellar feedback, and their black hole formation, growth, and energetic feedback processes, with the caveat that the fundamental differences between the methods will make modeling even simple structure formation not necessarily identical. By comparing these simulations with successively more complex baryonic physics, we have observed the following key points:
\begin{itemize}
\item Regardless of the treatment of baryonic processes, SPH consistently has a lower central entropy profile than AMR, with the sole exception being directly after 
an energetic feedback event. While this has been seen before for non-radiative simulations, comparisons with moving mesh codes have suggested that SPH codes may
have higher central entropy profiles.
\item Gas that is heated by the virial shock can efficiently cool at high redshift in simulations, although this cooling may not be captured numerically if an artificial viscosity is employed or if the shock width is not well resolved. Future work studying the resolution dependence of this feature will shed further light on this issue. 
\item AGN feedback reduces the baryonic fraction of halos by roughly 30\%, regardless of the numerical method. The baryon fraction is highly dependent on the location of the virial shock, as it sets the bottleneck for subsequent baryonic accretion.
\item AGN feedback leads to better agreement between the two methods on the thermal state of the halo gas, consistent with a picture of self-regulation. Regardless of the numerical method, the AGN will accrete matter until
moving most of the gas to entropies of roughly 100 keV cm$^2$, at which point subsequent accretion diminishes.
\item We see hints of AGN feedback impacting the termination point and orientation of filamentary cold flows in AMR, acting to push these streams beyond the impact radius of the AGN. In SPH, the filaments are more discrete, allowing the AGN heated gas to escape around the flows, with little impact on their location
or orientation. This may also explain the decrease in angular momentum of the central halo in AMR.
\item AGN feedback leads to a reduction in the total star formation rate of the cluster halo, by up to an order of magnitude by $z=3$. Future work comparing the two methods at lower redshift is needed.
\end{itemize}

In summary, AGN clearly play an important role in the evolution and regulation of cluster growth. Their possible observational impact is becoming clearer as surveys become larger, and hydrodynamic simulations become more complex. Further work exploring the detailed physical implications of AGN feedback and its interaction with the environments is essential for understanding the cosmic history of the Universe.

\ 

\ 

This research has been supported by the Balzan Foundation via the University of Oxford. M.L.A.R. was supported by NSF grant AST11-03608 and the National Science and Engineering Research Council of Canada. E.S. was also supported by the National Science Foundation under grant AST11-03608 and NASA theory grants NNX09AD106 and NNX15AK826. R.J.T. is supported by a Discovery Grant from NSERC, the Canada Foundation for Innovation, the Nova Scotia Research and Innovation Trust, and the Canada Research Chairs Program. Simulations were conducted with the CFI-NSRIT funded St Mary's Computational Astrophysics Laboratory, and the High Performance Computing environment at the University of Leicester. M.L.A.R would also like to thank Debora Sijacki for very insightful discussion, and the International Balzan Prize Foundation for contributing to the breadth of this work.

\end{document}